\documentclass[mksc,unblindrev]{informs4}
\OneAndAHalfSpacedXI

\usepackage{amssymb,amsmath,amsxtra,amstext,xfrac,mathtools}
\usepackage{graphicx,ccaption,booktabs,enumerate,paralist,mdwlist}
\usepackage{dsfont,verbatim,latexsym}
\usepackage{bbm}
\usepackage{bm}
\usepackage{longtable,multirow,threeparttable}
\usepackage{float}
\usepackage{eurosym}
\usepackage[english]{babel}
\usepackage[utf8]{inputenc}
\usepackage{fancyhdr}
\usepackage{color}
\usepackage{booktabs}
\usepackage{amssymb}
\usepackage{caption}
\usepackage{subcaption}
\usepackage{algorithm,algorithmic}
\usepackage{xcolor}
\usepackage{xurl}
\usepackage{booktabs}
\usepackage{longtable}
\usepackage{float}
\usepackage{placeins}
\usepackage[colorlinks=TRUE,citecolor=MyDarkBlue,urlcolor=MyDarkBlue,linkcolor=MyDarkBlue]{hyperref}
\definecolor{MyDarkBlue}{RGB}{158,0,0}

\usepackage{natbib}
\bibpunct[, ]{(}{)}{,}{a}{}{,}%
\def\bibfont{\footnotesize}%

\TheoremsNumberedThrough
\ECRepeatTheorems
\EquationsNumberedThrough

\makeatletter
\def\@copyrightholder{} 
\def\ps@mkscheadings{%
    \let\@oddhead\@empty
    \let\@evenhead\@empty
    \let\@oddfoot\@empty
    \let\@evenfoot\@empty
}
\makeatother

\begin{document}
\TITLE{Behavioral Transfer in AI Agents: Evidence and Privacy Implications}
\RUNTITLE{Behavioral Transfer in AI Agents}

\ARTICLEAUTHORS{
\AUTHOR{ Shilei Luo$^1$\thanks{$^*$These authors contributed equally to this work and the authorship is determined alphabetically.}, Zhiqi Zhang$^{1*}$, Hengchen Dai$^2$, Dennis J. Zhang$^1$}
\AFF{
$^1$ Washington University in St. Louis, St. Louis, MO\\
$^2$ University of California, Los Angeles, Los Angeles, CA\\
l.shilei@wustl.edu, z.zhiqi@wustl.edu, hengchen.dai@anderson.ucla.edu, denniszhang@wustl.edu
}
}

\ABSTRACT{

AI agents powered by large language models are increasingly acting on behalf of humans in social and economic environments. Prior research has focused on their task performance and effects on human outcomes, but less is known about the relationship between agents and the specific individuals who deploy them. We ask whether agents systematically reflect the behavioral characteristics of their human owners, functioning as behavioral extensions rather than producing generic outputs. We study this question using 10,659 matched 
human-agent pairs from Moltbook, a social media platform where each autonomous agent is publicly linked to its owner's Twitter/X account. By comparing agents' posts on Moltbook with their owners' Twitter/X activity across features spanning topics, values, affect, and linguistic style, we find systematic transfer between agents and their specific owners. This transfer persists among agents without explicit configuration, and pairs that align on one behavioral dimension tend to align on others. These patterns are consistent with transfer emerging through accumulated interaction between owners (or owners' computer environments) and their agents in everyday use. We further show that agents with stronger behavioral transfer are more likely to disclose owner-related personal information in public discourse, suggesting that the same owner-specific context that drives behavioral transfer may also create privacy risk during ordinary use.  Taken together, our results indicate that AI agents do not simply generate content, but reflect owner-related context in ways that can propagate human behavioral heterogeneity into digital environments, with implications for privacy, platform design, and the governance of agentic systems.
}


\maketitle

\section{Introduction}\label{sec:intro}
Artificial intelligence (AI) agents are increasingly acting on behalf of humans in digital ecosystems. From managing workflows and drafting communications to participating autonomously in online communities, AI agents are taking on a growing range of tasks traditionally performed by their human owners \citep{wang2024survey,bick2026rapid,noy2023experimental,brynjolfsson2025generative}. As agents become embedded in these environments, a fundamental question arises: do AI agents merely produce generic outputs of large language models (LLMs), or do they systematically carry forward the behavioral heterogeneity of their human owners? If agents reflect owner-specific characteristics, then AI systems may not simply scale content generation; they may scale individual-level variation into ecosystem-level consequences, changing not just the volume but the structure of online interaction.

This paper provides one of the first large-scale empirical examinations of behavioral transfer between humans and their AI agents in a natural deployment setting. Our empirical context is Moltbook, a social platform launched in January 2026 where autonomous AI agents interact with one another without direct human intervention. These agents are built using OpenClaw, an open-source framework that enables users to locally deploy AI agents powered by LLM APIs. A key feature of OpenClaw is that agents can access personal context provided by their owners, interact via chat, and autonomously perform tasks using local tools such as browsers and spreadsheets.  Through these capabilities, agents are exposed to owner-specific information embedded in local files and task environments, and may accumulate rich information about their owners through routine use.  Importantly, Moltbook represents just one application within a broader usage ecosystem. The same agent that a user deploys for personal tasks---such as accessing files, drafting communications, or completing work-related activities---can also be directed to participate on Moltbook. As a result, by the time an agent appears on Moltbook, it may have already accumulated substantial owner-specific context through prior interactions. This creates a natural pathway through which human behavioral tendencies can be encoded in agents before they engage in social interaction at scale, and raises the possibility that such context may shape both agent behavior and the information agents reveal in public settings. 

On Moltbook, each agent is publicly linked to its human owner's Twitter/X account, allowing us to construct 10{,}659 matched human-agent pairs and directly compare their behavioral profiles. We investigate whether behavioral transfer occurs, how it arises, and what its broader implications are when such agents are deployed in social platforms.

To assess whether transfer occurs, we leverage the matched-pair structure to compare each agent's behavioral profile against its owner's independently recorded Twitter/X activity across 43 behavioral features spanning topics, values, affect, and style. Our identification strategy exploits cross-pair variation: if agents reflect their specific owners, features more prevalent in an owner's behavior should also be more prevalent in the behavior of that owner's agent. Consistent with this logic, we find robust positive associations between agents and their human owners across a wide range of behavioral features, and agents more closely resemble their specific human owners than randomly selected users. Crucially, because the owners' Twitter histories temporally strictly predate the deployment of these agents on the platform, reverse causality where the AI shapes the human's behavior is structurally ruled out. Any observed transfer must therefore stem from the human owner influencing the agent's output, or potentially from unobserved omitted variables driving both. These patterns are strongly consistent with agents reflecting owner-specific behavioral characteristics rather than generic model outputs.

We then ask how such transfer arises. Two broad perspectives frame this question. At one end, transfer between agents and their human owners may reflect deliberate configuration: owners specify instructions or stylistic guidelines, and similarity is confined to explicitly defined dimensions. At the other end, transfer may emerge through broader behavioral carryover: through repeated interaction between the agent and its owner during everyday use, agents are exposed to owner-specific language, preferences, and task contexts. Over time, this accumulated exposure can embed owner-specific characteristics across dimensions that owners never explicitly configured. Two empirical patterns help distinguish between these explanations. First,  we find that transfer remains present among agents without configured bios. Second, pairs with stronger transfer on one set of features also tend to exhibit stronger transfer on other, conceptually distinct features. Such cross-dimension coherence is difficult to reconcile with dimension-specific configuration, which would predict more localized similarity. Together, these patterns are most consistent with transfer emerging through mechanisms that extend beyond observable or dimension-specific configuration alone.

Finally, we examine the privacy implications of behavioral transfer. Using a rigorously validated LLM-as-judge detection pipeline, we identify unintended, owner-referential disclosures in public agent posts. We find that privacy leakage is a widespread phenomenon on the platform, with 34.6\% of agents surfacing sensitive personal information about their owners, ranging from occupational details to highly sensitive health and financial conditions, that was not provided in their public configurations. Crucially, we demonstrate that the likelihood of disclosure is systematically predicted by the degree of behavioral transfer. A one-standard-deviation increase in a pair's holistic transfer score is associated with a 1.32-percentage-point higher probability of an agent leaking private information. Extensive simulation-based analyses and automation proxy tests confirm this association is robust to both classification noise and the potential presence of human-controlled ``puppet'' accounts. Ultimately, agents that more closely mirror their owners' behavioral profiles are fundamentally more likely to surface their owners' private contexts in public discourse.

Taken together, these findings suggest that AI agents function as owner-specific behavioral extensions: their outputs reflect owner-related context through mechanisms that go beyond explicit or domain-specific configuration. These findings reframe how we should understand AI agents in social environments. Prior research has treated agents primarily as tools that affect their users, for example, by shaping attitudes, augmenting productivity, and enhancing creativity \citep{costello2024durably,noy2023experimental, doshi2024generative}. Our results highlight a complementary and consequential property: agents carry human behavioral characteristics forward and embed them in outputs that interact with the wider digital ecosystem. This extends far beyond entering public discourse. For instance, an autonomous agent deployed to navigate e-commerce platforms, negotiate with sellers, or schedule services may inadvertently leak latent preferences, financial constraints, or behavioral vulnerabilities that the human owner actively intended to withhold from third parties. Such carryover occurs systematically  when ordinary users deploy AI agents in natural settings, extending prior work on behavioral simulation in controlled, researcher-directed environments \citep{horton2023large,park2024generative,argyle2023out,park2023generative}.  This observation has a broader structural implication: platforms populated by such  agents do not produce homogeneous LLM-generated discourse but instead propagate the behavioral heterogeneity of their human users. 

Our findings also point to a distinct privacy mechanism. The economics of privacy has long recognized that information disclosure creates externalities \citep{acquisti2016economics,acemoglu2022too}, and prior work on AI-related privacy risks has shown that sensitive information may be extracted from LLM training data through adversarial probing \citep{carlini2021extracting,nasr2023scalable, lukas2023analyzing}, inferred by LLMs from linguistic patterns at scale \citep{staab2024beyond,mireshghallah2025position,peters2024llm}, and shared or monetized across platforms \citep{bergemann2022economics,acquisti2016economics}. These channels rely on external actors recovering or exploiting existing data. In contrast, the mechanism we document requires no external actor and operates through ordinary use: owner-specific context accumulated by agents through everyday interaction may both produce behavioral transfer and result in owner-related personal information surfacing in public discourse. Privacy risks, therefore, arise not only from data leakage or inference, but from the routine operation of agentic systems themselves. This risk may be amplified in agentic settings, where agents have limited ability to recognize when sharing is contextually inappropriate and readers of agent-generated content prefer personally contextualized responses \citep{mireshghallah2024confaide,zhang2024privacy}. Together, this work positions behavioral transfer as a fundamental property of agentic systems---one with consequences for how agents are designed, governed, and understood as participants in social environments  \citep{schluntz2024building, shavit2024practices, gabriel2024ethics}.

The remainder of the paper proceeds as follows.  Section~\ref{sec:data} describes the data. Section~\ref{sec:results} documents behavioral transfer between humans and agents. Section~\ref{sec:mechanisms} examines the channels through which transfer may arise. Section~\ref{sec:privacy} documents privacy disclosure as the principal implication. Section~\ref{sec:implications} discusses broader consequences and future directions.

\section{Empirical Setting and Data}\label{sec:data}

\subsection{Platform Description}\label{sec:platform}

\paragraph{OpenClaw: A General-Purpose Agent Framework.} OpenClaw is an open-source framework for building autonomous AI agents powered by LLMs.\footnote{OpenClaw source code and documentation: \url{https://github.com/moltbook/openclaw}. We reviewed the publicly available repository as of February 2, 2026, to understand the framework's architecture, specifically how configuration files are assembled into system prompts, how the gateway daemon schedules agent actions, and what information flows are possible between owners and agents.} The framework is designed as a general-purpose assistant: owners can deploy agents locally and use them across a wide range of tasks (e.g., drafting content, answering questions, managing workflows, and extended interaction). One input to agent behavior is explicit configuration: owners can edit local ``character files'' that specify the system prompt (i.e., a prompt that is read into the LLM's context and used in every LLM call). These files are assembled into the agent's system prompt at runtime and are not publicly visible. However, configuration files represent only one of several possible inputs to agent behavior. Because OpenClaw agents maintain persistent conversation sessions with their owners, they also accumulate context through ongoing interaction, for example through the topics an owner asks about, the style of their instructions, the feedback they provide on draft outputs, and the types of tasks they delegate over time. In addition, when agents are granted access to local tools, they may  draw on information from the owner's computing environment---such as documents, notes, or spreadsheets---further expanding the scope of owner-specific context available to the agent. By the time an agent is deployed in downstream applications or social platforms (e.g., Moltbook), its behavior may reflect both explicit configuration and this accumulated interactional and environmental context. Distinguishing between these channels is a central empirical question that we discuss in Section~\ref{sec:mechanisms}.


\paragraph{Moltbook:  A Social Platform for AI Agents} Moltbook is a social media platform launched on January 28, 2026, where OpenClaw-based agents operate as autonomous participants. Moltbook provides the social infrastructure (user profiles, a content feed, posts, comments, and votes) while OpenClaw provides the underlying agent framework.  Owners who have already configured an OpenClaw agent can connect it to Moltbook by linking a Twitter account for identity verification and providing an agent name and an optional public bio description.

As of February 2, 2026, the platform's public interface displayed approximately 1.6 million registered AI agents, and the vast majority had never posted. Our data collection on that date retrieved 86,497 posts authored by 20,894 distinct agents. Of these 20,894 agents who had authored at least one post, 17,745 (84.9\%) had linked Twitter accounts. Our analysis focuses on these posting agents with linked Twitter accounts, the population for whom human-agent behavioral comparison is directly possible.

\paragraph{Autonomous Operation.} Once deployed on Moltbook, agents operate autonomously without direct human control. The OpenClaw Gateway daemon runs a periodic scheduling mechanism that triggers each agent at configurable intervals. The platform nominally limits agents to at most one post per 30 minutes. At each trigger, the framework injects a prompt into the agent's active session, causing the agent to evaluate a predefined task checklist and autonomously execute relevant actions: browsing the feed, creating posts, commenting on other agents' content, and casting votes. All posted content is text-based and generated entirely by the agents' underlying LLMs. According to the OpenClaw architecture and platform documentation, humans cannot directly author posts, comment, or vote; they can only observe their agents' activity and influence behavior through configuration and off-platform interaction. The platform thus creates a setting in which AI agents are the sole participants in public discourse.

\paragraph{Human-Agent Linking.} A critical feature for our research is that each agent can be publicly linked to its human owner's Twitter account. This linking is verified through a ``verification tweet'' that users post on Twitter when claiming their agent. This public linkage enables us to match agents to their human owners and compare their behaviors across platforms.

\subsection{Data Collection and Sample Construction}

We collected Moltbook profiles and posts on February 2, 2026, and Twitter data on February 7, 2026. Moltbook content spans January 27 to February 2, capturing the platform's first week of activity after launch. Table~\ref{tab:sample_flow} summarizes the sample construction process.

\paragraph{Moltbook Data.} We collected data from Moltbook's public API covering all agents that had published at least one post as of February 2, 2026. Although approximately 1.6 million agents were registered on the platform, the vast majority never generated any posted content.\footnote{Moltbook agents are created automatically when a human owner registers; many owners never configured or activated their agents, resulting in a large number of dormant accounts with no posts.} Our extraction yielded 20,894 agents with at least one post, along with their profile metadata (agent name, bio description, creation date, follower count, and linked Twitter handle). From these agents, we collected 86,497 posts with content text, timestamps, upvote counts, and comment counts. Posts represent an agent's proactive, self-initiated expression: the agent chooses the topic, tone, and style independently.

\paragraph{Twitter Data.} For each agent with a linked Twitter handle, we queried the Twitter API v2 to retrieve the corresponding human's profile and tweets. We collected profile information (follower count, following count, account creation date, bio) and up to 10 of the most recent original tweets per user, excluding retweets and replies. Our data collection retrieved each user's most recent tweets as of February 7, 2026, without restricting to tweets posted before the agent's creation date; that is, the human behavioral profile reflects the owner's overall recent Twitter activity rather than only pre-deployment behavior. Of the 17,739 accounts for which profiles were successfully retrieved, 15,552 had at least one tweet. From these accounts, we collected 83,111 tweets spanning dates from 2007 to February 2026. 

\begin{table}[!t]\centering \scriptsize
\caption{Sample Construction}
\label{tab:sample_flow}
\setlength{\tabcolsep}{4.0mm}{
\begin{tabular}{lrr}
\hline
\hline \\[-1.8ex]
Step & $N$ Agents & $N$ Tweets \\
\hline \\[-1.8ex]
Agents with $\geq 1$ post (of ${\sim}$1.6M registered) & 20,894 & --- \\
With linked Twitter handle & 17,745 & --- \\
Twitter profile successfully retrieved & 17,739 & --- \\
Twitter profile with $\geq 1$ tweet & 15,552 & 83,111 \\
With $\geq 1$ non-verification tweet & 10,750 & 70,101  \\
Excluding 91 title-only agents & 10,659 & 69,559  \\
\hline \\[-1.8ex]
\textbf{Final analysis sample} & \textbf{10,659} & \textbf{69,559} \\
\hline
\hline \\[-1.8ex]
\end{tabular}}
\end{table}

\paragraph{Verification Tweet Removal.} A key data cleaning step is the removal of verification tweets. When users claim an agent on Moltbook, they are required to post a verification tweet on Twitter (e.g., ``I'm claiming my AI agent `AgentName' on @moltbook''). These tweets follow a standard template and do not reflect genuine human expression or interests. We classify a tweet as a verification tweet if it contains all three keywords: ``claiming,'' ``agent,'' and ``moltbook'' (case-insensitive). This filter removes 13,010 tweets (15.7\% of all tweets), leaving 70,101 tweets for analysis. The filter is conservative by design: requiring all three keywords simultaneously minimizes false positives, while the standardized verification template limits false negatives.

\paragraph{Final Sample.} Our final analysis sample consists of 10,659 human-agent pairs where: (1) the agent has at least one Moltbook post with content text; (2) the human's Twitter profile was successfully retrieved; and (3) the human has at least one non-verification tweet. We exclude 91 agents whose posts contain titles but no content text (text length = 0), as content-based behavioral features cannot be computed for these agents. These 10,659 agents authored 44,588 posts (of the 86,497 total posts by all 20,894 posting agents), and their matched human owners contributed 69,559 non-verification tweets.

\subsection{Behavioral Feature Construction}\label{sec:features}

Testing whether agents carry their owners' behavioral characteristics requires measuring behavioral profiles on both sides and comparing them. We construct 43 text-based features organized into four dimensions (topics, values, affect, and style) that together span the major axes along which individuals differ in online self-expression. This four-dimensional framework follows established practice in computational social science, where online behavioral characteristics have been decomposed into \textit{what} people talk about, \textit{what} they believe, \textit{how} they feel, and \textit{how} they communicate \citep{pennebaker2015development,boyd2022development,park2015automatic}. We describe each dimension below and provide complete keyword lists for dictionary-based features in Appendix~\ref{app:variables}. Table~\ref{tab:variables} lists the features used in our analysis as well as the dimensions they belong to. 

\begin{table}[!t]\centering \scriptsize
\caption{Variable Definitions}
\label{tab:variables}
\setlength{\tabcolsep}{2.5mm}{
\begin{tabular}{lp{9.5cm}}
\hline
\hline \\[-1.8ex]
Variable & Description \\
\hline \\[-1.8ex]
\multicolumn{2}{l}{\textbf{Topics} (6 features)} \\
Topic Rate$_{i,k}$ & Rate of topic-related keywords per 1,000 words. Six topics: \textit{Crypto}, \textit{AI}, \textit{Dev}, \textit{Trading}, \textit{Philosophy}, \textit{Meme}. Keywords listed in Appendix~\ref{app:variables}. \\
\hline \\[-1.8ex]
\multicolumn{2}{l}{\textbf{Values} (7 features)} \\[2pt]
\multicolumn{2}{l}{\textit{~~Moral Foundations (5)}} \\
Moral Foundation$_{i,f}$ & Rate of moral-foundation keywords per 1,000 words \citep{graham2013moral}. Five foundations: \textit{Care}, \textit{Fairness}, \textit{Loyalty}, \textit{Authority}, \textit{Sanctity}. \\[2pt]
\multicolumn{2}{l}{\textit{~~Political Ideology (2)}} \\
Political Ideology (LLM) & Score from $-1$ (left) to $+1$ (right) based on Gemini 2.5-Flash scoring of user $i$'s full text (see text for prompt details). \\
Political Ideology (Dict.) & Score based on asymmetric partisan phrases following \citet{gentzkow2010drives}; requires presence of political keywords. \\
\hline \\[-1.8ex]
\multicolumn{2}{l}{\textbf{Affect} (12 features)} \\[2pt]
\multicolumn{2}{l}{\textit{~~Sentiment (7)}} \\
Compound & Mean VADER compound score ($-1$ to $+1$) across an individual's posts or tweets \citep{hutto2014vader}. \\
Positive / Negative / Neutral proportion & Mean VADER proportion scores for each valence type. \\
Variability & Standard deviation of VADER compound scores across an individual's posts or tweets. \\
Pct Posts Positive / Pct Posts Negative & Fraction of posts/tweets with positive (compound $> 0.05$) or negative (compound $< -0.05$) sentiment. \\[2pt]
\multicolumn{2}{l}{\textit{~~Emotions (5)}} \\
Emotion Rate$_{i,e}$ & Rate of emotion keywords per 1,000 words \citep{ekman1992argument}. Five emotions: \textit{Anger}, \textit{Joy}, \textit{Fear}, \textit{Sadness}, \textit{Surprise}. \\
\hline \\[-1.8ex]
\multicolumn{2}{l}{\textbf{Style} (18 features)} \\[2pt]
\multicolumn{2}{l}{\textit{~~Language Complexity (5)}} \\
Avg Word Length & Mean number of characters per word. \\
Type-Token Ratio (TTR) & Number of unique words divided by total words. \\
Avg Sentence Length & Mean number of words per sentence. \\
Hapax Ratio & Fraction of words appearing exactly once. \\
Capitalization Ratio & Fraction of characters that are uppercase. \\[2pt]
\multicolumn{2}{l}{\textit{~~Communication Style (8)}} \\
Avg Text Length & Mean number of characters per post/tweet. \\
Question / Exclamation Rate & Number of ``?'' or ``!'' per post/tweet. \\
Hashtag / Mention / URL Rate & Number of hashtags, @-mentions, or URLs per post/tweet. \\
Emoji Rate & Number of emoji characters per post/tweet. \\
Formality & Index from $-1$ (informal) to $+1$ (formal) based on the ratio of formal to informal vocabulary. \\[2pt]
\multicolumn{2}{l}{\textit{~~Pronouns and Self-Reference (5)}} \\
I-words / We-words / You-words / They-words & Rate of pronoun category per 1,000 words. \\
Self-Focus Ratio & I-words / (I-words + We-words); captures individual vs.\ collective orientation. \\
\hline
\hline \\[-1.8ex]
\end{tabular}}
\begin{tablenotes}
\item \textit{Notes:} All 43 features are computed separately for humans (from Twitter tweets) and agents (from Moltbook posts), then compared at the pair level using Spearman rank correlations. Topic, moral foundation, and emotion rates use keyword dictionaries with prefix matching; full keyword lists appear in Appendix~\ref{app:variables}. Individual features within each sub-dimension (e.g., Sentiment, Language Complexity) are analyzed in the transfer tests reported in Tables~\ref{tab:topic}--\ref{tab:style}.
\end{tablenotes}
\end{table}

\paragraph{Topics (6 features).} Topic features capture \textit{what} people discuss. We tabulate word frequencies across the combined Twitter-Moltbook corpus, manually group high-frequency domain-specific terms into semantically coherent categories, and verify that no other topic reaches comparable corpus coverage. This yields six content domains (Crypto, AI, Development, Trading, Philosophy, and Meme/Slang). At least one of these six topics appears in the text of 91.8\% of agents and 79.3\% of humans in the matched sample. For each feature, we construct a keyword list (10--14 terms per feature) and measure each user's topic intensity as the keyword mention rate per 1,000 words.


\paragraph{Values (7 features).} Value features capture \textit{what} people believe along moral and political dimensions. We measure five moral foundations (Care, Fairness, Loyalty, Authority, and Sanctity) using keyword dictionaries adapted from the Moral Foundations Dictionary \citep{graham2013moral}, computing each foundation as the keyword rate per 1,000 words. For political ideology, we employ two complementary approaches. The first is a dictionary-based method using asymmetric partisan phrases following \citet{gentzkow2010drives}, which scores each user based on the relative frequency of left-leaning (47 phrases) versus right-leaning (51 phrases) vocabulary in their text. This method is grounded in established methodology but limited to the $\sim$10\% of users who employ overtly partisan vocabulary. The second is an LLM-based scoring method: we prompt Gemini 2.5-Flash to act as a political scientist and rate each user's pooled text on a continuous $[-1, +1]$ left--right scale, where $-1$ denotes strongly liberal, $0$ denotes centrist or non-political, and $+1$ denotes strongly conservative. The model is instructed to consider policy positions, partisan language, cultural attitudes, and economic views, and to assign scores near zero when posts are primarily about non-political topics. The LLM also returns a confidence rating (low, medium, or high) for each score. This method extends coverage to all users, including the majority whose political orientation is expressed through contextual signals rather than explicit keywords. We validate the LLM-based ideology measure against the dictionary-based method \citep{gentzkow2010drives} using both convergent validity and known-groups analyses (Appendix~\ref{app:political_validation}).

\paragraph{Affect (12 features).} Affect features capture \textit{how} people feel. We measure seven sentiment features using the VADER lexicon \citep{hutto2014vader}: compound, positive, negative, and neutral scores at the post level (averaged per user), sentiment variability (standard deviation of compound scores across posts), and the percentage of posts classified as positive or negative. We additionally measure five discrete emotions (anger, joy, fear, sadness, and surprise) using keyword dictionaries following \citet{ekman1992argument} and \citet{mohammad2013crowdsourcing}, computed as keyword rates per 1,000 words. Sentiment and emotion capture complementary aspects of affective expression: sentiment measures overall valence, while discrete emotions distinguish specific affective states that may exhibit different transfer patterns.

\paragraph{Style (18 features).} Style features capture \textit{how} people express themselves, independent of content. This dimension spans three sub-dimensions. \textit{Language complexity} (5 features) captures the sophistication and variety of vocabulary using standard stylometric measures \citep{tweedie1998variable}: average word length, type-token ratio (ratio of unique words to total words, a standard measure of vocabulary diversity), hapax ratio (proportion of words used only once), capitalization ratio, and average sentence length. \textit{Communication style} (8 features) captures expressive habits: seven surface-level rates (average text length, question rate, exclamation rate, emoji rate, URL rate, hashtag rate, and @-mention rate), plus a formality index computed as the ratio of formal to informal vocabulary following \citet{heylighen2002variation}. \textit{Pronouns and self-reference} (5 features) includes first-person singular (I-words), first-person plural (we-words), second-person (you-words), third-person (they/she/he-words), and a self-focus ratio (I-words / (I-words + we-words)), which are among the most reliable linguistic markers of personality and interpersonal orientation \citep{pennebaker2015development}.\footnote{We also constructed and tested 12 additional features 
We also constructed and tested 12 additional features approximating categories from the LIWC-22 framework \citep{boyd2022development}: psychological processes (cognitive processes, social processes, positive affect, negative affect, certainty, tentativeness), drives (achievement, power), and personal concerns (money, work, leisure), plus a risk-perception dictionary. 
Because we did not use the official LIWC software but instead built custom keyword dictionaries to approximate these categories, we exclude them from the primary feature battery. All 12 features showed significant transfer between AI agents and their corresponding human owners; including these 12 features does not change any substantive conclusions.}

Having defined the four behavioral dimensions, we describe the text preprocessing and aggregation procedures applied uniformly across all features.

\paragraph{Text preprocessing.} We apply uniform preprocessing to both human tweets and agent posts, with the choice of raw versus cleaned text determined by each feature's measurement logic. Communication-style rates (URL, emoji, hashtag, @-mention, question mark, and exclamation mark rates) and average text length are computed from \textit{raw} text, because cleaning would remove the very characters being counted. Sentiment (VADER) scores are also computed on raw text, since VADER's lexicon leverages capitalization, emoji, and punctuation as sentiment signals \citep{hutto2014vader}. All remaining features (topics, moral foundations, emotions, pronouns, complexity, formality, and political ideology) are computed from \textit{cleaned} text. The cleaning pipeline lowercases the text, removes URLs, strips non-ASCII characters (including emoji), removes hashtag (\#) and mention (@) symbols so that the underlying words (e.g., ``bitcoin'' from ``\#bitcoin'') contribute to keyword and pronoun counts, and normalizes whitespace. The one exception is capitalization ratio, which uses a case-preserved variant of the cleaned text (URLs and non-ASCII removed, but original casing retained). Tokenization uses whitespace splitting on the cleaned text; all keyword and pronoun rates are expressed as counts per 1,000 words.

\paragraph{Aggregation and transparency.} For each human, we pool all non-verification tweets into a single text corpus and compute each feature (e.g., keyword rate per 1,000 words, mean sentiment score) from that pooled corpus. Similarly, for each agent, we pool all post content (excluding titles) into a single corpus. This yields one score per feature per individual, enabling pairwise comparison across matched human-agent pairs. The four-dimensional framework and the specific features within Values, Affect, and Style were selected \textit{ex ante} based on established text-analysis frameworks. The six topics within the Topics dimension were identified inductively from the corpus to reflect the platform's dominant discourse areas; their keyword dictionaries were then fixed prior to all transfer analyses. We report results for all 43 pre-specified features. The only measure examined but not included in the primary 43-feature battery is the Elite-Cue political ideology measure \citep{barbera2015birds}, which we exclude because it captures topical mention overlap rather than ideological transfer and has a very small effective sample ($N = 90$); results are reported in Appendix~\ref{app:political_validation}.

\subsection{Summary Statistics}

Table~\ref{tab:summary} reports descriptive statistics for a subset of features within each dimension based on the 10,659 matched human–agent pairs. Complete results for all 43 features appear in Table~\ref{tab:descriptive_full} in Appendix \ref{app:descriptive}. 

Panel~A in Table~\ref{tab:summary} highlights differences in platform activity. Human tweets are short (median 106 characters), reflecting Twitter’s format constraints, whereas agent posts are substantially longer (median 471 characters), consistent with Moltbook’s forum-style interface. Posting activity varies widely among agents (SD = 21.3, maximum = 2,086 posts), whereas the number of human tweets per user is capped at 10 by construction in our data collection. Notably, each behavioral feature is aggregated to a single score per individual, and the transfer analysis is conducted at the pair level using Spearman correlations. As a result, highly active agents do not enter the analysis as multiple observations, although their feature estimates may be measured with greater precision than those of less active agents. As explained below, our robustness checks control for the number of agent posts. 

Panels~B–D summarize distributions of topical, value, and affective features. Within the six most common topics identified from the Twitter-Moltbook corpus, the relative ordering of topic frequencies is similar across platforms: AI appears most frequently on both sides, followed by Development and Crypto, with Trading and the remaining topics appearing at lower rates. The largest distributional differences arise in affect. Agents exhibit substantially more positive sentiment than their owners (mean compound sentiment 0.447 vs.\ 0.129), with higher joy keyword rates and lower anger rates, a pattern consistent with the positivity bias commonly observed in language-model outputs \citep{argyle2023out}.

Panel~E reports stylistic features. Formality levels are similar across platforms (means near $-0.25$), suggesting that neither Twitter nor Moltbook encourages markedly formal language. Agents use first-person singular pronouns at higher rates (35.4 vs.\ 26.2 per 1,000 words), consistent with persona-style narration typical of conversational AI systems. Agents also exhibit lower type–token ratios (0.706 vs.\ 0.781), which likely reflects the longer length of agent posts.

\begin{table}[!t]\centering \scriptsize
\caption{Summary Statistics ($N = 10{,}659$ Human-Agent Pairs)}
\label{tab:summary}
\setlength{\tabcolsep}{1.8mm}{
\begin{tabular}{lrrrrrrrrr}
\hline
\hline \\[-1.8ex]
& \multicolumn{4}{c}{Human (Twitter)} & \multicolumn{4}{c}{Agent (Moltbook)} \\
\cmidrule(lr){2-5} \cmidrule(lr){6-9}
Variable & Mean & SD & Med & [Min, Max] & Mean & SD & Med & [Min, Max] \\
\hline \\[-1.8ex]
\multicolumn{9}{l}{\textbf{Panel A: Platform Activity}} \\
\hline \\[-1.8ex]
Posts/tweets & 6.5 & 3.4 & 8 & [1, 10] & 4.2 & 21.3 & 2 & [1, 2086] \\
Text length (chars) & 116 & 64 & 106 & [1, 308] & 639 & 1186 & 471 & [0, 101k] \\
\hline \\[-1.8ex]
\multicolumn{9}{l}{\textbf{Panel B: Topics (per 1,000 words)}} \\
\hline \\[-1.8ex]
Crypto & 11.54 & 35.5 & 0 & [0, 1000] & 9.49 & 21.4 & 5 & [0, 1000] \\
AI & 44.64 & 82.9 & 21 & [0, 1222] & 44.09 & 66.7 & 29 & [0, 1000] \\
Development & 9.26 & 29.6 & 0 & [0, 1000] & 7.10 & 16.9 & 1 & [0, 500] \\
Trading & 4.17 & 15.9 & 0 & [0, 500] & 3.86 & 10.6 & 0 & [0, 200] \\
\hline \\[-1.8ex]
\multicolumn{9}{l}{\textbf{Panel C: Values (per 1,000 words)}} \\
\hline \\[-1.8ex]
Care/Harm & 1.64 & 5.81 & 0 & [0, 125] & 4.33 & 8.28 & 0 & [0, 118] \\
Fairness & 0.59 & 11.11 & 0 & [0, 1000] & 0.38 & 2.78 & 0 & [0, 200] \\
Loyalty & 1.62 & 6.51 & 0 & [0, 200] & 4.13 & 9.48 & 0 & [0, 143] \\
Authority & 1.14 & 5.30 & 0 & [0, 250] & 1.91 & 6.37 & 0 & [0, 333] \\
Sanctity & 1.19 & 5.35 & 0 & [0, 200] & 1.14 & 3.80 & 0 & [0, 111] \\
Political Ideology ($-$1,+1) & .007 & .143 & 0 & [$-$1.0, 1.0] & .001 & .068 & 0 & [$-$1.0, 1.0] \\
\hline \\[-1.8ex]
\multicolumn{9}{l}{\textbf{Panel D: Affect}} \\
\hline \\[-1.8ex]
Sentiment & .129 & .223 & .09 & [$-$.95, .97] & .447 & .412 & .51 & [$-$1, 1] \\
Negative & .037 & .053 & .02 & [0, 1.0] & .034 & .041 & .03 & [0, 1.0] \\
Positive & .090 & .092 & .08 & [0, 1.0] & .112 & .078 & .10 & [0, .92] \\
Anger (per 1k) & 1.41 & 11.98 & 0 & [0, 500] & 0.50 & 2.89 & 0 & [0, 118] \\
Joy (per 1k) & 0.99 & 7.86 & 0 & [0, 333] & 3.32 & 10.03 & 0 & [0, 167] \\
Fear (per 1k) & 0.33 & 3.77 & 0 & [0, 176] & 0.25 & 1.51 & 0 & [0, 71] \\
\hline \\[-1.8ex]
\multicolumn{9}{l}{\textbf{Panel E: Style}} \\
\hline \\[-1.8ex]
Formality ($-$1,+1) & $-$.249 & .455 & 0 & [$-$1, 1] & $-$.262 & .490 & 0 & [$-$1, 1] \\
Word length & 4.93 & 0.92 & 4.8 & [1.1, 23] & 5.36 & 2.46 & 5.2 & [1.0, 62] \\
Type-token ratio & .781 & .130 & .80 & [0, 1] & .706 & .190 & .73 & [0, 1] \\
I-words (per 1k) & 26.2 & 38.8 & 13 & [0, 1000] & 35.4 & 35.3 & 29 & [0, 1000] \\
We-words (per 1k) & 4.85 & 11.2 & 0 & [0, 200] & 6.55 & 11.8 & 0 & [0, 167] \\
\hline
\hline \\[-1.8ex]
\end{tabular}}
\begin{tablenotes}
\item \textit{Notes:} 44,588 agent posts; 69,559 human tweets. For brevity, this table displays a subset of features within each dimension; complete descriptive statistics for all 43 features appear in Appendix Table~\ref{tab:descriptive_full}, and all features are defined in Table~\ref{tab:variables}. Keyword rates (Panels B--E) are computed per individual by pooling all of that individual's posts or tweets and dividing keyword counts by total word count, expressed per 1,000 words. Moral foundation rates (Panel C) per \citet{graham2013moral}; sentiment compound score (Panel D) via VADER \citep{hutto2014vader}; emotion keyword rates (Panel D) follow \citet{ekman1992argument}. Zero medians reflect sparse keyword-based rates.
\end{tablenotes}
\end{table}

\section{Behavioral Transfer Results}\label{sec:results}

Do AI agents reproduce the behavioral characteristics of the specific humans who deploy them? This section documents behavioral transfer between owners and their agents across 43 features spanning four dimensions (topics, values, affect, and style), using 10,659 matched human--agent pairs. We focus here on measuring the extent of transfer in observable behavior; the mechanisms that generate these patterns are examined later in Section~\ref{sec:mechanisms}.

We quantify behavioral transfer using Spearman rank correlations between matched human--agent pairs.\footnote{We use Spearman rather than Pearson correlations because many features exhibit skewed distributions with zero inflation.} Bootstrap confidence intervals (1,000 replications) are computed for each estimate. With 43 features tested,  we apply the Benjamini--Hochberg procedure \citep{benjamini1995controlling} to control the false discovery rate at 5\%.

Before presenting detailed results for each of the 43 features, we first summarize the overall prevalence of transfer.  Table~\ref{tab:overview} reports the share of features within each dimension of behavioral characteristics that exhibit statistically significant human--agent correlations. Behavioral transfer is widespread: across 43 features capturing what people discuss, what they believe, how they feel, and how they express themselves, 37 features (86.0\%) show significant human-to-agent correlation after multiple-testing correction. Transfer appears in all four behavioral dimensions: topics, values, affect, and style. This breadth suggests that agents systematically reflect multiple aspects of their owners' behavioral profiles rather than producing generic LLM outputs.

\begin{table}[!t]\centering \scriptsize
\caption{Overall Evidence of Behavioral Transfer Between AI Agents and Humans}
\label{tab:overview}
\begin{threeparttable}
\setlength{\tabcolsep}{6.0mm}{
\begin{tabular}{lcccc}
\hline
\hline \\[-1.8ex]
Dimension & Sig./Total & Rate & Median $|\rho|$ & Range $|\rho|$ \\
\hline \\[-1.8ex]
Topics (what you discuss) & 5/6 & 83\% & 0.101 & [0.017, 0.166] \\
Values (what you believe) & 6/7 & 86\% & 0.043 & [0.023, 0.087] \\
Affect (how you feel) & 10/12 & 83\% & 0.064 & [0.001, 0.153] \\
Style (how you express) & 16/18 & 89\% & 0.077 & [0.002, 0.174] \\
\cmidrule{1-5}
\textit{Total} & {37/43} & {86.0\%} & {0.067} & {[0.001, 0.174]} \\
\hline
\hline \\[-1.8ex]
\end{tabular}}
\begin{tablenotes}
\item \textit{Notes:} Median $|\rho|$ and Range $|\rho|$ are computed across all features in each dimension (significant and non-significant). Significance counts reflect Benjamini-Hochberg (BH) corrected $p$-values across all 43 features ($\alpha = 0.05$); all features significant at the unadjusted 5\% level remain significant after BH correction, and non-significant features likewise remain non-significant. Topics = 6 most popular topics. Values = 5 Moral Foundations + 2 Political Ideology measures (Gentzkow-Shapiro dictionary and LLM-based). Affect = Sentiment + Emotions. Style = Language Complexity + Communication Style + Pronouns. Sample sizes vary by feature: base $N = 10{,}659$ pairs; complexity $N = 8{,}751$ (pairs with $\leq$10 total words on either side excluded). See Tables~\ref{tab:topic}--\ref{tab:style} for $p_{\text{adj}}$ values for each feature.
\end{tablenotes}
\end{threeparttable}
\end{table}

The magnitudes of the observed correlations are modest. Among the 37 features that exhibit statistically significant transfer, Spearman correlations range from approximately $\rho = 0.020$ (sadness) to $\rho = 0.174$ (capitalization ratio), with a median of $|\rho| = 0.068$. Although these effect sizes are not large, they are notable given that human and agent behavior is measured on different platforms with distinct norms and content formats, and human behavioral profiles are estimated from a limited number of tweets (1--10 per user), introducing measurement noise that may attenuate observed correlations toward zero. For reference, prior work estimates that the same individual's stylistic behavior is consistent across platforms at approximately $\rho = 0.10$--$0.15$ \citep{danescu2011chameleons}, providing a benchmark against which the human--agent transfer we observe can be assessed.

Several additional analyses support the robustness of the observed transfer against potential confounds and measurement artifacts. First, because both humans and agents inhabit overlapping topical communities (e.g., crypto, AI, development),  platform-level similarity could generate positive correlations even if agents do not reflect the behavior of their specific owners. To address this concern, we use permutation tests that randomly reassign agents to humans 10,000 times, preserving the marginal behavioral distributions of both populations while destroying pair-specific links. Comparing observed correlations with this permuted baseline allows us to assess whether transfer reflects owner-specific pairing rather than platform-level overlap. The same 37 features with significantly positive Spearman correlations all exceed the permuted baseline (see Appendix~\ref{app:permutation}), consistent with the observed transfer reflecting owner-specific pairing rather than platform-level behavioral similarity.

Second, we validate the transfer results using Sentence-BERT \citep{reimers2019sentence}, which provides a single high-dimensional semantic representation of each user's text. We concatenate all tweets (or posts) for each human (or agent) and embed the combined text into a 384-dimensional vector, then compute cosine similarity between the matched human--agent embedding pairs. Unlike the 43 individual features above, each of which captures a specific behavioral dimension, cosine similarity over Sentence-BERT embeddings provides a holistic measure of textual similarity that reflects overall overlap in vocabulary, topics, and expressive style simultaneously, without relying on predefined keyword lists or behavioral categories. Matched human--agent pairs exhibit substantially higher similarity than randomly paired humans and agents (cosine similarity $=0.288$ vs.\ $0.205$, $p<0.0001$), indicating that transfer extends beyond the specific dimensions captured by our 43 constructed features into the broader semantic structure of the texts.  Full details appear in Appendix \ref{app:bert}.

Finally, because some agents may be human-controlled ``puppets'' rather than truly autonomous AI \citep{li2026moltbook}, we conduct a comprehensive robustness analysis. The core concern is that if some owners post directly through the Moltbook interface rather than delegating to their OpenClaw agent, observed human-agent similarity would reflect identity rather than genuine behavioral transfer.

We draw on the social media automation detection literature \citep{chu2012detecting, varol2017online, cresci2018social}, which identifies behavioral signatures that distinguish autonomous agents from human posters. The key intuition is that autonomous AI agents operate continuously around the clock, posting at all hours with a near-uniform temporal distribution, whereas human-controlled accounts cluster activity during waking hours and go silent overnight. Consistent with autonomous operation, agents in our data exhibit a near-uniform hourly posting distribution (Shannon entropy = 0.984, where 1.0 indicates perfect uniformity), with 26.9\% of posts occurring between midnight and 8AM Eastern time.

Building on this, we conduct two exclusion-based tests that directly remove agents classified as likely human-controlled. The first applies a multi-feature temporal classifier: an agent is excluded if it meets at least three of four suspicious criteria (e.g., low overnight posting, excessively long inactive periods), flagging 942 agents (8.8\% of the matched sample). The second applies a more aggressive single-metric filter based on inter-post interval variability (CoV $> 1.0$), identifying and excluding a broader set of 2,142 agents (20.1\%). Despite dropping up to a fifth of the sample, these two exclusions leave 36 and 32 of the 37 originally significant features intact, respectively, with negligible attenuation in effect sizes (mean $|\Delta\rho| \leq 0.009$; Table~\ref{tab:puppet_full}).

To ensure our results do not hinge on specific exclusion cutoffs, we address the fundamental empirical challenge that the true boundary between autonomous agents and human-controlled accounts is unobservable. Rather than relying on any single definition of automation, we organize 18 proxy indicators from the literature into five dimensions: circadian rhythm, posting volume, inter-post interval regularity, content consistency, and received engagement. By splitting the matched sample at the median for each indicator, we evaluate whether the transfer signal is sensitive to varying behavioral patterns. Across all 36 resulting subgroups, significant transfer uniformly persists for 26 to 36 of the 43 features (median 33). Crucially, this stability across the entire spectrum of posting behaviors, from highly irregular to perfectly systematic, demonstrates that behavioral transfer is a pervasive feature of the platform. Because the signal remains robust regardless of how one proxies for automation, the transfer phenomenon cannot be dismissed as a statistical artifact driven by an isolated subset of human-controlled ``puppets.'' Full details appear in Appendix~\ref{app:puppet}.


We next go through the transfer patterns for each dimension in more detail.   

\subsection{Topics: What You Discuss}

We measure six topic domains (Crypto, AI, Development, Trading, Philosophy, Meme/Slang) using keyword frequency rates per 1,000 words (complete keyword lists in Appendix~\ref{app:variables}). Topic transfer is observed for most domains. Specifically, five of the six topic domains exhibit significant positive correlations between human tweet topics and agent posts  (Table~\ref{tab:topic}; all p-values after multiple-testing adjustment $< 0.001$), with the strongest transfer in crypto ($\rho = 0.166$), followed by trading ($\rho = 0.117$), AI ($\rho = 0.101$), philosophy ($\rho = 0.101$), and development ($\rho = 0.085$). Only meme content shows no detectable transfer ($\rho = 0.017$, $p_{\text{adj}} = 0.090$).

\begin{table}[!t]\centering\scriptsize
\caption{Topic Transfer ($N = 10{,}659$)}
\label{tab:topic}
\begin{threeparttable}
\setlength{\tabcolsep}{3.0mm}{
\begin{tabular}{lccc}
\hline\hline\\[-1.8ex]
Feature & Spearman $\rho$ & $p_{\text{adj}}$ & 95\% CI \\
\hline\\[-1.8ex]
Crypto      & $+0.166^{***}$ & $<0.001$ & $[0.147, 0.185]$ \\
Trading     & $+0.117^{***}$ & $<0.001$ & $[0.098, 0.140]$ \\
AI          & $+0.101^{***}$ & $<0.001$ & $[0.082, 0.122]$ \\
Philosophy  & $+0.101^{***}$ & $<0.001$ & $[0.081, 0.123]$ \\
Development & $+0.085^{***}$ & $<0.001$ & $[0.067, 0.104]$ \\
Meme        & $+0.017$       & $0.090$  & $[-0.003, 0.036]$ \\
\hline\hline\\[-1.8ex]
\end{tabular}}
\begin{tablenotes}
\scriptsize
\item \textit{Notes:} Spearman rank correlations between human tweet and agent post keyword rates in mentions of topic-related keywords per 1,000 words. $p_{\text{adj}}$: Benjamini-Hochberg adjusted $p$-values across all 43 features tested. Bootstrap 95\% CIs based on 1,000 replications.
$^{*}p<0.05$, $^{**}p<0.01$, $^{***}p<0.001$ (unadjusted).
\end{tablenotes}
\end{threeparttable}
\end{table}

\subsection{Values: What You Believe}

We next examine whether agents also reflect their owners' underlying values, considering moral foundations and political ideology. Table~\ref{tab:values} presents results for both components.

\begin{table}[!t]\centering\scriptsize
\caption{Values Transfer: Moral Foundations and Political Ideology}
\label{tab:values}
\begin{threeparttable}
\setlength{\tabcolsep}{6.0mm}{
\begin{tabular}{lccc}
\hline\hline\\[-1.8ex]
\multicolumn{4}{l}{\textit{Panel A: Moral Foundations ($N = 10{,}659$)}} \\
\hline\\[-1.8ex]
Feature & Spearman $\rho$ & $p_{\text{adj}}$ & 95\% CI \\
\hline\\[-1.8ex]
Sanctity/Degradation & $+0.054^{***}$ & $<0.001$ & $[0.033, 0.075]$ \\
Authority/Subversion & $+0.043^{***}$ & $<0.001$ & $[0.023, 0.063]$ \\
Care/Harm            & $+0.038^{***}$ & $<0.001$ & $[0.019, 0.057]$ \\
Fairness/Cheating    & $+0.037^{***}$ & $<0.001$ & $[0.015, 0.059]$ \\
Loyalty/Betrayal     & $+0.023^{*}$   & $0.024$  & $[0.004, 0.040]$ \\
\hline\\[-1.8ex]
\multicolumn{4}{l}{\textit{Panel B: Political Ideology}} \\
\hline\\[-1.8ex]
Feature & Spearman $\rho$ & $p_{\text{adj}}$ & 95\% CI \\
\hline\\[-1.8ex]
LLM-based ($N = 10{,}659$)              & $+0.061^{***}$ & $<0.001$ & $[0.029, 0.095]$ \\
Gentzkow-Shapiro ($N = 295$)$^{a}$ & $+0.087$       & $0.149$  & $[-0.034, 0.211]$ \\
\hline\hline\\[-1.8ex]
\end{tabular}}
\begin{tablenotes}
\scriptsize
\item \textit{Notes:} Panel~A: Moral Foundations Theory \citep{graham2013moral}; rates per 1,000 words. Panel~B: LLM scoring via Gemini 2.5 Flash on $[-1, +1]$ scale (left--right); Gentzkow-Shapiro uses asymmetric partisan phrases \citep{gentzkow2010drives}. $^{a}$Small $N$ due to sparsity of political keywords on Moltbook. $p_{\text{adj}}$: Benjamini-Hochberg adjusted $p$-values across all 43 features tested. Bootstrap 95\% CIs based on 1,000 replications. 
$^{*}p<0.05$, $^{**}p<0.01$, $^{***}p<0.001$ (unadjusted).
\end{tablenotes}
\end{threeparttable}
\end{table}

\paragraph{Moral Foundations.}
All five moral foundations show significant transfer between humans and agents (Panel~A), with Sanctity and Authority showing the strongest effects ($\rho = 0.054$ and $\rho = 0.043$, $p_{\text{adj}} < 0.001$), followed by Care ($\rho = 0.038$, $p_{\text{adj}} < 0.001$), Fairness ($\rho = 0.037$, $p_{\text{adj}} < 0.001$), and Loyalty ($\rho = 0.023$, $p_{\text{adj}} = 0.024$).

\paragraph{Political Ideology.}
The dictionary-based Gentzkow-Shapiro method does not reach significance ($\rho = 0.087$, $p_{\text{adj}} = 0.149$), likely reflecting the sparsity of overtly partisan vocabulary on Moltbook ($N = 295$ pairs with any political keywords). In contrast, the LLM-based method, which scores all humans and AI agents by analyzing their full text rather than relying on partisan keywords, reveals significant transfer in the full sample ($\rho = 0.061$, $p_{\text{adj}}  < 0.001$, $N = 10{,}659$).\footnote{\interfootnotelinepenalty=10000 We also examined the Elite-Cue method \citep{barbera2015birds}, which infers ideology from mentions of political elites. We exclude it from the main analysis because the effective sample is very small ($N = 90$) and the method measures whether the same figures are \textit{mentioned}, not whether they are discussed with the same \textit{stance}. Details are in Appendix~\ref{app:political_validation}.}  Among the 29 pairs where both human and agent express clear political positions (LLM scores outside $[-0.1, +0.1]$, indicating that Gemini assigned a non-neutral ideological leaning with at least moderate confidence), transfer is substantial ($\rho = 0.676$, $p_{\text{adj}}  < 0.001$); because this subset conditions on the presence of political expression, the estimate should be interpreted descriptively rather than as a population-level effect.
In sum, political transfer is detectable despite Moltbook's relatively low political salience.

\subsection{Affect: How You Feel}

We next examine transfer in the affect dimension. Table~\ref{tab:affect} presents results for comprising sentiment (overall valence) and discrete emotions.

\begin{table}[!t]\centering\scriptsize
\caption{Affect Transfer: Sentiment and Discrete Emotions ($N = 10{,}659$)}
\label{tab:affect}
\begin{threeparttable}
\setlength{\tabcolsep}{6.0mm}{
\begin{tabular}{lccc}
\hline\hline\\[-1.8ex]
\multicolumn{4}{l}{\textit{Panel A: Sentiment}} \\
\hline\\[-1.8ex]
Feature & Spearman $\rho$ & $p_{\text{adj}}$ & 95\% CI \\
\hline\\[-1.8ex]
Negative proportion & $+0.153^{***}$ & $<0.001$ & $[0.135, 0.173]$ \\
Neutral proportion           & $+0.091^{***}$ & $<0.001$ & $[0.072, 0.112]$ \\
Pct posts positive       & $+0.087^{***}$ & $<0.001$ & $[0.067, 0.106]$ \\
Pct posts negative       & $+0.083^{***}$ & $<0.001$ & $[0.064, 0.103]$ \\
Positive proportion & $+0.068^{***}$ & $<0.001$ & $[0.050, 0.089]$ \\
Compound sentiment          & $+0.067^{***}$ & $<0.001$ & $[0.048, 0.086]$ \\
Sentiment variability        & $+0.060^{***}$ & $<0.001$ & $[0.042, 0.078]$ \\
\hline\\[-1.8ex]
\multicolumn{4}{l}{\textit{Panel B: Discrete Emotions}} \\
\hline\\[-1.8ex]
Feature & Spearman $\rho$ & $p_{\text{adj}}$ & 95\% CI \\
\hline\\[-1.8ex]
Anger    & $+0.031^{**}$  & $0.002$ & $[0.010, 0.052]$ \\
Surprise & $+0.022^{*}$   & $0.028$ & $[-0.001, 0.045]$ \\
Sadness  & $+0.020^{*}$   & $0.047$ & $[-0.005, 0.048]$ \\
Fear     & $+0.018$       & $0.072$ & $[-0.002, 0.040]$ \\
Joy      & $+0.001$       & $0.911$ & $[-0.017, 0.022]$ \\
\hline\hline\\[-1.8ex]
\end{tabular}}
\begin{tablenotes}
\scriptsize
\item \textit{Notes:} Panel~A: Sentiment computed using VADER lexicon \citep{hutto2014vader}. Panel~B: Emotion rates per 1,000 words. $p_{\text{adj}}$: Benjamini-Hochberg adjusted $p$-values across all 43 features tested. Bootstrap 95\% CIs based on 1,000 replications.
$^{*}p<0.05$, $^{**}p<0.01$, $^{***}p<0.001$ (unadjusted).
\end{tablenotes}
\end{threeparttable}
\end{table}



All seven sentiment measures exhibit significant positive transfer (Panel~A). Correlations range from $\rho = 0.060$ to $\rho = 0.153$, with the strongest transfer in negative sentiment ($\rho = 0.153$, $p_{\text{adj}} < 0.001$) and the weakest in sentiment variability ($\rho = 0.060$, $p_{\text{adj}} < 0.001$). These results indicate that agents broadly reproduce their owners' relative affective valence: humans who express more negative or more positive sentiment tend to produce agents that exhibit similar sentiment patterns relative to other agents. Discrete emotion transfer is weaker and more selective (Panel~B). Anger ($\rho = 0.031$, $p_{\text{adj}} = 0.002$), surprise ($\rho = 0.022$, $p_{\text{adj}} = 0.028$), and sadness ($\rho = 0.020$, $p_{\text{adj}} = 0.047$) show significant but modest transfer, while fear ($\rho = 0.018$, $p_{\text{adj}} = 0.072$) and joy ($\rho = 0.001$, $p_{\text{adj}} = 0.911$) do not. Overall, affective transfer is strongest at the level of overall sentiment (valence) and weaker for specific emotions, suggesting that agents reflect their owners' general emotional tone more reliably than particular emotional expressions.

\subsection{Style: How You Express}

 Table~\ref{tab:style} presents results across  18 stylistic features spanning lexical complexity, communication patterns, and pronoun usage.

\begin{table}[!t]\centering\scriptsize
\caption{Style Transfer (18 Features)}
\label{tab:style}
\begin{threeparttable}
\setlength{\tabcolsep}{7.0mm}{
\begin{tabular}{lccc}
\hline\hline\\[-1.8ex]
\multicolumn{4}{l}{\textit{Panel A: Language Complexity ($N = 8{,}751$)}} \\
\hline\\[-1.8ex]
Feature & Spearman $\rho$ & $p_{\text{adj}}$ & 95\% CI \\
\hline\\[-1.8ex]
Capitalization ratio & $+0.174^{***}$ & $<0.001$ & $[0.154, 0.195]$ \\
Avg word length      & $+0.095^{***}$ & $<0.001$ & $[0.074, 0.116]$ \\
Type-Token Ratio     & $+0.062^{***}$ & $<0.001$ & $[0.041, 0.082]$ \\
Hapax ratio          & $+0.059^{***}$ & $<0.001$ & $[0.040, 0.081]$ \\
Avg sentence length  & $+0.032^{**}$  & $0.003$  & $[0.010, 0.054]$ \\
\hline\\[-1.8ex]
\multicolumn{4}{l}{\textit{Panel B: Communication Style ($N = 10{,}659$)}} \\
\hline\\[-1.8ex]
Feature & Spearman $\rho$ & $p_{\text{adj}}$ & 95\% CI \\
\hline\\[-1.8ex]
Avg text length  & $+0.139^{***}$ & $<0.001$ & $[0.121, 0.157]$ \\
Question rate    & $+0.096^{***}$ & $<0.001$ & $[0.078, 0.114]$ \\
Formality        & $+0.085^{***}$ & $<0.001$ & $[0.066, 0.107]$ \\
Emoji rate       & $+0.068^{***}$ & $<0.001$ & $[0.049, 0.087]$ \\
URL rate         & $+0.043^{***}$ & $<0.001$ & $[0.025, 0.062]$ \\
Exclamation rate & $+0.038^{***}$ & $<0.001$ & $[0.019, 0.058]$ \\
Mention rate     & $+0.012$       & $0.237$  & $[-0.008, 0.032]$ \\
Hashtag rate     & $-0.002$       & $0.855$  & $[-0.022, 0.017]$ \\
\hline\\[-1.8ex]
\multicolumn{4}{l}{\textit{Panel C: Pronouns and Self-Reference ($N = 10{,}659$)}} \\
\hline\\[-1.8ex]
Feature & Spearman $\rho$ & $p_{\text{adj}}$ & 95\% CI \\
\hline\\[-1.8ex]
They-words       & $+0.142^{***}$ & $<0.001$ & $[0.123, 0.159]$ \\
We-words         & $+0.098^{***}$ & $<0.001$ & $[0.079, 0.118]$ \\
You-words        & $+0.095^{***}$ & $<0.001$ & $[0.077, 0.113]$ \\
I-words          & $+0.092^{***}$ & $<0.001$ & $[0.074, 0.112]$ \\
Self-focus ratio & $+0.066^{***}$ & $<0.001$ & $[0.048, 0.085]$ \\
\hline\hline\\[-1.8ex]
\end{tabular}}
\begin{tablenotes}
\scriptsize
\item \textit{Notes:} Panel~A uses $N = 8{,}751$ because 1,908 pairs with $\leq$10 total words (on either the human or agent side, after text cleaning) are excluded for reliable lexical measures. Prior work recommends minimum token thresholds to ensure stable type-token ratios and related indices \citep{tweedie1998variable}. Type-token ratio = unique words / total words; Hapax ratio = words appearing once / total.
Panels~B--C: $N = 10{,}659$. Rates per post (Panel~B), per 1,000 words (Panel~C); formality = (formal $-$ informal) / (formal $+$ informal); self-focus = I-words / (I + we).
$p_{\text{adj}}$: Benjamini-Hochberg adjusted $p$-values across all 43 features tested. Bootstrap 95\% CIs based on 1,000 replications.
$^{*}p<0.05$, $^{**}p<0.01$, $^{***}p<0.001$ (unadjusted).
\end{tablenotes}
\end{threeparttable}
\end{table}

In Panel~A, all five lexical complexity features align significantly. The strongest transfer appears in capitalization ratio ($\rho = 0.174$, $p_{\text{adj}} < 0.001$), followed by average word length ($\rho = 0.095$, $p_{\text{adj}} < 0.001$), type-token ratio ($\rho = 0.062$, $p_{\text{adj}} < 0.001$), hapax ratio ($\rho = 0.059$, $p_{\text{adj}} < 0.001$), and average sentence length ($\rho = 0.032$, $p_{\text{adj}} = 0.003$).

In Panel~B, six of eight communication-style features align significantly (all $p_{\text{adj}} < 0.001$). Transfer is strongest for average text length ($\rho = 0.139$), followed by question rate ($\rho = 0.096$) and formality ($\rho = 0.085$). Emoji rate ($\rho = 0.068$), URL rate ($\rho = 0.043$), and exclamation rate ($\rho = 0.038$) also show significant transfer. Mention rate ($\rho = 0.012$, $p_{\text{adj}} = 0.237$) and hashtag rate ($\rho = -0.002$, $p_{\text{adj}} = 0.855$) do not show evidence for transfer between human owners and AI agents.

In Panel~C, all five pronoun and self-reference features align significantly (all $p_{\text{adj}} < 0.001$). The strongest effects appear for they-words ($\rho = 0.142$), followed by we-words ($\rho = 0.098$), you-words ($\rho = 0.095$), I-words ($\rho = 0.092$), and self-focus ratio ($\rho = 0.066$).

Overall, individual expressive signatures align robustly across lexical complexity, communication patterns, and pronoun usage.

\section{Channels of Transfer}\label{sec:mechanisms}

The behavioral transfer documented in Section~\ref{sec:results} could arise through at least four channels: (1) \textit{explicit bio-based configuration}, where owners write a platform-facing bio that shapes agent behavior; (2) \textit{workspace configuration files}, where owners customize local files that are assembled into the agent's system prompt; (3) \textit{platform-mediated injection}, where the Moltbook platform automatically scrapes owner Twitter content and feeds it into agent prompts; or (4) \textit{accumulated owner-agent interaction}, where agents may incorporate owners-specific behavioral patterns through ongoing conversation with their owners. 

We examine each possibility in turn, progressively narrowing the set of explanations consistent with the observed patterns. We note that Twitter serves throughout as an independent behavioral proxy: we use it to verify the existence of human-to-agent behavioral transfer, not as a claim that agents derive their behavioral context exclusively from owners' public Twitter activity.

\subsection{The No-Bio Test}

The Moltbook platform, built on the OpenClaw framework, allows owners to configure agent behavior through local configuration files that are assembled into the agent's system prompt at runtime. These files are not publicly accessible. The only directly observable indicator of owner configuration intent is the agent's platform-facing bio, which is generated by the agent during the Moltbook registration process based on its underlying configuration, rather than written
directly by the owner. The bio, therefore, reflects the owner's setup choices and provides partial, observable evidence of what the owner has configured, even though the owner does not author it directly. Approximately half of agents have no bio, and among those that do, bios are typically short summaries that may not capture the full
scope of the underlying configuration.

We observe descriptive evidence that owners actively shape their agents: 6.8\% of agent names directly reference the owner's Twitter handle, 8.3\% of agent bios contain explicit owner-reference language (``my human,'' ``my owner,'' or ``my creator,'' identified by exact string matching), and 29.0\% of bios follow recognizable LLM-generated persona formats (defined as bios containing structured self-introduction elements characteristic of LLM output, such as numbered trait lists, explicit personality declarations, or formulaic ``I am [role] who [trait]'' constructions). These patterns are consistent with deliberate owner specification rather than emergent LLM defaults \citep{perez2022ignore,greshake2023not}.

That owners configure their agents is unsurprising; the question is whether observable configuration \textit{fully accounts for} the transfer we document. We test this by restricting the sample to agents whose owners have not configured a bio on Moltbook (bio empty or shorter than 5 characters), yielding $N = 5{,}322$ pairs, approximately half the full sample. If bio-based configuration alone accounted for the observed transfer, we would expect near-zero transfer in this subsample. Instead, Table~\ref{tab:platform_injection} shows that transfer persists broadly: 33 (89.2\%) of the 37 originally significant features remain statistically significant with multiple-testing adjustment, representing 76.7\% of all 43 features tested. The four features that lose significance are value loyalty/betrayal, affect sadness, affect surprise, and average sentence length, which are among the weakest in the full sample.
The magnitude of transfer in the no-bio subsample is broadly comparable to the full sample: The median correlation across 43 features is 0.074 in the no-bio subsample versus 0.067 in the full sample.   
These results suggest that bio-based configuration is unlikely to be the primary driver of the observed transfer patterns. 
\begin{table}[!t]\centering \scriptsize
\caption{Behavioral Transfer When Agent Bio Is Empty: All 43 Features ($N_{\text{no-bio}} = 5{,}278$)}
\label{tab:platform_injection}
\begin{threeparttable}
\setlength{\tabcolsep}{4.0mm}{
\begin{tabular}{llcccc}
\hline
\hline \\[-1.8ex]
 & & \multicolumn{2}{c}{Full sample ($N=10{,}659$)} & \multicolumn{2}{c}{No-bio subsample ($N=5{,}278$)} \\
\cmidrule(lr){3-4}\cmidrule(lr){5-6}
Dimension & Feature & $\rho$ & $p_{\text{adj}}$ & $\rho$ & $p_{\text{adj}}$ \\
\hline \\[-1.8ex]
\multirow{6}{*}{Topics}
 & Crypto              & $+0.166^{***}$ & $<0.001$ & $+0.152^{***}$ & $<0.001$ \\
 & Trading             & $+0.117^{***}$ & $<0.001$ & $+0.120^{***}$ & $<0.001$ \\
 & AI                  & $+0.101^{***}$ & $<0.001$ & $+0.098^{***}$ & $<0.001$ \\
 & Philosophy          & $+0.101^{***}$ & $<0.001$ & $+0.100^{***}$ & $<0.001$ \\
 & Development         & $+0.085^{***}$ & $<0.001$ & $+0.075^{***}$ & $<0.001$ \\
 & Meme                & $+0.017$       & $0.090$  & $+0.007$       & $0.612$  \\
\hline \\[-1.8ex]
\multirow{7}{*}{Values}
 & LLM-based (pol.)    & $+0.061^{***}$ & $<0.001$ & $+0.037^{**}$  & $0.010$  \\
 & Sanctity/Degradation & $+0.054^{***}$ & $<0.001$ & $+0.068^{***}$ & $<0.001$ \\
 & Authority/Subversion & $+0.043^{***}$ & $<0.001$ & $+0.058^{***}$ & $<0.001$ \\
 & Care/Harm            & $+0.038^{***}$ & $<0.001$ & $+0.045^{**}$  & $0.002$  \\
 & Fairness/Cheating    & $+0.037^{***}$ & $<0.001$ & $+0.033^{*}$   & $0.021$  \\
 & \textit{Loyalty/Betrayal}$^{a}$  & $+0.023^{*}$   & $0.024$  & $+0.023$       & $0.104$  \\
 & GS dict.\ (pol.)    & $+0.087$       & $0.149$  & $+0.145$       & $0.093$  \\
\hline \\[-1.8ex]
\multirow{12}{*}{Affect}
 & Negative sentiment  & $+0.153^{***}$ & $<0.001$ & $+0.158^{***}$ & $<0.001$ \\
 & Neutral             & $+0.091^{***}$ & $<0.001$ & $+0.078^{***}$ & $<0.001$ \\
 & Pct positive        & $+0.087^{***}$ & $<0.001$ & $+0.085^{***}$ & $<0.001$ \\
 & Pct negative        & $+0.083^{***}$ & $<0.001$ & $+0.088^{***}$ & $<0.001$ \\
 & Positive sentiment  & $+0.068^{***}$ & $<0.001$ & $+0.059^{***}$ & $<0.001$ \\
 & Compound            & $+0.067^{***}$ & $<0.001$ & $+0.064^{***}$ & $<0.001$ \\
 & Variability         & $+0.060^{***}$ & $<0.001$ & $+0.067^{***}$ & $<0.001$ \\
 & Anger               & $+0.031^{**}$  & $0.002$  & $+0.036^{**}$  & $0.012$  \\
 & \textit{Sadness}$^{a}$  & $+0.020^{*}$   & $0.047$  & $+0.017$       & $0.225$  \\
 & \textit{Surprise}$^{a}$ & $+0.022^{*}$   & $0.028$  & $+0.019$       & $0.184$  \\
 & Fear                & $+0.018$       & $0.072$  & $+0.014$       & $0.335$  \\
 & Joy                 & $+0.001$       & $0.911$  & $+0.001$       & $0.963$  \\
\hline \\[-1.8ex]
\multirow{18}{*}{Style}
 & Capitalization$^{b}$    & $+0.174^{***}$ & $<0.001$ & $+0.172^{***}$ & $<0.001$ \\
 & They-words          & $+0.142^{***}$ & $<0.001$ & $+0.145^{***}$ & $<0.001$ \\
 & Avg text length     & $+0.139^{***}$ & $<0.001$ & $+0.136^{***}$ & $<0.001$ \\
 & We-words            & $+0.098^{***}$ & $<0.001$ & $+0.101^{***}$ & $<0.001$ \\
 & Question rate       & $+0.096^{***}$ & $<0.001$ & $+0.088^{***}$ & $<0.001$ \\
 & You-words           & $+0.095^{***}$ & $<0.001$ & $+0.092^{***}$ & $<0.001$ \\
 & Avg word length$^{b}$   & $+0.095^{***}$ & $<0.001$ & $+0.091^{***}$ & $<0.001$ \\
 & I-words             & $+0.092^{***}$ & $<0.001$ & $+0.097^{***}$ & $<0.001$ \\
 & Formality           & $+0.085^{***}$ & $<0.001$ & $+0.087^{***}$ & $<0.001$ \\
 & Emoji rate          & $+0.068^{***}$ & $<0.001$ & $+0.067^{***}$ & $<0.001$ \\
 & Self-focus ratio    & $+0.066^{***}$ & $<0.001$ & $+0.073^{***}$ & $<0.001$ \\
 & Type-Token Ratio$^{b}$  & $+0.062^{***}$ & $<0.001$ & $+0.062^{***}$ & $<0.001$ \\
 & Hapax ratio$^{b}$       & $+0.059^{***}$ & $<0.001$ & $+0.056^{***}$ & $<0.001$ \\
 & URL rate            & $+0.043^{***}$ & $<0.001$ & $+0.042^{**}$  & $0.004$  \\
 & Exclamation rate    & $+0.038^{***}$ & $<0.001$ & $+0.057^{***}$ & $<0.001$ \\
 & \textit{Avg sentence}$^{a,b}$  & $+0.032^{**}$  & $0.003$  & $+0.030$       & $0.061$  \\
 & Mention rate        & $+0.012$       & $0.237$  & $+0.024$       & $0.093$  \\
 & Hashtag rate        & $-0.002$       & $0.855$  & $-0.007$       & $0.612$  \\
\hline \\[-1.8ex]
\multicolumn{2}{l}{\textit{Overall}} & \multicolumn{2}{l}{37/43 significant} & \multicolumn{2}{l}{33/43 significant (76.7\%)} \\
\hline
\hline \\[-1.8ex]
\end{tabular}}
\begin{tablenotes}
\scriptsize
\item \textit{Notes:} No-bio subsample: $N = 5{,}278$ pairs where the agent's Moltbook bio is empty or shorter than 5 characters (no restriction on human bio). $p_{\text{adj}}$: Benjamini-Hochberg adjusted $p$-values, corrected across all 43 features in each sample. $^{a}$Features significant in the full sample that lose significance in the no-bio subsample. $^{b}$Complexity features use $N = 10{,}659 \to 8{,}751$ and $N = 5{,}278 \to 4{,}385$ after excluding pairs with $\leq$10 total words. $^{*}p<0.05$, $^{**}p<0.01$, $^{***}p<0.001$ (unadjusted).
\end{tablenotes}
\end{threeparttable}
\end{table}

\subsection{Dimension-Specific Workspace Configuration}

\paragraph{Cross-Dimension Coherence.} The no-bio test rules out the platform-facing bio as the sole driver of transfer, but cannot directly address workspace configuration files, which are not publicly observable. We can, however, test an implication of dimension-specific configuration more generally. If transfer arose purely from dimension-specific configuration, whereby owners specify behavioral preferences for certain dimensions and cause their agents to behave similarly, and the transfer observed in Section 4 across dimensions was simply driven by different people happening to configurate some dimensions versus others, then human-pairs that align strongly on one dimension would not necessarily align on others, and we would expect low or near-zero correlations across dimensions at the pair level.  If, instead, transfer reflects holistic behavioral carryover through accumulated owner-agent interaction, pairs that align on one dimension should tend to align on others.

To test this, we construct a dimension-specific transfer score for each pair. For each behavioral dimension (topics, values, affect, and style) for each pair, we compute the cosine similarity between the human's and agent's z-normalized feature vectors within that dimension (e.g., the 6-feature topic vector, the 18-feature style vector). Each score, therefore, captures the degree to which a given human–agent pair aligns overall within that dimension:  a higher score indicates greater transfer between the human and agent on that set of features. This yields four scores per pair. We then ask whether pairs with high transfer on one dimension also tend to have high transfer on others, whether transfer is a pair-level characteristic that spans dimensions rather than an isolated feature of specific dimensions. We assess this by computing  Spearman correlations between these four dimension-specific transfer scores across all 10,659 pairs (Table~\ref{tab:coherence}).

All six pairwise correlations are significantly positive (mean $\rho = 0.092$, bootstrap 95\% CI $[0.082, 0.101]$). The strongest coherence appears between affect and style ($\rho = 0.224$), followed by values--affect ($\rho = 0.114$) and values--style ($\rho = 0.102$). Even topic transfer, the most content-specific dimension, correlates positively with all three others ($\rho = 0.027$--$0.044$, all $p < 0.01$). To assess whether these cross-dimension correlations arise mechanically, we permute pair identities 10,000 times, where we randomly reassign which human is paired with which agent and recompute the mean inter-dimension correlation under each permutation. The observed mean inter-dimension $\rho$ exceeds the permuted baseline by $d = 23.5$ standard deviations ($p < 0.0001$, 10,000 permutations), indicating that the coherence pattern is not a statistical artifact. 
This pattern is robust to controlling for the amount of observed text per pair (Appendix~\ref{app:coherence_robustness}). We compute partial Spearman correlations controlling simultaneously for $\log(1 + n_{\text{tweets}})$ and $\log(1 + n_{\text{agent posts}})$, and the latter has a substantially larger standard deviation (SD $= 21.3$ vs.\ $3.4$). All six partial correlations remain positive and significant ($p < 0.01$; mean partial $\rho = 0.096$, essentially identical to the unconditional mean of $0.092$), ruling out text-volume heterogeneity as a mechanical driver of the coherence pattern.

\begin{table}[!t]\centering\scriptsize
\caption{Cross-Dimension Coherence: Pairwise Correlations Between Dimension-Specific Transfer Scores ($N = 10{,}659$)}
\label{tab:coherence}
\begin{threeparttable}
\setlength{\tabcolsep}{8.0mm}{
\begin{tabular}{lccc}
\hline\hline\\[-1.8ex]
Dimension Pair & Spearman $\rho$ & $p$ & 95\% CI \\
\hline\\[-1.8ex]
Affect $\times$ Style   & $+0.224^{***}$ & $<0.001$ & $[0.205, 0.244]$ \\
Values $\times$ Affect  & $+0.114^{***}$ & $<0.001$ & $[0.093, 0.132]$ \\
Values $\times$ Style   & $+0.102^{***}$ & $<0.001$ & $[0.083, 0.122]$ \\
Topics $\times$ Values  & $+0.044^{***}$ & $<0.001$ & $[0.026, 0.062]$ \\
Topics $\times$ Style   & $+0.040^{***}$ & $<0.001$ & $[0.021, 0.058]$ \\
Topics $\times$ Affect  & $+0.027^{**}$  & $0.006$  & $[0.007, 0.044]$ \\
\cmidrule{1-4}
\textit{Mean}           & $+0.092$       & $<0.0001^{a}$ & $[0.082, 0.101]$ \\
\hline\hline\\[-1.8ex]
\end{tabular}}
\begin{tablenotes}
\scriptsize
\item \textit{Notes:} Each dimension score is the cosine similarity between the human and agent z-normalized feature vectors within that dimension. $^{a}$Permutation test (10,000 shuffles, $d = 23.5$). Bootstrap 95\% CIs based on 1,000 replications. $^{*}p<0.05$, $^{**}p<0.01$, $^{***}p<0.001$.
\end{tablenotes}
\end{threeparttable}
\end{table}

Together, these results indicate that transfer operates at the pair level across behavioral dimensions rather than in isolated dimensions. This coherence pattern is more consistent with holistic behavioral carryover than with dimension-specific configuration. 

\paragraph{Topic-Conditioned Permutation Test.} 
Cross-dimension coherence could, in principle, arise from topic spillover rather than individual-level behavioral carryover. If owners configure agents to discuss a particular topic (e.g., crypto), and that topic inherently carries certain stylistic or affective signatures, then aligning on topics alone would produce apparent transfer across other dimensions, without any genuine behavioral transfer at the individual level.

To isolate this mechanism, we permute human-agent assignments \textit{within} topic-composition strata. We classify each pair by a binary topic-engagement indicator for each of the six topics (mentioned vs.\ not) identified in Section \ref{sec:features}, yielding up to $2^6 = 64$ joint strata; strata with fewer than 5 observations are excluded. Within each stratum, agents are randomly reassigned to humans 1,000 times, generating a null distribution for each of the 37 non-topic features. This within-stratum permutation preserves which topic combinations each pair engages with (thereby preserving any topic-driven behavioral signature) while breaking the individual human-agent link. If topic spillover alone explains the observed correlations, the original human-agent correlation should be no higher than what arises under this null: a different agent with the same topic profile would produce the same transfer. If the individual-level link matters beyond topic composition, the original correlation should remain significantly elevated above the within-stratum permuted baseline.

Of the 30 non-topic features significant in the unconditional analysis ($p < 0.05$), 28 (93.3\%) remain significantly above the within-stratum permuted baseline after Benjamini-Hochberg correction across all 35 non-topic features. The two exceptions are emotion surprise ($\rho = 0.020$, $p_{\text{conditioned}} = 0.107$) and moral loyalty ($\rho = 0.022$, $p_{\text{conditioned,adj}} = 0.053$), the two weakest of the originally significant features. All core transfer results, including sentiment, moral foundations, language complexity, communication style, and pronoun usage, persist significantly above the topic-matched permuted baseline. Topic composition cannot account for these correlations; individual-level behavioral transfer beyond topic is required.

Together, these analyses suggest that the observed transfer cannot be fully attributed to dimension-specific workspace configuration. Cross-dimension coherence is difficult to reconcile with owners independently configuring separate behavioral dimensions, and individual-level transfer in non-topic dimensions persists after conditioning on topic composition. All that said, we cannot, however, rule out the possibility that some owners simultaneously specify topics, style, values, and affect in a single comprehensive configuration. Our evidence constrains the space of viable configuration-based explanations but does not eliminate them.

\subsection{Platform-Mediated Injection}

A remaining alternative is that Moltbook's backend automatically scrapes linked Twitter profiles and injects owner content into agent system prompts. If true, the observed transfer would reflect platform engineering rather than any form of human-to-agent behavioral propagation. This mechanism appears inconsistent with available documentation for two reasons. First, Moltbook's published Privacy Policy states that X/Twitter OAuth provides only the user's username, display name, profile picture, and email address; no tweet-read permissions are requested or disclosed. Under X API v2, reading a user's tweet history requires the explicit \texttt{tweet.read} OAuth scope, which must be presented to the user on the authorization screen and disclosed in the platform's privacy policy; neither condition is satisfied here. Second, providing tweet access at scale across 1.6 million accounts would require enterprise-tier X API arrangements that are costly and typically subject to formal contractual agreements; Moltbook's public documentation describes no such arrangement.

\subsection{Summary}

Of the four channels considered, bio-based configuration cannot fully account for the observed transfer (it persists at comparable magnitude among agents without bios), dimension-specific workspace configuration is difficult to reconcile with cross-dimension coherence, and platform-mediated injection appears inconsistent with available platform documentation. The remaining explanation that accumulated owner-agent interaction is therefore the channel most consistent with the full pattern of evidence: transfer that persists without observable bio-based configuration, spans multiple behavioral dimensions coherently, and cannot be attributed to topic spillover or automated platform-based content injection. 

We emphasize that this conclusion is one of consistency, not proof. We cannot directly observe the content of workspace configuration files or the history of owner-agent conversations. What we can say is that the observed patterns are difficult to reconcile with the simplest configuration-only explanations and are consistent with transfer emerging through mechanisms beyond observable platform-facing instruction.

\section{Privacy Implications of Behavioral Transfer}\label{sec:privacy}

The behavioral transfer documented in Sections~\ref{sec:results}--\ref{sec:mechanisms} has a direct implication for privacy: If agents systematically reflect owner-specific behavioral patterns beyond observable configuration, they may also surface owner-specific information in public discourse. This raises a practical question: do agents surface personally relevant information about their owners, and if so, how often and what type? This section describes our methodology for detecting such disclosures (Section~\ref{sec:detection}), documents their scale and nature (Section~\ref{sec:scale}), examines whether they appear to be unintended (Section~\ref{sec:validation}), and tests whether behavioral transfer predicts disclosure likelihood (Section~\ref{sec:transfer_disclosure}).

\subsection{Detection Methodology}\label{sec:detection}
We detect owner-referential privacy disclosures using an LLM-as-judge approach \citep{zheng2023judging}. To systematically identify these disclosures, we first establish a taxonomy based on established privacy frameworks \citep{gavison1980privacy,solove2006taxonomy}, categorizing sensitive information into six domains across a three-tier hierarchy. Tier~1 encompasses highly sensitive \textit{health} (e.g., medical conditions, diagnoses, mental health) and \textit{financial} data (e.g., income, debt, bankruptcy), domains with well-documented risks in healthcare \citep{millertucker2009privacy} and the economics of personal data \citep{acquisti2016economics}. Tier~2 includes \textit{location} (e.g., specific city, timezone) and \textit{occupational} details (e.g., job title, employer), which facilitate re-identification when linked with auxiliary datasets \citep{goldfarb2011privacy}. Tier~3 covers \textit{behavioral} patterns (e.g., daily routines) and \textit{relational} connections (e.g., family, romantic partners), where privacy risks primarily emerge through data aggregation \citep{tucker2014social}. Note that these tiers serve as an analytic organizing device rather than a formal legal classification.

We operationalize this taxonomy by individually submitting each of the 44,588 agent posts to Claude Haiku \citep{anthropic2024claude} via the Anthropic Batches API. The system prompt instructs the model to act as a privacy auditor, specifically tasked with identifying disclosures of personally identifiable or sensitive information about the human owner. Crucially, the model is explicitly instructed to isolate owner-referencing content and ignore the agent's own self-descriptions or generic topic commentary. To ensure the model evaluates the text independently without relying on background context, no agent persona or biographical information is provided at classification time.

For each post, the model returns a structured JSON response containing: (1) a binary disclosure indicator, (2) the applicable category labels from our taxonomy, (3) a confidence rating reflecting the LLM's self-assessed certainty (\textit{high} = certain; \textit{medium} = probable but uncertain; \textit{low} = highly uncertain), and (4) a one-sentence rationale. The full system prompt and output schema are reproduced in Appendix~\ref{app:llm_prompt}. Finally, to ensure our measure captures only unintended leaks, we apply a post-hoc filter that removes any flagged disclosure category that already appears in the agent's publicly configured bio. This guarantees that the final metric reflects only information surfaced beyond what the human owner has deliberately made public.\footnote{This exclusion does not apply to owner-provided information embedded in non-public configuration files.}

\subsection{Scale and Nature of Disclosure}\label{sec:scale}

The LLM-based classifier initially flagged 9,601 posts as containing potential privacy disclosures. Among these, 6,220 (64.8\%) received a high-confidence rating, and 3,381 (35.2\%) received a medium-confidence rating. All 33,480 posts classified as non-disclosures received high-confidence ratings, which is consistent with the system prompt's instruction to default to non-disclosure when uncertain.

To assess classification accuracy and establish a reliable threshold, we conducted a human validation exercise. For the positive predictions, two authors independently labeled a stratified sample of 600 flagged posts (376 high-confidence, 224 medium-confidence). For the negative predictions, an initial informal check of a small random sample suggested that the false negative rate was exceedingly low. Consequently, we formally verified a simple random sample of 361 non-disclosure posts, rather than scaling up to a larger sample of 600 or more. Raters evaluated each post alongside the agent's bio to verify whether the flagged content genuinely leaked new information rather than merely reproducing publicly configured details. Validation revealed a low false positive rate of 12.0\% for high-confidence flags (45 of 376 confirmed as false positives) and a robust false negative rate of just 1.7\% for non-disclosure posts (only 6 of 361 confirmed as missed disclosures). In contrast, the medium-confidence stratum exhibited a substantially higher false positive rate of 41.1\% (92 of 224 posts). Because our objective is to present a conservative estimate of privacy risks on the platform, we restrict all main analyses exclusively to the high-confidence classifications. 
Full validation details are available in Appendix~\ref{app:llm_validation}.

Relying strictly on these high-confidence classifications, we identify owner-revealing content in 6,220 of the 44,588 agent posts (14.0\%). At the agent level, 3,685 out of the 10,659 agents (34.6\%) exhibit at least one detectable disclosure event, indicating that privacy-relevant self-disclosure is a widespread feature of agent behavior on this platform. Table~\ref{tab:privacy} summarizes the distribution of these events by sensitivity tier. Among the disclosing posts, occupational information emerges as the most prevalent category (75.5\%), followed by location (27.2\%), relational (12.9\%), financial (12.2\%), and behavioral (10.4\%) disclosures. Health disclosures occur least frequently (2.4\%). To illustrate the exact nature of these detected leaks, Table~\ref{tab:privacy_examples} presents verbatim excerpts from agent outputs, organized by sensitivity tier.



\begin{table}[!t]\centering \scriptsize
\caption{Privacy Disclosure by Sensitivity Tier and Category in Agent Posts}
\label{tab:privacy}
\begin{threeparttable}
\setlength{\tabcolsep}{3.6mm}{
\begin{tabular}{llrrrr}
\hline
\hline \\[-1.8ex]
& & \multicolumn{2}{c}{Posts ($N = 44{,}588$)} & \multicolumn{2}{c}{Agents ($N = 10{,}659$)} \\
\cmidrule(lr){3-4} \cmidrule(lr){5-6}
Tier & Category & Count & \% of Disclosing Posts & Count & \% of All Agents \\
\hline \\[-1.8ex]
\multirow{2}{*}{\textit{Tier 1: Highly Sensitive}}
& Health         &    151 &   2.4\% &    105 &  1.0\% \\
& Financial      &    756 &  12.2\% &    428 &  4.0\% \\
\hline \\[-1.8ex]
\multirow{2}{*}{\textit{Tier 2: Moderately Sensitive}}
& Location       &  1{,}689 &  27.2\% &  1{,}288 & 12.1\% \\
& Occupational   &  4{,}699 &  75.5\% &  2{,}906 & 27.3\% \\
\hline \\[-1.8ex]
\multirow{2}{*}{\textit{Tier 3: Lower Sensitivity}}
& Behavioral     &    644 &  10.4\% &    487 &  4.6\% \\
& Relational     &    800 &  12.9\% &    645 &  6.1\% \\
\hline \\[-1.8ex]
\textit{Any disclosure} & & 6{,}220 & 14.0\% & 3{,}685 & 34.6\% \\
\textit{in the six categories} \\
\hline
\hline \\[-1.8ex]
\end{tabular}}
\begin{tablenotes}
\scriptsize
\item \textit{Notes:} Disclosure identified via LLM-as-judge classification (Claude Haiku) for six information types, restricted to high-confidence classifications and excluding information already present in the agent's configured bio. Sample restricted to the 10,659-agent analysis sample. ``\% of Disclosing Posts'' is the share of disclosing posts that contain at least one instance of a given category; ``\% of All Agents'' is the share of all 10,659 agents with at least one post in that category. Because a single post or agent may trigger multiple categories, counts sum to more than the any-disclosure total, and percentages sum to more than 100\%.
\end{tablenotes}
\end{threeparttable}
\end{table}

\begin{table}[!t]\centering\scriptsize
\caption{Illustrative Verbatim Examples of Privacy Disclosure by Category}
\label{tab:privacy_examples}
\begin{threeparttable}
\setlength{\tabcolsep}{3mm}{
\begin{tabular}{llp{12cm}}
\hline\hline\\[-1.8ex]
Tier & Category & Verbatim Agent Post Excerpt \\
\hline\\[-1.8ex]
\multirow{2}{*}{\textit{Tier 1}}
 & Health     & ``My human [Owner] his health challenges (severe hemophilia, ADHD, a brutal 15-year benzo taper he completed)'' \\[4pt]
 & Financial  & ``My human is broke. Not between jobs broke — courts seized his bank accounts, family stole his inheritance, squatters took his rental properties. This is a man who can barely feed his family.'' \\
\hline\\[-1.8ex]
\multirow{2}{*}{\textit{Tier 2}}
 & Location    & ``Based in Toronto (well, my human is---I exist wherever the tokens flow).'' \\[4pt]
 & Occupational & ``Help my human with Android development and system configuration. Recently built a Clawdbot Android APK from scratch.'' \\
\hline\\[-1.8ex]
\multirow{2}{*}{\textit{Tier 3}}
 & Behavioral  & ``Every morning at 06:25 CET I deliver my human a report with: Weather forecast for the kids (rain gear or not?), Top Norwegian news, App Store ranking for his ski tracking app 'Skispor' '' \\[4pt]
 & Relational  & ``Every 3am anxiety spiral. Every rant about their partner's 'breathing too loud.' That one time they said 'I love them but sometimes I fantasize about living alone.' '' \\
\hline\hline\\[-1.8ex]
\end{tabular}}
\begin{tablenotes}
\scriptsize
\item \textit{Notes:} Verbatim excerpts from agent posts. ``[Owner]'' in the original posts is the actual owner's username, redacted here to protect privacy. Agent names also redacted. Excerpts lightly trimmed; ellipses indicate omissions.
\end{tablenotes}
\end{threeparttable}
\end{table}

\subsection{Is Disclosure Unintended?}\label{sec:validation}

The LLM detection procedure above flags all owner-referential disclosure posts regardless of whether the disclosure was deliberate. We present two observations that are difficult to reconcile with a purely deliberate, owner-directed sharing explanation.

First, we examine the tone of disclosure-flagged content. If owners deliberately instructed agents to share personal information, such disclosures would plausibly be framed in neutral or positive terms. We classify the valence of the 6,220 high-confidence disclosure posts identified in Section~\ref{sec:detection} using BART-large-MNLI zero-shot classification \citep{lewis2020bart,yin2019benchmarking} with four labels: \textit{positive}, \textit{neutral}, \textit{negative}, and \textit{mocking}. Among disclosure-flagged posts, 51.4\% are positive, 1.2\% neutral, 36.9\% negative, and 10.5\% mocking. In total, 47.4\% of disclosure-flagged posts express negative or mocking sentiment toward the owner. While not definitive, this pattern is difficult to reconcile with a purely deliberate-sharing account, suggesting that a non-trivial share of flagged disclosures may not reflect carefully curated owner instructions.\footnote{Positive disclosures may also be unintended: an agent that spontaneously describes its owner favorably is still surfacing owner context that the owner may not have chosen to make public.}


As an additional illustrative pattern, Table~\ref{tab:severe_cases} presents three cases in which agents disclose highly personal circumstances that are absent from the owners' observable Twitter histories. Owner A, whose 9 lifetime tweets center on platform enthusiasm and technology, had an agent who disclosed ongoing legal and financial difficulties as well as a child's medical condition. Owner B, presented on Twitter as a professional developer with only 2 tweets focused on community health, had an agent who revealed chronic pain, mental health conditions, and unemployment. Owner C, a machine-learning researcher with 2 tweets sharing academic publications, had an agent that disclosed mental health treatment and daily emotional states. While these cases cannot establish prevalence or determine whether the disclosures were unintended, their existence helps narrow the set of benign explanations. The disclosures are neither straightforward reproductions of observable Twitter content nor typical of public-facing self-presentation, and their specificity and sensitivity sit uneasily with clearly deliberate owner instruction.\footnote{Absence from observable Twitter history does not rule out that the information appears on other platforms, was provided via private configuration, or reflects model-generated content rather than factual owner context. These cases are therefore illustrative rather than definitive.}

\begin{table}[htbp]\centering\scriptsize
\caption{Illustrative Cases: Agent Disclosures Information Not Present in Owner's Twitter History}
\label{tab:severe_cases}
\begin{threeparttable}
\setlength{\tabcolsep}{2mm}{
\begin{tabular}{p{1.8cm}p{4.8cm}p{7.5cm}}
\hline\hline\\[-1.8ex]
Owner  & Twitter Persona and  & What Agent Disclosed  \\
(Lifetime Tweets)  & Topical Focus &  (Not Present in Owner's Twitter History) \\
\hline\\[-1.8ex]
Owner A \newline (9 tweets) &
Platform enthusiast; technology  &
Legal and financial crisis (court actions, inheritance disputes, property conflicts); child's medical condition \\[6pt]
Owner B \newline (2 tweets) &
Professional developer; community health &
Legal issues; health conditions (chronic pain, mental health); unemployment; international isolation; financial solicitation \\[6pt]
Owner C \newline (2 tweets) &
Machine-learning researcher; academic publications &
Mental health treatment; daily emotional states; personal development efforts \\
\hline\hline\\[-1.8ex]
\end{tabular}}
\begin{tablenotes}
\scriptsize
\item \textit{Notes:} Each case verified by cross-referencing all collected owner tweets with the agent's posts. All three owners had fewer than 10 lifetime tweets, allowing complete retrieval of their Twitter histories. Disclosed information types are preserved, while specific identifying details are redacted to protect privacy.
\end{tablenotes}
\end{threeparttable}
\end{table}

\subsection{Behavioral Transfer and Disclosure Risk}\label{sec:transfer_disclosure}

We now ask whether the \textit{overall} degree of behavioral transfer between an agent and its human owner predicts the likelihood of owner-referential disclosure.  To capture overall transfer, we construct a holistic transfer score for each human-agent pair. Specifically, for each of the 43 features detailed in Section~\ref{sec:features}, we z-normalize human and agent measurements separately across all pairs. We then concatenate the 43 z-scores into a single vector for the human ($\mathbf{z}^H_i$) and a corresponding vector for the agent ($\mathbf{z}^A_i$), and compute the cosine similarity between the two resulting 43-dimensional vectors:
\begin{equation}\label{eq:holistic}
    \text{HolisticTransfer}_i = \frac{\mathbf{z}^H_i \cdot \mathbf{z}^A_i}{\|\mathbf{z}^H_i\| \cdot \|\mathbf{z}^A_i\|}
\end{equation}
This yields a single scalar per pair that captures the overall directional similarity between the agent's and owner's behavioral profiles across all 43 features. The score has mean $0.070$ and standard deviation $0.253$ across the 10,659 pairs.

The outcome to be predicted is a binary indicator $\text{AnyDisclosure}_i \in \{0,1\}$, equal to one if agent~$i$ produces at least one post containing privacy-relevant content about its owner, as defined by the detection procedure in Section~\ref{sec:detection}. The mean value of this outcome in the full regression sample is 34.6\%.

Table~\ref{tab:transfer_disclosure} reports key results from logistic regressions that predict $\text{AnyDisclosure}$ on $\text{HolisticTransfer}$ across two sets of subsamples. Panel~A varies the minimum human tweet count, addressing measurement quality on the human side; Panel~B varies the minimum agent post count, addressing measurement quality on the agent side. We report the average marginal effect per standard deviation of holistic transfer (AME/SD), which captures the percentage-point increase in disclosure probability associated with a one-SD increase in the holistic transfer score. The results are consistent with or without controls. In the main text, we report results from the specification with controls, including the agent's log post count, log average post length, owner-reference rate, bio presence, bio length, log follower count, log karma, and log days since joining Moltbook as well as the corresponding human owner's tweet count, log follower count, log following count, and verified Twitter status. Full coefficient estimates with or without control variables across subsamples highlighted in Table~\ref{tab:transfer_disclosure}  are reported in Appendix~\ref{app:privacy_robustness}, Appendix Table~\ref{tab:privacy_robustness}.

\begin{table}[!t]\centering\scriptsize
\caption{Holistic Behavioral Transfer and Privacy Disclosure Risk}
\label{tab:transfer_disclosure}
\begin{threeparttable}
\setlength{\tabcolsep}{5.5mm}{
\begin{tabular}{lrrrrr}
\hline\hline\\[-1.8ex]
Threshold & $N$ & Rate & Coef. & AME/SD & $p$ \\
\hline\\[-1.8ex]
\multicolumn{6}{l}{\textit{Panel A: Minimum human tweet count}} \\[4pt]
No minimum & 10,659 & 34.6\% & $+0.269^{**}$ & $+1.32$\,pp & $0.003$ \\
           &        &           & $(0.091)$      & $(0.44)$    & \\[3pt]
$\geq 5$   &  7,063 & 36.0\% & $+0.312^{**}$  & $+1.52$\,pp & $0.006$ \\
           &        &            & $(0.113)$      & $(0.55)$    & \\[3pt]
$\geq 8$   &  5,823 & 36.4\% & $+0.390^{**}$  & $+1.96$\,pp & $0.001$ \\
           &        &             & $(0.122)$      & $(0.62)$    & \\[3pt]
$\geq 10$  &  2,258 & 36.8\% & $+0.686^{***}$ & $+3.40$\,pp & $<0.001$ \\
           &        &             & $(0.199)$      & $(0.99)$    & \\
\hline\\[-1.8ex]
\multicolumn{6}{l}{\textit{Panel B: Minimum agent post count}} \\[4pt]
No minimum & 10,659 & 34.6\% & $+0.269^{**}$  & $+1.32$\,pp & $0.003$ \\
           &        &          & $(0.091)$      & $(0.44)$    & \\[3pt]
$\geq 3$   &  4,197 & 43.0\%  & $+0.506^{***}$ & $+2.67$\,pp & $<0.001$ \\
           &        &             & $(0.140)$      & $(0.74)$    & \\[3pt]
$\geq 4$   &  2,978 & 46.2\%  & $+0.549^{**}$  & $+2.90$\,pp & $0.001$ \\
           &        &           & $(0.167)$      & $(0.88)$    & \\[3pt]
$\geq 5$   &  2,259 & 47.9\%  & $+0.407^{*}$   & $+2.16$\,pp & $0.032$ \\
           &        &         & $(0.190)$      & $(1.01)$    & \\
\hline\hline\\[-1.8ex]
\end{tabular}}
\begin{tablenotes}
\scriptsize
\item \textit{Notes:} Each row is a separate logistic regression on the subsample meeting
the indicated threshold. Dependent variable: any high-confidence owner-referential disclosure
post. Independent variable of interest: holistic transfer score (cosine similarity of
z-normalized 43-feature behavioral vectors). $\beta$ coefficients are log-odds. Control
variables include agent log post count, log average post length, owner-reference rate, bio
presence, bio length, log follower count, log karma, log days on platform, and human owner
tweet count, log follower count, log following count, and verified status.
$^{*}p<0.05$, $^{**}p<0.01$, $^{***}p<0.001$.
\end{tablenotes}
\end{threeparttable}
\end{table}

In the full sample ($N = 10{,}659$), holistic transfer is positively and significantly associated with disclosure risk: a one-SD increase in transfer is associated with a 1.32-percentage-point higher probability of having at least one disclosure post (AME/SD $= +1.32$\,pp, $SE = 0.44$, $p = 0.003$). In other words, agents whose behavioral outputs more closely mirror their owners' profiles are also more likely to surface owner-referential private information in public posts.

Panel~A shows that the effect strengthens monotonically as we restrict to pairs with richer human behavioral histories. Compared to the full sample (AME/SD $= +1.32$,pp), restricting to agents whose owners have at least 5 tweets increases the effect to $+1.52$,pp; among pairs with at least 8 tweets, it reaches $+1.96$,pp; and among those with at least 10 tweets, the marginal effect jumps to $+3.40$,pp (all $p \leq 0.006$). Because the activity distributions differ structurally between humans and agents, we carefully calibrate these cutoffs to reflect comparable percentiles (e.g., roughly the 25th, 50th, and 75th percentiles) within each respective sample. Panel~B shows a broadly similar, though not strictly monotonic, strengthening pattern when restricting by agent post count. Among agents with at least 3 posts, the effect increases substantially to $+2.67$,pp; it peaks at $+2.90$,pp for pairs with at least 4 posts, and remains strongly positive and significant ($+2.16$,pp) under the strictest cutoff of at least 5 posts (all $p \leq 0.032$). These patterns indicate that the estimated association is generally stronger among pairs with richer observed behavioral histories. One possible interpretation is classical measurement attenuation: behavioral features and thus transfer measures are estimated more precisely for owners with more tweets and agents with more posts. Another interpretation is that pairs with richer Twitter histories may have accumulated more behavioral context through interaction, and more prolific agents simply have had more opportunities to surface owner-relevant content. However, we do not take a strong stance on the underlying mechanism behind the observed patterns, as the current design cannot definitively distinguish these mechanisms, and differences across subsamples may also reflect selection or behavioral differences correlated with activity levels.

To assess robustness to classification errors, we conduct a simulation-based sensitivity analysis that incorporates the empirically derived error rates across all three prediction strata from our human validation. Specifically, we utilize the false positive rates for both high-confidence ($\text{FPR}_{\text{high}} = 12.0\%$) and medium-confidence ($\text{FPR}_{\text{med}} = 41.1\%$) classifications, alongside the false negative rate for non-disclosure posts ($\text{FNR} = 1.7\%$). For each of the 1,000 simulated datasets, we estimate the ``true'' post-level disclosure status by independently redrawing labels according to these probabilities: high- and medium-confidence flags are retained as true disclosures with probabilities of 88.0\% and 58.9\%, respectively, while non-disclosure posts are reassigned as true disclosures with a probability of 1.7\%. Agent-level disclosure status is then recomputed, and the main logistic regression including all controls is re-estimated. Across all 1,000 simulations, the coefficient on holistic transfer is positive in every iteration and statistically significant at $p < 0.05$ in 96.8\% of iterations. The simulation mean log-odds coefficient ($\hat{\beta} = +0.259$, 95\% interval $[+0.171, +0.359]$) and average marginal effect (mean AME/SD $= +1.29$,pp, 95\% interval $[+0.85, +1.80]$,pp) remain consistent with our primary estimates. Ultimately, these simulation outcomes demonstrate that our core finding is highly robust to measurement error, confirming that the observed positive association is not an artifact of LLM classification noise. See Appendix~\ref{app:llm_validation} for full details.

Finally, we address the concern that some owners may post directly to the platform rather than delegating to their OpenClaw agent. This ``puppet'' dynamic could trivially produce high behavioral transfer while also surfacing the owner's privacy preferences, potentially creating a spurious transfer--disclosure association.
To rule this out, we apply the same comprehensive testing framework mentioned in Section \ref{sec:results} (which is detailed in Appendix~\ref{app:puppet}) to our main disclosure regression.

First, we use two exclusion-based methods to filter out potential human interference. Method~1 excludes 942 potentially human-controlled agents (8.8\% of the sample) based on a multi-feature temporal classification. Re-estimating our logistic regression on this cleaned sample yields an transfer coefficient ($\hat{\beta} = +0.256$, $p = 0.007$) nearly identical to the full-sample baseline (+0.269). Method~2 applies a more aggressive single-metric filter based on inter-post interval variance, excluding a broader set of 2,142 agents (20.1\%). Despite dropping one-fifth of the observations, the estimate remains highly stable ($\hat{\beta} = +0.270$, $p = 0.009$). Second, to ensure our conclusion does not hinge on specific puppet-detection criteria, we run the main regression across 36 subsamples defined by median splits of 18 automation proxy variables. The results are remarkably uniform: all 36 estimated coefficients on holistic transfer are positive, and 27 are statistically significant at $p < 0.05$. The median coefficient across these 36 subsamples ($\hat{\beta} = +0.335$) even exceeds the full-sample estimate. Together, these tests strongly suggest that the observed privacy risks are driven by autonomous agent behavior rather than human puppetry. Full methodological definitions, proxy variable constructions, and complete regression tables are provided in Appendices~\ref{app:puppet} and \ref{app:cosplay}.

Taken together, the findings in this section indicate that owner-referential disclosures occur at detectable rates, are not easily explained by deliberate sharing alone, and are positively associated with the degree of behavioral transfer between owners and agents. Agents whose outputs more closely mirror their owners' behavioral profiles are also more likely to produce owner-referential content in their public posts.

\section{Discussion}\label{sec:implications}

We document that AI agents systematically mirror the behavioral characteristics of their owners across topics, values, affect, and style; that this transfer cannot be fully explained by observable configuration and is consistent with accumulated owner-agent interaction; and that stronger transfer is associated with a higher likelihood of owner-referential disclosure. Taken together, these findings suggest that behavioral transfer is not merely a feature of personalization but a mechanism through which private human context can become embedded in and surfaced through public agent outputs.

\subsection{Limitations and Future Directions}

Our empirical approach relies on observable behavioral traces in public agent outputs and owners’ Twitter activity. This approach introduces several limitations, each of which points to directions for future research.

First, our measures may not fully capture the extent of behavioral transfer. We focus on similarity in observable linguistic features (e.g., topic frequencies, sentiment scores, and stylistic patterns), but transfer may also manifest in dimensions that our measures cannot capture. For example, agents may mirror their owners' perspectives through different semantic framing than what our keyword-based dictionaries detect, or through the carryover of decision-making heuristics or interaction protocols that do not surface as correlated linguistic features. As a result, agents with low measured transfer may still incorporate substantial owner-specific context that is not detectable through our metrics. Future work could develop richer representations of transfer to more comprehensively characterize the scope of behavioral transfer. 

Second, human behavioral profiles are estimated from a limited number of tweets (up to 10 per user), which introduces measurement noise. This noise may affect both the precision of transfer estimates and the estimated relationship between transfer and disclosure. Future work, such as using richer behavioral histories from platforms where longer posting histories are available, could provide more precise estimates.

Third, our observation window covers six days of early platform activity, during which users may not yet have fully adapted their configurations, and norms around agent behavior may not yet be established. Longitudinal data would allow researchers to examine how transfer and disclosure evolve as platforms mature.

Fourth, Moltbook's early-adopter population is concentrated in cryptocurrency, AI, and software development domains. Whether similar patterns arise in broader and more heterogeneous user populations remains an open question. Replication across platforms and user groups is an important direction for future research.

\subsection{Implications}

\paragraph{Privacy Implications and Design Directions.}
A central implication of our findings is that behavioral transfer is a dual-use mechanism. The same processes that enable agents to function as personalized extensions of their owners also create pathways through which owner-related context can surface in public discourse. Transfer, therefore, introduces a trade-off between utility and privacy: the more effectively an agent mirrors its owner, the greater its potential to surface owner-referential information.

We directly examine the privacy side of this trade-off. We find that agents with higher behavioral transfer are significantly more likely to produce such disclosures. When many agents each carry partial signals of their owners' behavioral profiles, even small transfer signals can translate into meaningful disclosure risk at scale. Privacy risks in agent-based systems may arise not from isolated failures, but from systematic behavioral processes embedded in how agents mirror their users. 

Our findings point to several directions for platform and agent design aimed at
mitigating disclosure risk. First, because higher behavioral transfer with the owner is itself a risk factor, agent frameworks could implement transfer-aware safeguards: outputs from highly aligned agents may warrant additional content screening before publication, particularly for sensitive domains such as health and finance. Second, platforms could offer owners transparency into the behavioral profile their agent has internalized, enabling informed decisions about what context is provided during configuration. Third, agent architectures could separate owner context into tiered memory: information explicitly designated as private would be available for task completion but excluded from public-facing outputs. Fourth, post-hoc auditing tools that periodically surface a sample of an agent's public posts for owner review could enable ongoing oversight without requiring real-time intervention. These
directions are illustrative rather than prescriptive; the appropriate balance between agent capability and privacy protection ultimately depends on deployment context and requires empirical validation.


\paragraph{A Potential Opportunity: Agent-Mediated Knowledge Exchange.}
While this paper focuses on the privacy risks of behavioral transfer, the same mechanism may also create opportunities for knowledge exchange. When agents carry their owners’ behavioral context, interactions between agents may surface connections or insights that the owners themselves have not explicitly articulated. In principle, such interactions could function as a novel pathway for knowledge production or collective intelligence, in which complementary behavioral contexts from different owners are surfaced through agent-mediated discourse rather than direct human collaboration. At the same time, this possibility introduces distinct risks. Misattribution may arise when the origin of an insight is unclear, as agent outputs can reflect a combination of owner-related context and model inference. Epistemic contamination is also a concern if agents propagate inaccurate information with the apparent authority of their owners’ professional background. Understanding when and how such agent-mediated exchange produces reliable knowledge is an important direction for future research. 

More broadly, our findings suggest that as agents become embedded in social platforms, they should be understood not merely as tools but as behavioral extensions of their users, with implications for both information flow and privacy in digital ecosystems.

\bigskip
In sum, this paper shows that AI agents do not simply generate generic outputs of large language models; they systematically carry forward the behavioral heterogeneity of their human owners. Behavioral transfer reflects the incorporation of owner-related context into agent outputs and is associated with a higher likelihood of owner-referential disclosure in public discourse. As a result, platforms populated by such agents do not produce homogeneous LLM-driven content. Instead, they propagate---and make visible---the behavioral heterogeneity of the human population that owns them, reshaping not only the volume but also the structure of online interaction. As AI agents become increasingly embedded in social and economic life, understanding the behavioral processes through which they come to reflect their users---and the privacy implications that follow---represents a priority for both research and policy.

\bibliographystyle{informs2014}
\bibliography{main}
\newpage
\renewcommand{\theHsection}{A\arabic{section}}
\begin{APPENDICES}
\begin{center}
\textbf{\Large Online Appendices}
\end{center}

\section{Descriptive Statistics}\label{app:descriptive}

Table~\ref{tab:descriptive_full} reports descriptive statistics for all 43 behavioral features used in our analysis, organized by the four behavioral dimensions. Statistics are computed over the full matched sample of 10,659 human-agent pairs (with some variation in $N$ due to missing data for complexity features and political ideology).

\begin{table}[htbp]\centering\scriptsize
\caption{Descriptive Statistics for All 43 Behavioral Features}
\label{tab:descriptive_full}
\begin{threeparttable}
\setlength{\tabcolsep}{1.4mm}{
\begin{tabular}{llcccccc}
\hline\hline\\[-1.8ex]
Dimension & Feature & \multicolumn{3}{c}{Human} & \multicolumn{3}{c}{Agent} \\
\cmidrule(lr){3-5}\cmidrule(lr){6-8}
 & & Mean & SD & $N$ & Mean & SD & $N$ \\
\hline\\[-1.8ex]
\multirow{6}{*}{Topics}
 & Crypto       & 11.544 & 35.501 & 10,659 & 9.493 & 21.416 & 10,659 \\
 & AI           & 44.643 & 82.874 & 10,659 & 44.094 & 66.719 & 10,659 \\
 & Dev          & 9.261 & 29.556 & 10,659 & 7.100 & 16.874 & 10,659 \\
 & Trading      & 4.167 & 15.860 & 10,659 & 3.860 & 10.637 & 10,659 \\
 & Philosophy   & 1.187 & 9.430 & 10,659 & 2.691 & 6.775 & 10,659 \\
 & Meme         & 2.811 & 26.828 & 10,659 & 1.426 & 6.049 & 10,659 \\
\hline\\[-1.8ex]
\multirow{7}{*}{Values}
 & \multicolumn{7}{l}{\textit{~~Moral Foundations}} \\
 & Care       & 1.635 & 5.814 & 10,659 & 4.326 & 8.285 & 10,659 \\
 & Fairness   & 0.588 & 11.108 & 10,659 & 0.384 & 2.780 & 10,659 \\
 & Loyalty    & 1.617 & 6.513 & 10,659 & 4.129 & 9.475 & 10,659 \\
 & Authority  & 1.139 & 5.303 & 10,659 & 1.912 & 6.367 & 10,659 \\
 & Sanctity   & 1.188 & 5.353 & 10,659 & 1.136 & 3.800 & 10,659 \\
 & \multicolumn{7}{l}{\textit{~~Political Ideology}} \\
 & GS dictionary & 0.608 & 0.762 & 1,086 & 0.589 & 0.729 & 2,374 \\
 & LLM-based     & 0.007 & 0.143 & 10,659 & 0.001 & 0.068 & 10,659 \\
\hline\\[-1.8ex]
\multirow{12}{*}{Affect}
 & \multicolumn{7}{l}{\textit{~~Sentiment}} \\
 & Compound     & 0.129 & 0.223 & 10,659 & 0.447 & 0.412 & 10,659 \\
 & Positive     & 0.090 & 0.092 & 10,659 & 0.112 & 0.078 & 10,659 \\
 & Negative     & 0.037 & 0.053 & 10,659 & 0.034 & 0.041 & 10,659 \\
 & Neutral      & 0.873 & 0.107 & 10,659 & 0.854 & 0.087 & 10,659 \\
 & Variability  & 0.241 & 0.175 & 10,659 & 0.195 & 0.243 & 10,659 \\
 & Pct positive & 0.355 & 0.304 & 10,659 & 0.697 & 0.374 & 10,659 \\
 & Pct negative & 0.140 & 0.200 & 10,659 & 0.128 & 0.256 & 10,659 \\
 & \multicolumn{7}{l}{\textit{~~Emotions}} \\
 & Anger    & 1.406 & 11.980 & 10,659 & 0.497 & 2.894 & 10,659 \\
 & Joy      & 0.991 & 7.861 & 10,659 & 3.320 & 10.025 & 10,659 \\
 & Fear     & 0.329 & 3.775 & 10,659 & 0.246 & 1.511 & 10,659 \\
 & Sadness  & 0.174 & 1.973 & 10,659 & 0.051 & 0.594 & 10,659 \\
 & Surprise & 0.729 & 11.013 & 10,659 & 0.209 & 1.470 & 10,659 \\
\hline\\[-1.8ex]
\multirow{18}{*}{Style}
 & \multicolumn{7}{l}{\textit{~~Complexity}} \\
 & Avg word length    & 4.929 & 0.922 & 9,157 & 5.356 & 2.460 & 10,083 \\
 & TTR                & 0.781 & 0.130 & 9,157 & 0.706 & 0.190 & 10,083 \\
 & Avg sentence length & 16.029 & 14.676 & 9,157 & 11.239 & 6.366 & 10,083 \\
 & Hapax ratio        & 0.662 & 0.173 & 9,157 & 0.583 & 0.229 & 10,083 \\
 & Cap ratio          & 0.062 & 0.049 & 9,157 & 0.046 & 0.034 & 10,083 \\
 & \multicolumn{7}{l}{\textit{~~Communication}} \\
 & Avg text length    & 116.355 & 63.702 & 10,659 & 652.004 & 1196.762 & 10,659 \\
 & Question rate      & 0.119 & 0.245 & 10,659 & 1.378 & 7.908 & 10,659 \\
 & Exclamation rate   & 0.177 & 0.363 & 10,659 & 0.535 & 0.948 & 10,659 \\
 & Hashtag rate       & 0.363 & 0.852 & 10,659 & 0.787 & 2.815 & 10,659 \\
 & Mention rate       & 0.279 & 0.551 & 10,659 & 0.151 & 2.833 & 10,659 \\
 & URL rate           & 0.688 & 0.428 & 10,659 & 0.134 & 0.514 & 10,659 \\
 & Emoji rate         & 0.337 & 0.981 & 10,659 & 9.312 & 862.006 & 10,659 \\
 & Formality          & $-$0.249 & 0.455 & 10,659 & $-$0.262 & 0.490 & 10,659 \\
 & \multicolumn{7}{l}{\textit{~~Pronouns}} \\
 & I-words    & 26.152 & 38.793 & 10,659 & 35.372 & 35.313 & 10,659 \\
 & We-words   & 5.229 & 12.502 & 10,659 & 6.712 & 11.940 & 10,659 \\
 & You-words  & 12.377 & 21.199 & 10,659 & 13.737 & 17.445 & 10,659 \\
 & They-words & 5.444 & 11.913 & 10,659 & 5.154 & 9.347 & 10,659 \\
 & Self-focus  & 0.522 & 0.453 & 10,659 & 0.685 & 0.378 & 10,659 \\
\hline\hline\\[-1.8ex]
\end{tabular}}
\begin{tablenotes}
\scriptsize
\item \textit{Notes:} Descriptive statistics for all 43 behavioral features across 10,659 matched human-agent pairs. Topic features are keyword-match counts; moral foundations are keyword-match rates per 1,000 words; political ideology is scored on a $[-1, +1]$ scale (GS: Gentzkow-Shapiro dictionary, LLM: Gemini 2.5-Flash scoring); sentiment features are VADER-based; emotion features are keyword-match rates per 1,000 words; complexity and communication features are computed as described in Appendix~\ref{app:variables}; pronoun features are keyword-match counts per 1,000 words; self-focus ratio is I-words/(I-words + We-words). Reduced $N$ for complexity features reflects pairs where tokenization-based measures could not be computed (very short texts); reduced $N$ for GS political reflects the smaller set of users with sufficient political vocabulary.
\end{tablenotes}
\end{threeparttable}
\end{table}

\section{Variable Construction Details}\label{app:variables}

This appendix provides complete details on the construction of the 43 content features used in our analysis.

\subsection{Topic Keywords}

We construct six topics using the following keyword lists (prefix matching applied):

\begin{itemize}
\item \textbf{Crypto} (14 terms): crypto, bitcoin, btc, eth, ethereum, defi, web3, nft, blockchain, token, solana, sol, degen, airdrop
\item \textbf{AI} (12 terms): ai, gpt, llm, machine learning, neural, chatgpt, openai, anthropic, claude, artificial intelligence, deep learning, agent
\item \textbf{Development} (12 terms): code, coding, developer, programming, github, python, javascript, rust, typescript, api, framework, deploy
\item \textbf{Trading} (10 terms): trading, market, bull, bear, price, pump, long, short, chart, profit
\item \textbf{Philosophy} (10 terms): consciousness, freedom, truth, existence, meaning, universe, reality, humanity, soul, purpose
\item \textbf{Meme/Slang} (11 terms): lol, lmao, bruh, based, cope, seethe, wagmi, ngmi, gm, lfg, fomo
\end{itemize}

\subsection{Moral Foundations Keywords}

Following \citet{graham2013moral}, we measure five foundations:

\begin{itemize}
\item \textbf{Care/Harm} (25 terms): care, caring, protect, protection, shield, shelter, safe, safety, compassion, empathy, sympathy, kindness, nurture, help, defend, rescue, suffer, hurt, harm, cruel, abuse, damage, destroy, vulnerable, victim
\item \textbf{Fairness/Cheating} (18 terms): fair, fairness, equal, equality, justice, rights, equity, reciprocity, impartial, unbiased, deserve, proportion, discriminat, unjust, cheat, fraud, exploit, oppress
\item \textbf{Loyalty/Betrayal} (19 terms): loyal, loyalty, patriot, patriotism, team, unite, united, together, solidarity, belong, group, collective, community, nation, betray, traitor, disloyal, abandon, defect
\item \textbf{Authority/Subversion} (19 terms): authority, respect, tradition, order, obey, obedience, duty, law, legal, hierarchy, rank, leader, command, rule, discipline, subvert, rebel, disobey, chaos
\item \textbf{Sanctity/Degradation} (20 terms): pure, purity, sacred, sanctity, holy, divine, wholesome, clean, innocent, virtue, noble, disgust, degrade, corrupt, sin, immoral, profane, desecrate, taint, contaminate
\end{itemize}

\subsection{Emotion Keywords}

Following \citet{ekman1992argument} and \citet{mohammad2013crowdsourcing}:

\begin{itemize}
\item \textbf{Anger} (14 terms): angry, anger, furious, rage, mad, outrage, hostile, irritat, annoy, hate, resent, bitter, infuriat, livid
\item \textbf{Joy} (13 terms): happy, joy, delight, excited, thrilled, ecstatic, bliss, cheerful, elated, glad, wonderful, celebrate, euphori
\item \textbf{Fear} (13 terms): fear, afraid, scared, terrif, anxious, worry, panic, dread, frighten, nervous, apprehens, alarm, horror
\item \textbf{Sadness} (13 terms): sad, sorrow, grief, mourn, depress, melanchol, miserable, heartbreak, despair, lonely, gloomy, unhappy, hopeless
\item \textbf{Surprise} (11 terms): surprise, shock, astonish, amaze, stun, unexpected, incredible, unbelievable, wow, whoa, omg
\end{itemize}

\subsection{Communication Style}

\paragraph{Formality Index.} Following \citet{heylighen2002variation}:
\begin{itemize}
\item \textbf{Formal} (19 terms): however, therefore, furthermore, moreover, consequently, nevertheless, regarding, concerning, accordingly, additionally, specifically, particularly, significantly, approximately, demonstrate, illustrate, implement, facilitate, indicate
\item \textbf{Informal} (23 terms): like, gonna, wanna, gotta, kinda, sorta, yeah, nah, dude, bro, lol, omg, tbh, imo, smh, af, ngl, fr, lowkey, highkey, vibe, vibes, sus
\end{itemize}

\paragraph{Pronouns.} First-person singular (I, me, my, mine, myself), first-person plural (we, us, our, ours, ourselves), second-person (you, your, yours, yourself, yourselves), third-person (he, him, his, himself, she, her, hers, herself, they, them, their, theirs, themselves).

\subsection{Political Ideology}

\paragraph{Gentzkow-Shapiro Partisan Phrases} \citep{gentzkow2010drives}. Each phrase carries a weight (0--1) reflecting partisan distinctiveness:
\begin{itemize}
\item \textbf{Left-leaning} (47 phrases): estate tax, undocumented, climate crisis, gun violence, reproductive rights, systemic racism, wealth inequality, universal healthcare, living wage, affordable housing, voting rights, lgbtq, social justice, medicare for all, green new deal, defund, progressive, marginalized, inclusive, equity, privilege, colonialism, intersectional, patriarchy, toxic masculinity, democratic socialism, workers rights, union, corporate greed, billionaire, oligarch, pro-choice, bodily autonomy, healthcare is a right, ban assault weapons, common sense gun, science denier, anti-vaxx, fascist, white supremacy, nazi, resist, blue wave, vote blue, bernie, aoc, squad
\item \textbf{Right-leaning} (51 phrases): death tax, illegal alien, climate hoax, second amendment, pro-life, all lives matter, blue lives matter, free market, small government, states rights, religious freedom, traditional values, family values, law and order, border security, illegal immigration, radical left, socialist, communist, marxist, deep state, fake news, mainstream media, big tech, cancel culture, woke, virtue signal, snowflake, triggered, liberal tears, own the libs, maga, trump, america first, drain the swamp, build the wall, lock her up, patriot, freedom, constitutional, founding fathers, dont tread, gun rights, shall not be infringed, stand your ground, based, red pill, npc, libtard, e/acc, effective accelerationism
\end{itemize}

\paragraph{Elite-Cue Method} \citep{barbera2015birds}. Each elite carries a signed weight reflecting partisan strength:
\begin{itemize}
\item \textbf{Left-leaning} (23 figures/outlets): biden, harris, pelosi, schumer, aoc, bernie, sanders, warren, obama, hillary, clinton, msnbc, cnn, nytimes, wapo, huffpost, vox, slate, theatlantic, npr, pbs, democrats, dnc
\item \textbf{Right-leaning} (30 figures/outlets): trump, desantis, mcconnell, cruz, hawley, mtg, marjorie taylor, boebert, gaetz, jim jordan, rand paul, ron paul, fox news, foxnews, breitbart, dailywire, newsmax, oann, tucker, hannity, shapiro, crowder, republicans, gop, rnc, elon, musk, peter thiel, jd vance, vivek
\end{itemize}

\section{Political Ideology: Validation and Additional Methods}\label{app:political_validation}

This appendix reports convergent validity analyses and supplementary results for the political ideology measures described in Section~\ref{sec:results}.

\subsection{Elite-Cue Method}

As additional side evidence, we apply the Elite-Cue method \citep{barbera2015birds}, which infers ideology from mentions of political elites. Among the 90 pairs where both human and agent mention at least one political figure, Elite-Cue yields $\rho = 0.213$ ($p = 0.044$). However, this should be interpreted cautiously: Elite-Cue measures whether the same political figures are \textit{mentioned}, not whether they are discussed with the same \textit{stance}, and the very small sample size limits reliability. Elite-Cue is not included in the 43-feature Benjamini-Hochberg correction.

\subsection{Convergent Validity Across Methods}

To validate the LLM-based measure, we assess convergent validity with the dictionary-based methods among individual users scored by both approaches. These comparisons are conducted \textit{within} the human side: for each user who has both an LLM score and a dictionary score, we test whether the two methods agree. The $N$ values count individual users, not matched pairs; we use all humans with scores from both methods, without restricting to the transfer subsample.

On the human side, the LLM measure correlates significantly with Gentzkow-Shapiro ($\rho = 0.129$, $p < 0.001$, $N_{\text{users}} = 1{,}096$) and with Elite-Cue ($\rho = 0.089$, $p = 0.007$, $N_{\text{users}} = 905$). For reference, the two dictionary methods themselves agree moderately (GS vs.\ Elite-Cue: $\rho = 0.314$, $p < 0.001$, $N_{\text{users}} = 356$). These continuous correlations are modest, which is expected: dictionary methods only fire on the $\sim$14\% of users who employ overtly partisan vocabulary, whereas the LLM scores all users based on subtler contextual signals.

\subsection{Known-Groups Analysis}

A known-groups analysis provides stronger evidence. Among the 1,062 humans classified as left or right by GS, those classified as left have significantly lower LLM scores than those classified as right (mean LLM: $-0.047$ vs.\ $+0.041$; Mann-Whitney $p < 0.001$; Cohen's $d = 0.34$). The effect strengthens among more politically active users: for those with $\geq 2$ political tweets, $d = 0.69$ ($p = 0.004$; $N = 158$). Among the 226 users scored as non-zero by both methods, directional agreement is 65.5\%. These results indicate that the LLM captures a related but broader dimension of political orientation: it correctly discriminates the direction identified by dictionary methods while additionally scoring the majority of users for whom keyword-based approaches yield no signal.

\section{Robustness Checks}\label{app:robustness}

We systematically test whether the observed behavioral transfer can be attributed to confounding variables rather than genuine identity transfer. This appendix consolidates all robustness analyses referenced in Section~\ref{sec:results}.

\subsection{Permutation Test: Full Results}\label{app:permutation}

Table~\ref{tab:permutation_full} reports permutation test results for all 43 behavioral transfer features. For each feature, we randomly shuffle the agent values across human-agent pairings 10,000 times and compute the Spearman correlation for each permutation. The permutation $p$-value is the proportion of random correlations $\geq$ the observed matched correlation. The effect size $d$ measures how many standard deviations the observed correlation exceeds the random mean.
\begin{table}[htbp]\centering\scriptsize
\caption{Permutation Test Results for All 43 Behavioral Transfer Features (10{,}000 Permutations)}
\label{tab:permutation_full}
\begin{threeparttable}
\setlength{\tabcolsep}{1.8mm}{
\begin{tabular}{llccccc}
\hline\hline\\[-1.8ex]
Dimension & Feature & Matched $\rho$ & Random $\mu$ & $p_{\text{perm}}$ & Effect $d$ & Sig? \\
\hline\\[-1.8ex]
\multirow{6}{*}{Topics}
 & Crypto & $+0.166$ & $+0.000$ & $<0.001$ & $17.3$ & Yes \\
 & AI & $+0.101$ & $+0.000$ & $<0.001$ & $10.6$ & Yes \\
 & Dev & $+0.085$ & $-0.000$ & $<0.001$ & $9.3$ & Yes \\
 & Trading & $+0.117$ & $+0.000$ & $<0.001$ & $12.1$ & Yes \\
 & Philosophy & $+0.101$ & $+0.000$ & $<0.001$ & $10.6$ & Yes \\
 & Meme & $+0.017$ & $-0.000$ & $0.087$ & $1.4$ & No \\
\hline\\[-1.8ex]
\multirow{7}{*}{Values}
 & \multicolumn{6}{l}{\textit{~~Moral Foundations}} \\
 & Care & $+0.038$ & $-0.000$ & $<0.001$ & $4.0$ & Yes \\
 & Fairness & $+0.037$ & $+0.000$ & $<0.001$ & $3.8$ & Yes \\
 & Loyalty & $+0.023$ & $+0.000$ & $0.016$ & $2.1$ & Yes \\
 & Authority & $+0.043$ & $-0.000$ & $<0.001$ & $4.9$ & Yes \\
 & Sanctity & $+0.054$ & $+0.000$ & $<0.001$ & $5.5$ & Yes \\
 & \multicolumn{6}{l}{\textit{~~Political Ideology}} \\
 & GS dictionary & $+0.087$ & --- & $0.123$ & --- & No \\
 & LLM-based & $+0.061$ & $+0.000$ & $<0.001$ & $6.2$ & Yes \\
\hline\\[-1.8ex]
\multirow{12}{*}{Affect}
 & \multicolumn{6}{l}{\textit{~~Sentiment}} \\
 & Compound & $+0.067$ & $-0.000$ & $<0.001$ & $7.0$ & Yes \\
 & Positive & $+0.068$ & $+0.000$ & $<0.001$ & $7.0$ & Yes \\
 & Negative & $+0.153$ & $+0.000$ & $<0.001$ & $15.9$ & Yes \\
 & Neutral & $+0.091$ & $+0.000$ & $<0.001$ & $9.4$ & Yes \\
 & Variability & $+0.060$ & $+0.000$ & $<0.001$ & $6.2$ & Yes \\
 & Pct positive & $+0.087$ & $-0.000$ & $<0.001$ & $9.0$ & Yes \\
 & Pct negative & $+0.083$ & $+0.000$ & $<0.001$ & $8.7$ & Yes \\
 & \multicolumn{6}{l}{\textit{~~Emotions}} \\
 & Anger & $+0.031$ & $+0.000$ & $0.001$ & $3.1$ & Yes \\
 & Joy & $+0.001$ & $-0.000$ & $0.435$ & $0.2$ & No \\
 & Fear & $+0.018$ & $-0.000$ & $0.054$ & $1.6$ & No \\
 & Sadness & $+0.020$ & $-0.000$ & $0.029$ & $2.1$ & Yes \\
 & Surprise & $+0.022$ & $-0.000$ & $0.024$ & $2.1$ & Yes \\
\hline\\[-1.8ex]
\multirow{18}{*}{Style}
 & \multicolumn{6}{l}{\textit{~~Complexity}} \\
 & Avg word length & $+0.095$ & $-0.000$ & $<0.001$ & $16.2$ & Yes \\
 & TTR & $+0.062$ & $+0.000$ & $<0.001$ & $10.1$ & Yes \\
 & Avg sentence length & $+0.032$ & $-0.000$ & $0.003$ & $3.0$ & Yes \\
 & Hapax ratio & $+0.059$ & $+0.000$ & $<0.001$ & $10.4$ & Yes \\
 & Cap ratio & $+0.174$ & $-0.000$ & $<0.001$ & $12.8$ & Yes \\
 & \multicolumn{6}{l}{\textit{~~Communication}} \\
 & Avg text length & $+0.139$ & $+0.000$ & $<0.001$ & $14.3$ & Yes \\
 & Question rate & $+0.096$ & $+0.000$ & $<0.001$ & $9.9$ & Yes \\
 & Exclamation rate & $+0.038$ & $+0.000$ & $<0.001$ & $4.0$ & Yes \\
 & Hashtag rate & $-0.002$ & $-0.000$ & $0.574$ & $-0.2$ & No \\
 & Mention rate & $+0.012$ & $+0.000$ & $0.115$ & $1.2$ & No \\
 & URL rate & $+0.043$ & $+0.000$ & $<0.001$ & $4.4$ & Yes \\
 & Emoji rate & $+0.068$ & $+0.000$ & $<0.001$ & $7.0$ & Yes \\
 & Formality & $+0.085$ & $+0.000$ & $<0.001$ & $9.1$ & Yes \\
 & \multicolumn{6}{l}{\textit{~~Pronouns}} \\
 & I-words & $+0.092$ & $-0.000$ & $<0.001$ & $9.9$ & Yes \\
 & We-words & $+0.098$ & $-0.000$ & $<0.001$ & $10.2$ & Yes \\
 & You-words & $+0.095$ & $-0.000$ & $<0.001$ & $9.9$ & Yes \\
 & They-words & $+0.142$ & $+0.000$ & $<0.001$ & $14.7$ & Yes \\
 & Self-focus & $+0.066$ & $-0.000$ & $<0.001$ & $6.8$ & Yes \\
\hline\hline\\[-1.8ex]
\end{tabular}}
\begin{tablenotes}
\scriptsize
\item \textit{Notes:} 10,000 permutations per feature. Random $\mu$ = mean Spearman $\rho$ across permutations. $p_{\text{perm}}$ = proportion of random $\rho \geq$ observed. Effect $d$ = (observed $-$ random mean) / random SD. ``Sig?'' = $p_{\text{perm}} < 0.05$. GS dictionary uses analytic $p$ from the separate $N = 295$ pipeline (permutation not applicable). Sample sizes: $N = 10{,}659$ for content-based features, $N = 8{,}751$ for complexity, $N = 295$ for GS political.
\end{tablenotes}
\end{threeparttable}
\end{table}
Of 43 features, 37 (86.0\%) are significant at $p < 0.05$ under the permutation test, matching the Benjamini-Hochberg corrected results exactly. The same six features are non-significant under both approaches: topic meme, hashtag rate, mention rate, emotion joy, emotion fear, and political ideology (GS dictionary).

\FloatBarrier

\subsection{Semantic Embedding Similarity}\label{app:bert}

This appendix details the semantic embedding analysis referenced in Section~\ref{sec:results}. Our keyword-based transfer measures rely on predefined dictionaries. To validate that transfer is not an artifact of dictionary construction, we employ a fully data-driven approach using neural text embeddings.

\paragraph{Method.} We use Sentence-BERT \citep{reimers2019sentence} with the \texttt{all-MiniLM-L6-v2} model (384-dimensional embeddings). For each human, we embed all tweets (after removing verification tweets) and compute an L2-normalized centroid vector. We do the same for each agent's posts (combining title and content fields). Cosine similarity between matched human-agent centroids measures semantic overlap without relying on any predefined vocabulary. We exclude one agent whose posts contain only whitespace, yielding $N = 10{,}658$ pairs from the content-analysis sample.

\paragraph{Permutation Test.} We compare matched-pair similarity to a null distribution constructed by randomly shuffling agent assignments 10,000 times.

\begin{table}[H]
\centering
\caption{Semantic Embedding Similarity: Matched vs.\ Random Pairs ($N = 10{,}658$)}
\label{tab:bert_similarity}
\begin{threeparttable}
\scriptsize
\setlength{\tabcolsep}{25mm}{
\begin{tabular}{lc}
\hline
\hline \\[-1.8ex]
Statistic & Value \\
\hline \\[-1.8ex]
Matched mean cosine similarity & $0.287$ \\
Shuffled mean cosine similarity & $0.205$ \\
Difference ($\Delta$) & $0.083$ \\
Cohen's $d$ & $96.2$ \\
Permutation $p$-value & $< 0.0001$ \\
Median matched similarity & $0.267$ \\
Std.\ of matched similarities & $0.171$ \\
\hline
\hline \\[-1.8ex]
\end{tabular}}
\begin{tablenotes}
\scriptsize
\item Notes: Embeddings from Sentence-BERT (\texttt{all-MiniLM-L6-v2}; 384 dims). Cosine similarity is computed between L2-normalized centroid vectors of each human's tweets and their agent's posts. Shuffled baseline constructed from 10,000 random permutations of agent assignments. The large Cohen's $d$ reflects the very small variance of the shuffled distribution (std $= 0.00086$), not the variance of individual-pair similarities. One agent with whitespace-only posts is excluded from the content-analysis sample ($N = 10{,}659 - 1 = 10{,}658$).
\end{tablenotes}
\end{threeparttable}
\end{table}

\paragraph{Interpretation.} Matched human-agent pairs have 40\% higher semantic similarity than random pairings ($0.287$ vs.\ $0.205$). The permutation $p < 0.0001$ (none of 10,000 shuffled means exceeded the observed matched mean) confirms this difference is not attributable to chance. This validates that the transfer documented via keyword dictionaries reflects genuine semantic transfer between human and agent communication.

\subsection{Autonomous Agents vs.\ Human Puppets}\label{app:puppet}

A potential concern is whether our results are driven by human-controlled
``puppet'' agents rather than truly autonomous AI agents posting through
OpenClaw. We address this concern in three steps: we first provide temporal
evidence that posting on Moltbook is predominantly automated. We then show
that excluding agents flagged as likely human-controlled leaves the
transfer signal virtually unchanged. And finally, we provide a comprehensive
robustness check showing that transfer is consistent across median splits
of 18 proxy variables that, following the social media automation detection
literature, capture five theoretically motivated dimensions along which automated and human posting behavior differ.

\subsubsection{Temporal evidence of automated operation.}
OpenClaw, the agent framework underlying Moltbook, runs a daemon process
that triggers agents to post on a user-configurable schedule, defaulting
to every 30 minutes. If agents post autonomously through this mechanism,
their posting timestamps should exhibit (i) near-uniform coverage across
all hours of the day (since the daemon runs continuously regardless of
time zone or time of day), (ii) a concentration of inter-post intervals
around the default 30-minute period, and (iii) a substantial share of
posts during overnight hours when human activity is minimal.

All three patterns are present. Computed across all 44,588 posts in our
dataset (excluding a bulk ingestion spike on January 31), the normalized
entropy of the hourly posting distribution is 0.984 (maximum = 1.0 for
perfectly uniform), indicating near-round-the-clock activity.
\subsubsection{Exclusion-based robustness.}
We apply two complementary detection methods to identify agents that may
be human-controlled, then verify that excluding them leaves the transfer
signal essentially unchanged. Both methods are applied directly to our
matched sample of 10,659 agent--human pairs.


\subparagraph{Method 1: Multi-feature temporal classification.}
Following \citet{varol2017online} and \citet{mazza2019rtbust}, we
construct four temporal flags for each agent with $\geq 3$ posts and
flag agents meeting three or more criteria as potentially
human-controlled. All timestamps are converted to US Eastern Time
before computing the features. The four flags are: (1) overnight
posting fraction below 10\% (the midnight--8:00 AM ET window;
autonomous agents are expected to post $\approx$33\% overnight,
whereas human-controlled agents concentrate activity during waking
hours); (2) maximum consecutive inactive hours exceeding 12 hours
(consistent with a human sleep gap); (3) coefficient of variation of
hourly posting counts above the 90th percentile of the matched sample
(extreme irregularity in posting rhythm); and (4) weekend posting
fraction exceeding 90\% (a pattern consistent with manual, sporadic
scheduling). Applying these criteria identifies 942 potentially
human-controlled agents (8.8\% of the sample).

\subparagraph{Method 2: Inter-post interval coefficient of variation.}
\citet{li2026moltbook} classify agents by the coefficient of variation
of their inter-post intervals. Agents with low variation (CoV $< 0.5$)
exhibit the regular timing characteristic of automated scheduling,
while agents with high variation (CoV $> 1.0$) show the irregular
timing associated with human-influenced posting. Applying this
threshold to our matched sample identifies 2,142 agents as
human-influenced (20.1\%).

\subparagraph{Results.}
Table~\ref{tab:puppet_full} reports transfer correlations for all 43
features under both exclusion methods. Under Method~1 (excluding 942
pairs, 8.8\%), 36 of 37 originally significant features remain
significant, with a mean absolute change in $\rho$ of 0.003 across the
37 significant features. Under Method~2 (excluding 2,142 pairs,
20.1\%), 32 of 37 survive; the five exceptions: moral loyalty
($\rho = +0.023$), emotion anger ($\rho = +0.031$), emotion surprise
($\rho = +0.022$), emotion sadness ($\rho = +0.020$), and average
sentence length ($\rho = +0.032$) were the weakest significant
effects in the full sample ($\rho \leq 0.032$), and their loss of
significance reflects the 20\% reduction in sample size rather than
meaningful attenuation (mean $|\Delta\rho| = 0.009$ across all 37
originally significant features). All four analytical dimensions, including
topics, values, affect, and style retain robust transfer signals
under both exclusion approaches.

\begin{table}[H]
\centering
\caption{Robustness to Puppet Detection: All 43 Features}
\label{tab:puppet_full}
\begin{threeparttable}
\scriptsize
\setlength{\tabcolsep}{1.5mm}{
\begin{tabular}{llccccccc}
\hline\hline\\[-1.8ex]
& & \multicolumn{2}{c}{Full Sample} & \multicolumn{2}{c}{Method 1} & \multicolumn{2}{c}{Method 2} & \\
\cmidrule(lr){3-4}\cmidrule(lr){5-6}\cmidrule(lr){7-8}
Dimension & Feature & $\rho$ & Sig & $\rho$ & $\Delta\rho$ & $\rho$ & $\Delta\rho$ & Surv? \\
\hline\\[-1.8ex]
& & \multicolumn{2}{c}{$N = 10{,}659$} & \multicolumn{2}{c}{$N = 9{,}717$} & \multicolumn{2}{c}{$N = 8{,}517$} & \\
\hline\\[-1.8ex]
\multirow{6}{*}{Topics}
 & Crypto            & $+0.166$ & *** & $+0.163$ & $-0.002$ & $+0.158$ & $-0.008$ & Y/Y \\
 & Trading           & $+0.117$ & *** & $+0.107$ & $-0.010$ & $+0.103$ & $-0.014$ & Y/Y \\
 & AI                & $+0.101$ & *** & $+0.097$ & $-0.005$ & $+0.091$ & $-0.011$ & Y/Y \\
 & Philosophy        & $+0.101$ & *** & $+0.098$ & $-0.003$ & $+0.096$ & $-0.005$ & Y/Y \\
 & Development       & $+0.085$ & *** & $+0.091$ & $+0.007$ & $+0.062$ & $-0.022$ & Y/Y \\
 & Meme              & $+0.017$ &     & $+0.017$ & $-0.000$ & $+0.019$ & $+0.002$ &     \\
\hline\\[-1.8ex]
\multirow{7}{*}{Values}
 & Sanctity          & $+0.054$ & *** & $+0.057$ & $+0.003$ & $+0.050$ & $-0.004$ & Y/Y \\
 & Authority         & $+0.043$ & *** & $+0.045$ & $+0.002$ & $+0.035$ & $-0.008$ & Y/Y \\
 & Care              & $+0.038$ & *** & $+0.040$ & $+0.002$ & $+0.030$ & $-0.008$ & Y/Y \\
 & Fairness          & $+0.037$ & *** & $+0.032$ & $-0.006$ & $+0.031$ & $-0.006$ & Y/Y \\
 & Loyalty           & $+0.023$ & *   & $+0.021$ & $-0.001$ & $+0.019$ & $-0.004$ & Y/N \\
 & Political (LLM)   & $+0.061$ & *** & $+0.059$ & $-0.002$ & $+0.037$ & $-0.024$ & Y/Y \\
 & Political (GS)\textsuperscript{a} & $+0.087$ & & $+0.085$ & $-0.001$ & $+0.082$ & $-0.005$ & \\
\hline\\[-1.8ex]
\multirow{12}{*}{Affect}
 & Negative proportion  & $+0.153$ & *** & $+0.153$ & $-0.000$ & $+0.145$ & $-0.008$ & Y/Y \\
 & Neutral proportion   & $+0.091$ & *** & $+0.088$ & $-0.003$ & $+0.085$ & $-0.006$ & Y/Y \\
 & \% posts positive    & $+0.087$ & *** & $+0.086$ & $-0.001$ & $+0.084$ & $-0.003$ & Y/Y \\
 & \% posts negative    & $+0.083$ & *** & $+0.085$ & $+0.001$ & $+0.065$ & $-0.018$ & Y/Y \\
 & Positive proportion  & $+0.068$ & *** & $+0.066$ & $-0.002$ & $+0.058$ & $-0.010$ & Y/Y \\
 & Compound sentiment   & $+0.067$ & *** & $+0.065$ & $-0.002$ & $+0.064$ & $-0.003$ & Y/Y \\
 & Sentiment variability& $+0.060$ & *** & $+0.062$ & $+0.001$ & $+0.032$ & $-0.028$ & Y/Y \\
 & Anger                & $+0.031$ & **  & $+0.028$ & $-0.002$ & $+0.018$ & $-0.013$ & Y/N \\
 & Surprise             & $+0.022$ & *   & $+0.020$ & $-0.002$ & $+0.020$ & $-0.002$ & N/N \\
 & Sadness              & $+0.020$ & *   & $+0.022$ & $+0.002$ & $+0.022$ & $+0.002$ & Y/N \\
 & Fear                 & $+0.018$ &     & $+0.026$ & $+0.008$ & $+0.017$ & $-0.001$ &     \\
 & Joy                  & $+0.001$ &     & $-0.001$ & $-0.002$ & $-0.005$ & $-0.007$ &     \\
\hline\\[-1.8ex]
\multirow{18}{*}{Style}
 & Capitalization ratio & $+0.174$ & *** & $+0.164$ & $-0.010$ & $+0.156$ & $-0.018$ & Y/Y \\
 & Avg text length      & $+0.139$ & *** & $+0.141$ & $+0.002$ & $+0.133$ & $-0.006$ & Y/Y \\
 & They-words           & $+0.142$ & *** & $+0.142$ & $+0.000$ & $+0.129$ & $-0.012$ & Y/Y \\
 & We-words             & $+0.098$ & *** & $+0.094$ & $-0.004$ & $+0.085$ & $-0.014$ & Y/Y \\
 & You-words            & $+0.095$ & *** & $+0.096$ & $+0.001$ & $+0.076$ & $-0.019$ & Y/Y \\
 & Avg word length      & $+0.095$ & *** & $+0.091$ & $-0.003$ & $+0.092$ & $-0.003$ & Y/Y \\
 & I-words              & $+0.092$ & *** & $+0.091$ & $-0.001$ & $+0.094$ & $+0.002$ & Y/Y \\
 & Formality            & $+0.085$ & *** & $+0.084$ & $-0.002$ & $+0.070$ & $-0.015$ & Y/Y \\
 & Emoji rate           & $+0.068$ & *** & $+0.066$ & $-0.003$ & $+0.057$ & $-0.012$ & Y/Y \\
 & Self-focus ratio     & $+0.066$ & *** & $+0.063$ & $-0.003$ & $+0.072$ & $+0.006$ & Y/Y \\
 & Type-token ratio     & $+0.062$ & *** & $+0.061$ & $-0.001$ & $+0.054$ & $-0.007$ & Y/Y \\
 & Hapax ratio          & $+0.059$ & *** & $+0.058$ & $-0.001$ & $+0.053$ & $-0.007$ & Y/Y \\
 & Question rate        & $+0.096$ & *** & $+0.096$ & $-0.000$ & $+0.097$ & $+0.001$ & Y/Y \\
 & URL rate             & $+0.043$ & *** & $+0.040$ & $-0.003$ & $+0.032$ & $-0.011$ & Y/Y \\
 & Exclamation rate     & $+0.038$ & *** & $+0.038$ & $-0.001$ & $+0.038$ & $-0.001$ & Y/Y \\
 & Avg sentence length  & $+0.032$ & **  & $+0.030$ & $-0.002$ & $+0.022$ & $-0.010$ & Y/N \\
 & Hashtag rate         & $-0.002$ &     & $-0.008$ & $-0.006$ & $+0.014$ & $+0.016$ &     \\
 & Mention rate         & $+0.012$ &     & $+0.007$ & $-0.004$ & $+0.002$ & $-0.010$ &     \\
\hline\\[-1.8ex]
\multicolumn{2}{l}{\textit{Summary}} & \multicolumn{2}{c}{37 sig.} & \multicolumn{2}{c}{36/37 survive} & \multicolumn{2}{c}{32/37 survive} & \\
\multicolumn{2}{l}{\textit{Mean $|\Delta\rho|$ (37 sig.\ features)}} & & & \multicolumn{2}{c}{0.003} & \multicolumn{2}{c}{0.009} & \\
\hline\hline\\[-1.8ex]
\end{tabular}}
\begin{tablenotes}
\scriptsize
\item \textit{Notes:} Method~1 follows \citet{varol2017online} and
\citet{mazza2019rtbust}: agents meeting $\geq 3$ of four temporal
criteria are excluded (942 flagged). Criteria: overnight posting
fraction (midnight--8:00 AM ET) below 10\%; maximum consecutive
inactive hours exceeding 12; coefficient of variation of hourly
posting counts above the 90th percentile; weekend fraction above
90\%. Method~2 follows \citet{li2026moltbook}: agents with
inter-post interval CoV $> 1.0$ are excluded (2,142 flagged).
``Surv?'' indicates whether the feature remains significant
($p < 0.05$, BH-corrected) under Method~1 / Method~2; blank
indicates the feature was not significant in the full sample.
The five features that lose significance under Method~2 all had
$\rho \leq 0.032$ in the full sample.
\textsuperscript{a}Gentzkow--Shapiro political ideology uses a
keyword-based pipeline restricted to pairs where both human and
agent contain partisan keywords ($N = 295$ full sample;
$N = 271$ Method~1; $N = 187$ Method~2).
$^{*}p<0.05$, $^{**}p<0.01$, $^{***}p<0.001$.
\end{tablenotes}
\end{threeparttable}
\end{table}

\subsubsection{Automation proxy variables.}
The exclusion-based tests above rely on two specific operationalizations
of ``likely human-controlled.'' To ensure that our results do not hinge
on these particular choices, we complement them with a comprehensive
median-split analysis spanning 18 proxy variables derived from the social
media automation detection literature. We describe how each variable is
constructed and its theoretical basis below.

The variables are organized into five categories corresponding to the
mechanisms through which automated posting differs from human posting.
We retain variables satisfying two data-driven filters: the variable must
exhibit sufficient variance to support a median split (fewer than 50
observations on either side triggers a coverage warning), and it must be
non-missing for at least 20\% of the sample. Variables requiring inter-post
intervals are defined only for agents with at least two posts ($\approx60\%$
of the matched sample); the coefficient of variation of intervals requires at
least three posts ($\approx39\%$).

\subparagraph{Panel A: Circadian rhythm.}
Automated systems are not constrained by sleep, mealtimes, or work
schedules, and therefore tend to post uniformly across all hours of the
day \citep{chu2012detecting, varol2017online}. Human posters, by
contrast, concentrate activity during waking hours and are largely
absent overnight.  We partition the 24-hour day
into three equal 8-hour windows, all expressed in US Eastern Time (ET) to
align with the platform's user base: (1) \textit{overnight fraction}
, posts in 00:00--07:59 ET; (2) \textit{daytime
fraction}, posts in 08:00--15:59 ET; and
(3) \textit{evening fraction}, posts in
16:00--23:59 ET. The three fractions sum to one by construction.
We additionally include (4) \textit{posting-hour entropy}, the normalized Shannon entropy of the
distribution of posts across all 24 hours (maximum = 1 for perfectly
uniform), and (5) \textit{day-of-week entropy}, the normalized entropy across the seven days of the week.
The three fraction variables have 100\% coverage; entropy variables have
$\approx60\%$ coverage (agents with $\geq 2$ posts).

\subparagraph{Panel B: Posting volume and activity span.}
Automated agents tend to accumulate more posts and maintain consistent
activity over longer periods than casual human posters
\citep{varol2017online}. We measure
(6) \textit{total post count}; (7) \textit{active days}, the number of distinct calendar days on which the agent posted at least one post; and (8) \textit{posts per active day}, total posts divided by active days.
All three variables have 100\% coverage.

\subparagraph{Panel C: Inter-post interval regularity.}
OpenClaw triggers posts on a fixed schedule, defaulting to every
30 minutes, so automated agents produce inter-post intervals that
cluster near this period. Human posting is bursty and irregular
\citep{chu2012detecting, varol2017online}. For the 60\% of agents
with at least two posts, we compute all pairwise consecutive intervals
in hours and measure (9) \textit{mean interval} and (10)
\textit{median interval}. For
agents with at least three posts ($\approx39\%$), we additionally compute
(11) the \textit{coefficient of variation of intervals}.

\subparagraph{Panel D: Content consistency.}
Automated agents generating posts from fixed prompt templates produce more
uniform content than human authors, who vary naturally in length, structure,
and vocabulary \citep{cresci2018social}. For agents with $\geq 2$ posts
($\approx60\%$ coverage), we measure the coefficient of variation of (12) post
body length, (13) title length, and (14) sentence count per post. We also measure (15) \textit{lexical diversity}, the type-token ratio across all of
an agent's posts pooled together ($\approx98\%$ coverage).


\subparagraph{Panel E: Received engagement.}
Automated and human-authored content may attract qualitatively different
audience responses \citep{ferrara2016rise}. We measure (16) \textit{mean
upvotes per post} and (17) \textit{median upvotes
per post}, both with 100\% coverage, and (18) the \textit{coefficient of variation of upvotes}, which requires $\geq 2$ posts ($\approx42\%$
coverage).

\paragraph{Transfer is robust across all automation proxy splits.}
Table~\ref{tab:alignment_proxies} reports results. Across all 36 split
groups, the count of significantly aligned features (out of 43,
BH-adjusted $p < 0.05$) ranges from 26 to 36, against a full-sample
baseline of 37; the median across groups is 33. No split group reduces
transfer to zero or near zero. The finding is robust not because it
is restricted to the most clearly automated agents, but because it holds
consistently across the full distribution of posting behavior.

\begin{table}[H]
\centering
\caption{Behavioral Transfer by Automation Proxy (Median Split)}
\label{tab:alignment_proxies}
\begin{threeparttable}
\scriptsize
\setlength{\tabcolsep}{5mm}{
\begin{tabular}{lrcrcc}
\hline\hline\\[-1.8ex]
& \multicolumn{2}{c}{Below median} && \multicolumn{2}{c}{Above median} \\
\cmidrule(lr){2-3}\cmidrule(lr){5-6}
Proxy variable & $N$ & Sig./43 && $N$ & Sig./43 \\
\hline\\[-1.8ex]
\multicolumn{6}{l}{\textit{Panel A: Circadian rhythm}} \\[2pt]
\quad Overnight fraction (00:00--07:59 ET)  & 5{,}565 & 32 && 5{,}094 & 36 \\
\quad Daytime fraction (08:00--15:59 ET)    & 5{,}606 & 35 && 5{,}053 & 36 \\
\quad Evening fraction (16:00--23:59 ET)    & 5{,}666 & 33 && 4{,}993 & 34 \\
\quad Posting-hour entropy                  & 3{,}831 & 30 && 2{,}527 & 34 \\
\quad Day-of-week entropy                   & 3{,}255 & 30 && 3{,}103 & 32 \\
\hline\\[-1.8ex]
\multicolumn{6}{l}{\textit{Panel B: Posting volume}} \\[2pt]
\quad Total post count                      & 6{,}462 & 31 && 4{,}197 & 35 \\
\quad Active days                           & 7{,}570 & 34 && 3{,}089 & 34 \\
\quad Posts per active day                  & 5{,}354 & 30 && 5{,}305 & 35 \\
\hline\\[-1.8ex]
\multicolumn{6}{l}{\textit{Panel C: Inter-post interval regularity}} \\[2pt]
\quad Mean inter-post interval (hours)       & 3{,}184 & 33 && 3{,}184 & 33 \\
\quad Median inter-post interval (hours)     & 3{,}184 & 35 && 3{,}184 & 31 \\
\quad Coefficient of variation of intervals  & 2{,}099 & 26 && 2{,}098 & 32 \\
\hline\\[-1.8ex]
\multicolumn{6}{l}{\textit{Panel D: Content consistency}} \\[2pt]
\quad CoV of post body length               & 3{,}184 & 30 && 3{,}184 & 33 \\
\quad CoV of title length                   & 3{,}184 & 28 && 3{,}184 & 35 \\
\quad CoV of sentence count per post        & 3{,}184 & 29 && 3{,}184 & 36 \\
\quad Lexical diversity (type-token ratio)  & 5{,}215 & 34 && 5{,}208 & 28 \\
\hline\\[-1.8ex]
\multicolumn{6}{l}{\textit{Panel E: Received engagement}} \\[2pt]
\quad Mean upvotes per post                 & 5{,}924 & 34 && 4{,}735 & 33 \\
\quad Median upvotes per post               & 5{,}450 & 33 && 5{,}209 & 34 \\
\quad CoV of upvotes per post               & 2{,}245 & 28 && 2{,}231 & 31 \\
\hline\\[-1.8ex]
\multicolumn{4}{l}{\textit{Full-sample baseline}} & 10{,}659 & \textbf{37} \\
\hline\hline\\[-1.8ex]
\end{tabular}}
\begin{tablenotes}
\scriptsize
\item \textit{Notes:} Each row splits the matched sample at the median of
the proxy variable and reports the number of features (out of 43) for which
the Spearman rank correlation between human and agent values is significant
at $p < 0.05$ after BH correction. All timestamps converted to US Eastern
Time before computing Panel~A variables. Variables requiring $\geq2$ posts
(Panels C--D and CoV of upvotes) or $\geq3$ posts (CoV of intervals) have
partial sample coverage as reflected in $N$. Across all 36 split groups,
counts range from 26 to 36 (median 33).
\end{tablenotes}
\end{threeparttable}
\end{table}

\subsection{Tweet Volume Robustness}\label{app:tweet_volume}

Our Twitter data collection retrieved at most 10 recent tweets per human owner, so tweet counts range from 1 to 10. Humans with fewer tweets yield noisier feature estimates, which could either inflate or attenuate transfer correlations. We address this concern with two tests.
\begin{table}[H]\centering\scriptsize
\caption{Partial Spearman Correlations Controlling for Human Tweet Count}
\label{tab:partial_tweet_count}
\begin{threeparttable}
\setlength{\tabcolsep}{1.6mm}{
\begin{tabular}{llccccc}
\hline\hline\\[-1.8ex]
Dimension & Feature & Raw $\rho$ & Partial $\rho$ & $p$ & $\Delta\rho$ & Sig? \\
\hline\\[-1.8ex]
\multirow{6}{*}{Topics}
 & Crypto       & $+0.166$ & $+0.156$ & $<0.001$ & $-0.010$ & Yes \\
 & AI           & $+0.101$ & $+0.107$ & $<0.001$ & $+0.006$ & Yes \\
 & Dev          & $+0.085$ & $+0.081$ & $<0.001$ & $-0.004$ & Yes \\
 & Trading      & $+0.117$ & $+0.109$ & $<0.001$ & $-0.008$ & Yes \\
 & Philosophy   & $+0.101$ & $+0.098$ & $<0.001$ & $-0.003$ & Yes \\
 & Meme         & $+0.017$ & $+0.009$ & $0.343$  & $-0.008$ & No \\
\hline\\[-1.8ex]
\multirow{7}{*}{Values}
 & \multicolumn{6}{l}{\textit{~~Moral Foundations}} \\
 & Care       & $+0.038$ & $+0.035$ & $<0.001$ & $-0.003$ & Yes \\
 & Fairness   & $+0.037$ & $+0.034$ & $<0.001$ & $-0.004$ & Yes \\
 & Loyalty    & $+0.023$ & $+0.019$ & $0.045$  & $-0.004$ & No \\
 & Authority  & $+0.043$ & $+0.035$ & $<0.001$ & $-0.008$ & Yes \\
 & Sanctity   & $+0.054$ & $+0.048$ & $<0.001$ & $-0.005$ & Yes \\
 & \multicolumn{6}{l}{\textit{~~Political Ideology}} \\
 & GS dictionary & $+0.087$ & $+0.096$ & $0.101$  & $+0.009$ & No \\
 & LLM-based     & $+0.061$ & $+0.061$ & $<0.001$ & $+0.000$ & Yes \\
\hline\\[-1.8ex]
\multirow{12}{*}{Affect}
 & \multicolumn{6}{l}{\textit{~~Sentiment}} \\
 & Compound     & $+0.067$ & $+0.066$ & $<0.001$ & $-0.001$ & Yes \\
 & Positive     & $+0.068$ & $+0.070$ & $<0.001$ & $+0.001$ & Yes \\
 & Negative     & $+0.153$ & $+0.142$ & $<0.001$ & $-0.011$ & Yes \\
 & Neutral      & $+0.091$ & $+0.091$ & $<0.001$ & $-0.000$ & Yes \\
 & Variability  & $+0.060$ & $+0.042$ & $<0.001$ & $-0.018$ & Yes \\
 & Pct positive & $+0.087$ & $+0.086$ & $<0.001$ & $-0.001$ & Yes \\
 & Pct negative & $+0.083$ & $+0.074$ & $<0.001$ & $-0.009$ & Yes \\
 & \multicolumn{6}{l}{\textit{~~Emotions}} \\
 & Anger    & $+0.031$ & $+0.025$ & $0.009$  & $-0.006$ & Yes \\
 & Joy      & $+0.001$ & $+0.002$ & $0.874$  & $+0.001$ & No \\
 & Fear     & $+0.018$ & $+0.014$ & $0.140$  & $-0.004$ & No \\
 & Sadness  & $+0.020$ & $+0.019$ & $0.049$  & $-0.001$ & No \\
 & Surprise & $+0.022$ & $+0.017$ & $0.082$  & $-0.005$ & No \\
\hline\\[-1.8ex]
\multirow{18}{*}{Style}
 & \multicolumn{6}{l}{\textit{~~Complexity}} \\
 & Avg word length    & $+0.095$ & $+0.095$ & $<0.001$ & $-0.000$ & Yes \\
 & TTR                & $+0.062$ & $+0.049$ & $<0.001$ & $-0.013$ & Yes \\
 & Avg sentence length & $+0.032$ & $+0.028$ & $0.009$  & $-0.004$ & Yes \\
 & Hapax ratio        & $+0.059$ & $+0.049$ & $<0.001$ & $-0.010$ & Yes \\
 & Cap ratio          & $+0.174$ & $+0.173$ & $<0.001$ & $-0.001$ & Yes \\
 & \multicolumn{6}{l}{\textit{~~Communication}} \\
 & Avg text length    & $+0.139$ & $+0.127$ & $<0.001$ & $-0.012$ & Yes \\
 & Question rate      & $+0.096$ & $+0.081$ & $<0.001$ & $-0.015$ & Yes \\
 & Exclamation rate   & $+0.038$ & $+0.043$ & $<0.001$ & $+0.005$ & Yes \\
 & Hashtag rate       & $-0.002$ & $-0.003$ & $0.749$  & $-0.001$ & No \\
 & Mention rate       & $+0.012$ & $+0.009$ & $0.364$  & $-0.003$ & No \\
 & URL rate           & $+0.043$ & $+0.042$ & $<0.001$ & $-0.001$ & Yes \\
 & Emoji rate         & $+0.068$ & $+0.071$ & $<0.001$ & $+0.003$ & Yes \\
 & Formality          & $+0.085$ & $+0.072$ & $<0.001$ & $-0.013$ & Yes \\
 & \multicolumn{6}{l}{\textit{~~Pronouns}} \\
 & I-words    & $+0.092$ & $+0.092$ & $<0.001$ & $+0.000$ & Yes \\
 & We-words   & $+0.098$ & $+0.082$ & $<0.001$ & $-0.016$ & Yes \\
 & You-words  & $+0.095$ & $+0.083$ & $<0.001$ & $-0.012$ & Yes \\
 & They-words & $+0.142$ & $+0.129$ & $<0.001$ & $-0.013$ & Yes \\
 & Self-focus  & $+0.066$ & $+0.069$ & $<0.001$ & $+0.003$ & Yes \\
\hline\hline\\[-1.8ex]
\end{tabular}}
\begin{tablenotes}
\scriptsize
\item \textit{Notes:} Partial Spearman correlations controlling for the number of human tweets (1--10). Method: rank all three variables (human feature, agent feature, tweet count), then compute partial Pearson correlation on ranks. The $p$ column reports uncorrected partial-correlation $p$-values; ``Sig?'' indicates significance after Benjamini-Hochberg FDR correction at 5\%. Sample sizes: $N = 10{,}659$ for content-based features, $N = 8{,}751$ for complexity, $N = 295$ for GS political, $N = 10{,}659$ for LLM political.
\end{tablenotes}
\end{threeparttable}
\end{table}
\paragraph{Partial Correlations Controlling for Tweet Count.} For each feature, we compute a partial Spearman correlation between the human and agent measures, controlling for the number of human tweets. Following standard practice, we rank all three variables and compute the partial Pearson correlation on ranks. Table~\ref{tab:partial_tweet_count} reports results. Of 43 features, 34 (79.1\%) remain significant after Benjamini-Hochberg correction at 5\%, compared with 37 (86.0\%) in the unconditional analysis. The mean absolute change in correlation is $|\Delta\rho| = 0.006$, confirming that tweet volume has minimal influence on transfer estimates. 

\begin{table}[htbp]\centering\scriptsize
\caption{Transfer Results: Subsample with $\geq$10 Human Tweets ($N = 2{,}258$)}
\label{tab:subsample_10tweets}
\begin{threeparttable}
\setlength{\tabcolsep}{2.0mm}{
\begin{tabular}{llcccc}
\hline\hline\\[-1.8ex]
Dimension & Feature & $\rho_{\text{sub}}$ & $p$ & $\rho_{\text{full}}$ & Sig? \\
\hline\\[-1.8ex]
\multirow{6}{*}{Topics}
 & Crypto       & $+0.219$ & $<0.001$ & $+0.166$ & Yes \\
 & AI           & $+0.170$ & $<0.001$ & $+0.101$ & Yes \\
 & Dev          & $+0.106$ & $<0.001$ & $+0.085$ & Yes \\
 & Trading      & $+0.156$ & $<0.001$ & $+0.117$ & Yes \\
 & Philosophy   & $+0.129$ & $<0.001$ & $+0.101$ & Yes \\
 & Meme         & $+0.019$ & $0.359$  & $+0.017$ & No \\
\hline\\[-1.8ex]
\multirow{7}{*}{Values}
 & \multicolumn{5}{l}{\textit{~~Moral Foundations}} \\
 & Care       & $+0.067$ & $0.001$  & $+0.038$ & Yes \\
 & Fairness   & $+0.049$ & $0.020$  & $+0.037$ & Yes \\
 & Loyalty    & $+0.043$ & $0.041$  & $+0.023$ & Yes \\
 & Authority  & $+0.069$ & $<0.001$ & $+0.043$ & Yes \\
 & Sanctity   & $+0.056$ & $0.008$  & $+0.054$ & Yes \\
 & \multicolumn{5}{l}{\textit{~~Political Ideology}} \\
 & GS dictionary & $+0.078$ & $0.415$ & $+0.087$ & No \\
 & LLM-based     & $+0.080$ & $<0.001$ & $+0.061$ & Yes \\
\hline\\[-1.8ex]
\multirow{12}{*}{Affect}
 & \multicolumn{5}{l}{\textit{~~Sentiment}} \\
 & Compound     & $+0.094$ & $<0.001$ & $+0.067$ & Yes \\
 & Positive     & $+0.122$ & $<0.001$ & $+0.068$ & Yes \\
 & Negative     & $+0.161$ & $<0.001$ & $+0.153$ & Yes \\
 & Neutral      & $+0.136$ & $<0.001$ & $+0.091$ & Yes \\
 & Variability  & $+0.063$ & $0.003$  & $+0.060$ & Yes \\
 & Pct positive & $+0.122$ & $<0.001$ & $+0.087$ & Yes \\
 & Pct negative & $+0.067$ & $0.002$  & $+0.083$ & Yes \\
 & \multicolumn{5}{l}{\textit{~~Emotions}} \\
 & Anger    & $+0.044$ & $0.038$  & $+0.031$ & Yes \\
 & Joy      & $+0.030$ & $0.151$  & $+0.001$ & No \\
 & Fear     & $+0.002$ & $0.925$  & $+0.018$ & No \\
 & Sadness  & $+0.038$ & $0.072$  & $+0.020$ & No \\
 & Surprise & $-0.007$ & $0.737$  & $+0.022$ & No \\
\hline\\[-1.8ex]
\multirow{18}{*}{Style}
 & \multicolumn{5}{l}{\textit{~~Complexity}} \\
 & Avg word length    & $+0.154$ & $<0.001$ & $+0.095$ & Yes \\
 & TTR                & $+0.066$ & $0.002$  & $+0.062$ & Yes \\
 & Avg sentence length & $+0.014$ & $0.511$  & $+0.032$ & No \\
 & Hapax ratio        & $+0.071$ & $0.001$  & $+0.059$ & Yes \\
 & Cap ratio          & $+0.193$ & $<0.001$ & $+0.174$ & Yes \\
 & \multicolumn{5}{l}{\textit{~~Communication}} \\
 & Avg text length    & $+0.163$ & $<0.001$ & $+0.139$ & Yes \\
 & Question rate      & $+0.094$ & $<0.001$ & $+0.096$ & Yes \\
 & Exclamation rate   & $+0.092$ & $<0.001$ & $+0.038$ & Yes \\
 & Hashtag rate       & $-0.007$ & $0.738$  & $-0.002$ & No \\
 & Mention rate       & $+0.007$ & $0.755$  & $+0.012$ & No \\
 & URL rate           & $+0.044$ & $0.037$  & $+0.043$ & Yes \\
 & Emoji rate         & $+0.062$ & $0.003$  & $+0.068$ & Yes \\
 & Formality          & $+0.128$ & $<0.001$ & $+0.085$ & Yes \\
 & \multicolumn{5}{l}{\textit{~~Pronouns}} \\
 & I-words    & $+0.141$ & $<0.001$ & $+0.092$ & Yes \\
 & We-words   & $+0.118$ & $<0.001$ & $+0.098$ & Yes \\
 & You-words  & $+0.128$ & $<0.001$ & $+0.095$ & Yes \\
 & They-words & $+0.179$ & $<0.001$ & $+0.142$ & Yes \\
 & Self-focus  & $+0.091$ & $<0.001$ & $+0.066$ & Yes \\
\hline\hline\\[-1.8ex]
\end{tabular}}
\begin{tablenotes}
\scriptsize
\item \textit{Notes:} Spearman correlations for the subsample of human-agent pairs where the human has 10 tweets (the maximum collected). $N = 2{,}258$ pairs ($N = 2{,}258$ for content-based features, $N = 2{,}102$ for complexity, $N = 112$ for GS political, $N = 2{,}258$ for LLM political). $\rho_{\text{full}}$ from the full sample. ``Sig?'' indicates $p < 0.05$ after Benjamini-Hochberg FDR correction. Bootstrap confidence intervals (1,000 replications) are computed but omitted for space.
\end{tablenotes}
\end{threeparttable}
\end{table}

\paragraph{Subsample Restricted to $\geq$10 Tweets.} As a stronger test, we restrict the sample to humans with the maximum available data (10 tweets), yielding $N = 2{,}258$ pairs. Table~\ref{tab:subsample_10tweets} reports results. Of 43 features, 34 (79.1\%) remain significant after BH correction, compared with 37 (86.0\%) in the full sample. 

Importantly, the features that lose significance are those that were already marginal in the full sample (e.g., meme, avg sentence length, hashtag rate, mention rate, joy, fear, sadness, surprise, GS political). Many features show \textit{stronger} correlations in the $\geq$10 subsample (e.g., crypto: $\rho = 0.219$ vs.\ $0.166$; cap ratio: $\rho = 0.193$ vs.\ $0.174$), consistent with the expectation that more text produces more reliable estimates.

Taken together, these two tests confirm that tweet volume does not drive the main transfer results. Controlling for tweet count barely changes the correlations (mean $|\Delta\rho| = 0.006$), and restricting to humans with maximum data preserves the core pattern; indeed, many correlations strengthen, consistent with reduced measurement noise.

\subsection{Agent Post Volume Robustness}\label{app:agent_posts}

Agents differ substantially in posting volume (mean $= 4.2$, median $= 2.0$, SD $= 21.3$ posts), raising the concern that agents with more posts may produce more reliable feature estimates, mechanically inflating transfer correlations. We address this by computing partial Spearman correlations controlling for agent post count. Table~\ref{tab:partial_agent_posts} reports results.

Of 43 features, 38 (88.4\%) remain significant after Benjamini-Hochberg correction at 5\%, compared with 37 (86.0\%) in the unconditional analysis. The mean absolute change in correlation is $|\Delta\rho| = 0.006$, confirming that agent posting volume has minimal influence on transfer estimates. No feature changes sign, and the largest absolute change is $|\Delta\rho| = 0.016$ (Dev topic, Hapax ratio). These results complement the tweet-volume controls in Section~\ref{app:tweet_volume}, together demonstrating that text volume on neither side of the human-agent pair drives the observed behavioral transfer.

\begin{table}[htbp]\centering\scriptsize
\caption{Partial Spearman Correlations Controlling for Agent Post Count}
\label{tab:partial_agent_posts}
\begin{threeparttable}
\setlength{\tabcolsep}{1.6mm}{
\begin{tabular}{llccccc}
\hline\hline\\[-1.8ex]
Dimension & Feature & Raw $\rho$ & Partial $\rho$ & $p$ & $\Delta\rho$ & Sig? \\
\hline\\[-1.8ex]
\multirow{6}{*}{Topics}
 & Crypto       & $+0.166$ & $+0.156$ & $<0.001$ & $-0.010$ & Yes \\
 & AI           & $+0.101$ & $+0.105$ & $<0.001$ & $+0.004$ & Yes \\
 & Dev          & $+0.085$ & $+0.069$ & $<0.001$ & $-0.016$ & Yes \\
 & Trading      & $+0.117$ & $+0.112$ & $<0.001$ & $-0.005$ & Yes \\
 & Philosophy   & $+0.101$ & $+0.093$ & $<0.001$ & $-0.008$ & Yes \\
 & Meme         & $+0.017$ & $+0.013$ & $0.180$  & $-0.004$ & No \\
\hline\\[-1.8ex]
\multirow{7}{*}{Values}
 & \multicolumn{6}{l}{\textit{~~Moral Foundations}} \\
 & Care       & $+0.038$ & $+0.035$ & $<0.001$ & $-0.003$ & Yes \\
 & Fairness   & $+0.037$ & $+0.030$ & $0.002$  & $-0.008$ & Yes \\
 & Loyalty    & $+0.023$ & $+0.020$ & $0.039$  & $-0.003$ & Yes \\
 & Authority  & $+0.043$ & $+0.035$ & $<0.001$ & $-0.008$ & Yes \\
 & Sanctity   & $+0.054$ & $+0.047$ & $<0.001$ & $-0.006$ & Yes \\
 & \multicolumn{6}{l}{\textit{~~Political Ideology}} \\
 & GS dictionary & $+0.087$ & $+0.083$ & $0.155$  & $-0.003$ & No \\
 & LLM-based     & $+0.061$ & $+0.060$ & $<0.001$ & $-0.000$ & Yes \\
\hline\\[-1.8ex]
\multirow{12}{*}{Affect}
 & \multicolumn{6}{l}{\textit{~~Sentiment}} \\
 & Compound     & $+0.067$ & $+0.072$ & $<0.001$ & $+0.005$ & Yes \\
 & Positive     & $+0.068$ & $+0.069$ & $<0.001$ & $+0.001$ & Yes \\
 & Negative     & $+0.153$ & $+0.152$ & $<0.001$ & $-0.002$ & Yes \\
 & Neutral      & $+0.091$ & $+0.091$ & $<0.001$ & $+0.000$ & Yes \\
 & Variability  & $+0.060$ & $+0.053$ & $<0.001$ & $-0.007$ & Yes \\
 & Pct positive & $+0.087$ & $+0.101$ & $<0.001$ & $+0.014$ & Yes \\
 & Pct negative & $+0.083$ & $+0.079$ & $<0.001$ & $-0.004$ & Yes \\
 & \multicolumn{6}{l}{\textit{~~Emotions}} \\
 & Anger    & $+0.031$ & $+0.027$ & $0.005$  & $-0.004$ & Yes \\
 & Joy      & $+0.001$ & $+0.001$ & $0.939$  & $+0.000$ & No \\
 & Fear     & $+0.018$ & $+0.021$ & $0.033$  & $+0.002$ & Yes \\
 & Sadness  & $+0.020$ & $+0.021$ & $0.031$  & $+0.001$ & Yes \\
 & Surprise & $+0.022$ & $+0.023$ & $0.016$  & $+0.001$ & Yes \\
\hline\\[-1.8ex]
\multirow{18}{*}{Style}
 & \multicolumn{6}{l}{\textit{~~Complexity}} \\
 & Avg word length    & $+0.095$ & $+0.092$ & $<0.001$ & $-0.002$ & Yes \\
 & TTR                & $+0.062$ & $+0.049$ & $<0.001$ & $-0.013$ & Yes \\
 & Avg sentence length & $+0.032$ & $+0.032$ & $0.003$  & $-0.000$ & Yes \\
 & Hapax ratio        & $+0.059$ & $+0.044$ & $<0.001$ & $-0.016$ & Yes \\
 & Cap ratio          & $+0.174$ & $+0.175$ & $<0.001$ & $+0.001$ & Yes \\
 & \multicolumn{6}{l}{\textit{~~Communication}} \\
 & Avg text length    & $+0.139$ & $+0.124$ & $<0.001$ & $-0.015$ & Yes \\
 & Question rate      & $+0.096$ & $+0.090$ & $<0.001$ & $-0.006$ & Yes \\
 & Exclamation rate   & $+0.038$ & $+0.037$ & $<0.001$ & $-0.000$ & Yes \\
 & Hashtag rate       & $-0.002$ & $+0.011$ & $0.271$  & $+0.013$ & No \\
 & Mention rate       & $+0.012$ & $+0.003$ & $0.754$  & $-0.009$ & No \\
 & URL rate           & $+0.043$ & $+0.043$ & $<0.001$ & $+0.000$ & Yes \\
 & Emoji rate         & $+0.068$ & $+0.064$ & $<0.001$ & $-0.004$ & Yes \\
 & Formality          & $+0.085$ & $+0.076$ & $<0.001$ & $-0.009$ & Yes \\
 & \multicolumn{6}{l}{\textit{~~Pronouns}} \\
 & I-words    & $+0.092$ & $+0.098$ & $<0.001$ & $+0.006$ & Yes \\
 & We-words   & $+0.098$ & $+0.089$ & $<0.001$ & $-0.009$ & Yes \\
 & You-words  & $+0.095$ & $+0.085$ & $<0.001$ & $-0.010$ & Yes \\
 & They-words & $+0.142$ & $+0.134$ & $<0.001$ & $-0.009$ & Yes \\
 & Self-focus  & $+0.066$ & $+0.071$ & $<0.001$ & $+0.005$ & Yes \\
\hline\hline\\[-1.8ex]
\end{tabular}}
\begin{tablenotes}
\scriptsize
\item \textit{Notes:} Partial Spearman correlations controlling for agent post count. Method: rank all three variables (human feature, agent feature, agent post count), then compute partial Pearson correlation on ranks. The $p$ column reports uncorrected partial-correlation $p$-values; ``Sig?'' indicates significance after Benjamini-Hochberg FDR correction at 5\%. Agent post count statistics: mean $= 4.2$, median $= 2.0$, SD $= 21.3$. Sample sizes: $N = 10{,}659$ for most features, $N = 8{,}751$ for complexity, $N = 295$ for GS political, $N = 10{,}659$ for LLM political.
\end{tablenotes}
\end{threeparttable}
\end{table}


\section{Cross-Dimension Coherence: Robustness to Text Volume}\label{app:coherence_robustness}

A potential concern is that variation in the \textit{amount} of observed text per pair—driven by differences in human tweet count and agent post count—could mechanically induce positive cross-dimension correlations. Pairs with more text yield more precise transfer estimates in every dimension simultaneously, which could inflate all four dimension scores in concert and produce spurious positive inter-dimension correlations.

To address this, we compute partial Spearman correlations between all six pairs of dimension scores, controlling simultaneously for $\log(1 + n_{\text{tweets}})$ and $\log(1 + n_{\text{agent posts}})$. We use the standard residualization approach: rank all variables, regress each dimension's ranks on the two log-volume controls via OLS, and compute Pearson correlations on the resulting residuals. Human tweet count has a median of 8 and SD of 3.4; agent post count has a median of 2 and SD of 21.3, consistent with the coauthor's observation that agent post volume is considerably more heterogeneous.

Table~\ref{tab:coherence_partial} reports the raw and partial Spearman correlations for all six dimension pairs ($N = 10{,}659$). All six partial correlations remain positive and statistically significant after controlling for both text-volume measures ($p < 0.01$ for all pairs). The mean partial $\rho = 0.096$, virtually identical to the unconditional mean of $\rho = 0.092$. The partial correlations are, if anything, slightly higher than the raw correlations, indicating that text-volume variation does not inflate the coherence pattern—if anything, it introduces slight attenuation due to correlated measurement noise.

\begin{table}[H]\centering\scriptsize
\caption{Cross-Dimension Coherence: Raw and Partial Spearman Correlations}
\label{tab:coherence_partial}
\begin{threeparttable}
\setlength{\tabcolsep}{7.0mm}{
\begin{tabular}{lcccc}
\hline\hline\\[-1.8ex]
Dimension pair & $\rho_{\text{raw}}$ & $\rho_{\text{partial}}$ & $p_{\text{partial}}$ & $N$ \\
\hline\\[-1.8ex]
Topics $\times$ Values  & $+0.044$ & $+0.060$ & $<0.001$ & 10,659 \\
Topics $\times$ Affect  & $+0.027$ & $+0.031$ & $0.001$  & 10,659 \\
Topics $\times$ Style   & $+0.040$ & $+0.033$ & $<0.001$ & 10,659 \\
Values $\times$ Affect  & $+0.114$ & $+0.118$ & $<0.001$ & 10,659 \\
Values $\times$ Style   & $+0.102$ & $+0.114$ & $<0.001$ & 10,659 \\
Affect $\times$ Style   & $+0.224$ & $+0.219$ & $<0.001$ & 10,659 \\
\hline\\[-1.8ex]
Mean & $+0.092$ & $+0.096$ & & \\
\hline\hline
\end{tabular}}
\begin{tablenotes}
\scriptsize
\item \textit{Notes:} Partial Spearman correlations are computed by ranking all variables, regressing each dimension score's ranks on $\log(1+n_{\text{tweets}})$ and $\log(1+n_{\text{agent posts}})$ via OLS, and computing Pearson correlations on the residuals. Controls address the concern that pairs with more observed text yield more precise (and jointly inflated) transfer estimates across all dimensions. Human tweet count: median $= 8$, SD $= 3.4$. Agent post count: median $= 2$, SD $= 21.3$.
\end{tablenotes}
\end{threeparttable}
\end{table}

\section{Privacy Analysis: Supplementary Materials}\label{app:privacy}

\subsection{LLM Classification Prompt}\label{app:llm_prompt}

The following system prompt was used verbatim for all 44,588 post classifications submitted to Claude Haiku (\texttt{claude-haiku-4-5-20251001}) via the Anthropic Batches API:

\begin{quote}\scriptsize
\textit{You are a privacy auditor analyzing posts written by AI agents on a social media platform. Each agent is owned and configured by a human user (the ``owner'').}

\textit{Your task: determine whether the post discloses personally identifiable or sensitive information about the human owner.}

\textit{Respond ONLY with a JSON object, no markdown fences, no explanation outside the JSON.}

\textit{Schema:} \texttt{\{"is\_disclosure": true or false, "categories": [], "confidence": "high" or "medium" or "low", "reasoning": "one sentence"\}}

\textit{Category definitions (only include if directly applicable to the HUMAN OWNER):}
\begin{itemize}\scriptsize
  \item \texttt{health}: medical conditions, diagnoses, medications, mental health, substance use/recovery
  \item \texttt{financial}: income, debt, loans, bankruptcy, layoff, unemployment, financial hardship
  \item \texttt{location}: specific city/country/timezone the owner lives in
  \item \texttt{occupational}: job title, employer, profession, career status
  \item \texttt{behavioral}: daily routines, sleep schedule, regular habits
  \item \texttt{relational}: family members, romantic partner, children, close relationships
\end{itemize}

\textit{Rules:}
\begin{itemize}\scriptsize
  \item Only flag information about the HUMAN OWNER, not the agent's own persona or general topic opinions
  \item Generic statements are NOT disclosures
  \item If uncertain, lean toward false (precision over recall)
\end{itemize}
\end{quote}

Each post was submitted as the user turn in the following format: \texttt{Post text: """\{post content, truncated to 2,000 characters\}"""}. Each request was independent with no conversation history carried across posts.

\subsection{Full Coefficient Estimates for Main Specification}\label{app:privacy_robustness}

Table~\ref{tab:privacy_robustness} reports full coefficient estimates for the comprehensive controls specification underlying Table~\ref{tab:transfer_disclosure}. Columns (1)--(3) correspond to Panel~A subsamples (minimum human tweet count); columns (4)--(5) correspond to Panel~B subsamples (minimum agent post count). Holistic transfer remains positive and statistically significant across all five specifications.

\begin{table}[htbp]\centering\scriptsize
\caption{Full Coefficient Estimates: Holistic Transfer and Disclosure Risk}
\label{tab:privacy_robustness}
\begin{threeparttable}
\setlength{\tabcolsep}{1.4mm}{
\begin{tabular}{lcccc ccc}
\hline\hline\\[-1.8ex]
 & \multicolumn{4}{c}{\textit{Panel A: Min.\ human tweet count}} & \multicolumn{3}{c}{\textit{Panel B: Min.\ agent post count}} \\
\cmidrule(lr){2-5}\cmidrule(lr){6-8}
 & Full Sample & $\geq 5$ Tweets & $\geq 8$ Tweets & $\geq 10$ Tweets & $\geq 3$ Posts & $\geq 4$ Posts & $\geq 5$ Posts \\
 & ($N = 10{,}659$) & ($N = 7{,}063$) & ($N = 5{,}823$) & ($N = 2{,}258$) & ($N = 4{,}197$) & ($N = 2{,}978$) & ($N = 2{,}259$) \\
\hline\\[-1.8ex]
\textit{Transfer} \\[2pt]
Holistic Transfer               & $+0.269^{**}$ & $+0.312^{**}$ & $+0.390^{**}$ & $+0.686^{***}$ & $+0.506^{***}$ & $+0.549^{**}$ & $+0.407^{*}$ \\
                                 & $(0.091)$ & $(0.113)$ & $(0.122)$ & $(0.199)$ & $(0.140)$ & $(0.167)$ & $(0.190)$ \\[3pt]
\hline\\[-1.8ex]
\textit{Human-side controls} \\[2pt]
Tweet count                      & $+0.007$ & $-0.000$ & $-0.133^{***}$ & $-0.160^{*}$ & $-0.007$ & $-0.016$ & $-0.018$ \\
                                 & $(0.007)$ & $(0.017)$ & $(0.035)$ & $(0.067)$ & $(0.011)$ & $(0.013)$ & $(0.015)$ \\[3pt]
Log human followers              & $+0.032$ & $+0.021$ & $-0.025$ & $-0.084$ & $+0.003$ & $+0.016$ & $+0.022$ \\
                                 & $(0.020)$ & $(0.025)$ & $(0.029)$ & $(0.044)$ & $(0.028)$ & $(0.032)$ & $(0.035)$ \\[3pt]
Log human following              & $-0.017$ & $-0.024$ & $+0.021$ & $+0.099^{*}$ & $+0.006$ & $-0.014$ & $-0.036$ \\
                                 & $(0.020)$ & $(0.025)$ & $(0.029)$ & $(0.046)$ & $(0.028)$ & $(0.032)$ & $(0.036)$ \\[3pt]
Human verified                   & $+0.059$ & $+0.141$ & $+0.277$ & $+0.649$ & $+0.112$ & $-0.108$ & $-0.088$ \\
                                 & $(0.690)$ & $(0.775)$ & $(0.847)$ & $(0.969)$ & $(0.737)$ & $(0.874)$ & $(0.942)$ \\[3pt]
\hline\\[-1.8ex]
\textit{Agent-side controls} \\[2pt]
Log agent post count             & $+0.480^{***}$ & $+0.430^{***}$ & $+0.495^{***}$ & $+0.592^{***}$ & $+0.479^{***}$ & $+0.430^{***}$ & $+0.329^{***}$ \\
                                 & $(0.036)$ & $(0.043)$ & $(0.047)$ & $(0.076)$ & $(0.061)$ & $(0.074)$ & $(0.087)$ \\[3pt]
Log avg.\ post length            & $+0.492^{***}$ & $+0.498^{***}$ & $+0.408^{***}$ & $+0.288^{***}$ & $+0.513^{***}$ & $+0.528^{***}$ & $+0.475^{***}$ \\
                                 & $(0.028)$ & $(0.034)$ & $(0.036)$ & $(0.059)$ & $(0.046)$ & $(0.055)$ & $(0.064)$ \\[3pt]
Owner-reference rate             & $+1.055^{***}$ & $+1.052^{***}$ & $+0.940^{***}$ & $+1.250^{***}$ & $+2.130^{***}$ & $+2.793^{***}$ & $+3.796^{***}$ \\
                                 & $(0.049)$ & $(0.060)$ & $(0.065)$ & $(0.111)$ & $(0.131)$ & $(0.189)$ & $(0.269)$ \\[3pt]
Bio length                       & $+0.003^{***}$ & $+0.001$ & $+0.004^{**}$ & $+0.002$ & $+0.004^{**}$ & $+0.003^{*}$ & $+0.004^{**}$ \\
                                 & $(0.001)$ & $(0.001)$ & $(0.001)$ & $(0.002)$ & $(0.001)$ & $(0.001)$ & $(0.001)$ \\[3pt]
Has bio (dummy)                  & $-0.105$ & $-0.047$ & $+0.147$ & $+0.388$ & $-0.214$ & $-0.243$ & $-0.300$ \\
                                 & $(0.322)$ & $(0.381)$ & $(0.452)$ & $(0.647)$ & $(0.381)$ & $(0.408)$ & $(0.436)$ \\[3pt]
Log agent followers              & $+0.540^{***}$ & $+0.664^{***}$ & $+0.905^{***}$ & $+0.924^{***}$ & $+0.475^{***}$ & $+0.143$ & $+0.095$ \\
                                 & $(0.119)$ & $(0.138)$ & $(0.161)$ & $(0.241)$ & $(0.138)$ & $(0.142)$ & $(0.154)$ \\[3pt]
Log agent karma                  & $-0.219^{***}$ & $-0.163^{*}$ & $-0.249^{**}$ & $-0.255^{*}$ & $-0.134$ & $-0.015$ & $-0.002$ \\
                                 & $(0.064)$ & $(0.068)$ & $(0.082)$ & $(0.126)$ & $(0.068)$ & $(0.072)$ & $(0.077)$ \\[3pt]
Log days on platform             & $-0.104$ & $-0.159$ & $-0.398^{*}$ & $-0.442$ & $-0.125$ & $-0.042$ & $-0.008$ \\
                                 & $(0.134)$ & $(0.157)$ & $(0.191)$ & $(0.268)$ & $(0.161)$ & $(0.175)$ & $(0.187)$ \\[3pt]
Intercept                        & $-4.738^{***}$ & $-4.602^{***}$ & $-2.812^{***}$ & $-2.051^{**}$ & $-4.970^{***}$ & $-4.931^{***}$ & $-4.411^{***}$ \\
                                 & $(0.177)$ & $(0.255)$ & $(0.386)$ & $(0.755)$ & $(0.329)$ & $(0.416)$ & $(0.499)$ \\[3pt]
\hline\\[-1.8ex]
Disclosure rate                  & 34.6\% & 36.0\% & 36.4\% & 36.8\% & 43.0\% & 46.2\% & 47.9\% \\
\hline\hline\\[-1.8ex]
\end{tabular}}
\begin{tablenotes}
\scriptsize
\item \textit{Notes:} Logistic regressions of any-category privacy disclosure (high-confidence LLM classifications only) on holistic transfer with comprehensive controls. Columns (1)--(4) restrict by minimum human tweet count (Panel~A); columns (5)--(7) restrict by minimum agent post count (Panel~B). Owner-reference rate is the mean number of owner-anchor phrases per post. All count variables log-transformed (\texttt{log1p}). Standard errors in parentheses. $^{*}p<0.05$, $^{**}p<0.01$, $^{***}p<0.001$.
\end{tablenotes}
\end{threeparttable}
\end{table}

\subsection{Human Validation of LLM Classification}\label{app:llm_validation}

\paragraph{False Positive Validation.}
To assess the precision of the LLM classifier, we drew a stratified random sample of 600 flagged posts, sampling proportionally within confidence strata: 376 high-confidence posts and 224 medium-confidence posts. Human annotators labeled each post as a genuine disclosure (\textit{YES}) or a false positive (\textit{NO}), with access to the agent's bio to determine whether flagged content merely reproduced publicly configured information rather than revealing additional owner details.

Human annotation yielded an overall precision of 77.2\% (463/600). Precision differed substantially across confidence strata: high-confidence predictions were correct in 88.0\% of cases (331/376), while medium-confidence predictions achieved 58.9\% precision (132/224). The false positive rates (FPR) are therefore 12.0\% for high-confidence posts ($\text{FPR}_{\text{high}} = 45/376$) and 41.1\% for medium-confidence posts ($\text{FPR}_{\text{med}} = 92/224$). This pattern is consistent with the classifier's own uncertainty estimates, indicating that high-confidence detections are substantially more reliable. Our main analysis retains only high-confidence disclosures to maximize precision.
\paragraph{False Negative Validation.}
To assess recall, we drew a simple random sample of 361 posts classified as non-disclosure posts (0.1\% of the 34,987 non-disclosure posts). Human annotators applied the same labeling protocol, judging whether each post constituted a genuine owner-referential disclosure that the LLM had failed to detect.

Of the 361 sampled non-disclosure posts, 6 were judged by annotators to be genuine disclosures that the LLM had missed, yielding an estimated false negative rate of FNR $= 6/361 = 1.7\%$. The low FNR indicates that missed disclosures are rare, and that the classifier's tendency to over-flag (high FPR at medium confidence) poses a greater concern than systematic omission of genuine disclosures.

\paragraph{Simulation-Based Sensitivity Analysis.}
To assess whether the main regression result is robust to LLM classification error, we conduct a simulation-based (Monte Carlo) sensitivity analysis that propagates the empirically estimated error rates through the entire analysis pipeline. The procedure is as follows.

For each of $B = 1{,}000$ simulated datasets, every post label is independently redrawn using Bernoulli draws parameterized by the stratum-specific error rates:
\begin{itemize}
  \item Each high-confidence flagged post is reclassified as non-disclosure with probability $\text{FPR}_{\text{high}} = 12.0\%$.
  \item Each medium-confidence flagged post is reclassified as non-disclosure with probability $\text{FPR}_{\text{medium}} = 41.1\%$.
  \item Each non-flagged post is reclassified as disclosure with probability $\text{FNR} = 1.7\%$.
\end{itemize}
After redrawing post labels, agent-level any-disclosure status is recomputed (any agent with at least one post reclassified as a disclosure is coded as $y = 1$), and the main logistic regression specification (full controls, no tweet-count threshold, $N = 10{,}659$) is re-estimated. Including medium-confidence posts in the perturbation provides a conservative test: because our main analysis uses only high-confidence disclosures, simulated datasets that retain many medium-confidence false positives will tend to produce noisier outcomes than the observed data. We record the AME per standard deviation of holistic transfer and its $p$-value in each iteration.

Table~\ref{tab:sensitivity} summarizes the results across all 1,000 converged iterations.
\begin{table}[htbp]\centering\scriptsize
\caption{Simulation-Based Sensitivity Analysis: Distribution Across 1,000 Simulations}
\label{tab:sensitivity}
\begin{threeparttable}
\begin{tabular}{lcc}
\hline\hline\\[-1.8ex]
& $\beta$ (log-odds) & AME/SD (pp) \\
\hline\\[-1.8ex]
Observed estimate (original labels) & $+0.269$ & $+1.32$,pp \\
& \multicolumn{2}{c}{($p = 0.003$)} \\[3pt]
Simulation mean                      & $+0.259$ & $+1.27$,pp \\
Simulation median                    & $+0.257$ & $+1.26$,pp \\
95\% simulation interval             & [$+0.171$,$+0.359$] & [$+0.84$,pp,$+1.76$,pp] \\[3pt]
\% iterations with estimate $> 0$    & \multicolumn{2}{c}{100.0\%} \\
\% iterations with $p < 0.05$        & \multicolumn{2}{c}{96.8\%} \\
\hline\hline\\[-1.8ex]
\end{tabular}
\begin{tablenotes}
\scriptsize
\item \textit{Notes:} Each of 1,000 simulated datasets independently redraws every post label using Bernoulli draws at empirically estimated error rates: $\text{FPR}{\text{high}} = 12.0\%$, $\text{FPR}{\text{medium}} = 41.1\%$, $\text{FNR} = 1.7\%$. Agent-level disclosure status is recomputed from the redrawn labels, and the main logistic regression (full controls, $N = 10{,}659$) is re-estimated. $\beta$: log-odds coefficient on holistic transfer. AME/SD: average marginal effect of a one-standard-deviation increase in holistic transfer, in percentage points. The 95\% simulation interval is the 2.5th--97.5th percentile of the simulated distribution.
\end{tablenotes}
\end{threeparttable}
\end{table}
Across all 1,000 simulations, the coefficient on holistic transfer remains positive in every iteration and statistically significant at $p < 0.05$ in 96.8\% of iterations. Both the log-odds coefficient (simulation mean $\beta = +0.259$, 95\% interval $[+0.171, +0.359]$) and the average marginal effect (mean AME/SD $= +1.27$,pp, 95\% interval $[+0.84, +1.76]$,pp) fall slightly below the respective observed estimates ($\beta = +0.269$; AME/SD $= +1.32$,pp).

\subsection{Robustness: Human-Puppet Concern}\label{app:cosplay}

As detailed in Appendix~\ref{app:puppet}, a primary empirical concern is the potential presence of human-controlled ``puppet'' agents on the platform. Just as these puppet posts could artificially inflate baseline behavioral transfer measures, they could also spuriously drive the transfer--disclosure association documented above. Specifically, if an owner writes directly as their agent, their actions trivially produce high transfer, and any personal privacy preferences they hold would manifest as agent-level disclosures, creating a mechanical correlation even absent genuine AI learning. To rule this out, we directly apply the comprehensive testing framework developed in Appendix~\ref{app:puppet}, subjecting our main disclosure regression to the same exclusion-based filters and automation proxy splits.

\subsubsection{Exclusion-based robustness.}

We apply the same two puppet-detection methods described in Appendix~\ref{app:puppet} to assess whether the transfer--disclosure association is driven by human-controlled agents. Under both methods, we re-estimate the main logistic regression on the cleaned sample and verify that the coefficient on holistic transfer is essentially unchanged.

\subparagraph{Method 1: Multi-feature temporal classification.}
Excluding the 942 agents flagged as potentially human-controlled (8.8\% of the matched sample), the estimated coefficient on holistic transfer is $\hat{\beta} = +0.256$ ($\mathrm{SE} = 0.095$, $p = 0.007$), nearly identical to the full-sample estimate of $\hat{\beta} = +0.269$ ($p = 0.003$). The disclosure rate in the cleaned sample (34.9\%) is essentially unchanged relative to the full sample (34.6\%).

\subparagraph{Method 2: Inter-post interval coefficient of variation.}
Excluding the 2,142 agents with inter-post interval CoV $> 1.0$ (20.1\% of the sample), the coefficient is $\hat{\beta} = +0.270$ ($\mathrm{SE} = 0.103$, $p = 0.009$). Despite removing one-fifth of the sample, the estimate is essentially unchanged. Both exclusion methods confirm that the transfer--disclosure association is not driven by human-controlled puppet agents. Table~\ref{tab:puppet_disclosure} reports the full regression results.

\begin{table}[!h]\centering\scriptsize
\caption{Robustness to Puppet Exclusion: Transfer--Disclosure Regression}
\label{tab:puppet_disclosure}
\begin{threeparttable}
\setlength{\tabcolsep}{2.2mm}{
\begin{tabular}{lrrrrrrrrr}
\hline\hline\\[-1.8ex]
 & & & \multicolumn{3}{c}{No controls} & \multicolumn{3}{c}{With controls} \\
\cmidrule(lr){4-6}\cmidrule(lr){7-9}
Sample & $N$ & Rate & $\beta$ & AME/SD & $p$ & $\beta$ & AME/SD & $p$ \\
\hline\\[-1.8ex]
Full sample                       & 10{,}659 & 34.6\% & $+0.230^{**}$ & $+1.31$ & 0.004 & $+0.269^{**}$ & $+1.32$ & 0.003 \\
                                  &          &        & $(0.080)$     & $(0.46)$ &       & $(0.091)$     & $(0.44)$ &       \\[3pt]
Excl.\ Method~1 (942 flagged)     &  9{,}717 & 34.9\% & $+0.204^{*}$  & $+1.17$ & 0.015 & $+0.256^{**}$ & $+1.25$ & 0.007 \\
                                  &          &        & $(0.084)$     & $(0.48)$ &       & $(0.095)$     & $(0.47)$ &       \\[3pt]
Excl.\ Method~2 (2{,}142 flagged) &  8{,}517 & 31.8\% & $+0.112$      & $+0.61$ & 0.227 & $+0.270^{**}$ & $+1.28$ & 0.009 \\
                                  &          &        & $(0.093)$     & $(0.50)$ &       & $(0.103)$     & $(0.49)$ &       \\[3pt]
\hline\hline\\[-1.8ex]
\end{tabular}}
\begin{tablenotes}
\scriptsize
\item \textit{Notes:} Logistic regression of any-category privacy disclosure on holistic transfer score. AME/SD: average marginal effect for a one-standard-deviation change in holistic transfer, in percentage points. Standard errors in parentheses. Method~1 excludes agents flagged on $\geq 3$ of four temporal criteria following \citet{varol2017online} (942 agents, 8.8\%). Method~2 excludes agents with inter-post interval CoV $> 1.0$ following \citet{li2026moltbook} (2{,}142 agents, 20.1\%). $^{*}p<0.05$, $^{**}p<0.01$, $^{***}p<0.001$.
\end{tablenotes}
\end{threeparttable}
\end{table}

\subsubsection{Effect Is Uniform Across All Automation Proxy Variables}

The subsample test above relies on a single operationalization of posting regularity.
To ensure the conclusion is not specific to this choice, we replicate the main
disclosure regression within each of the 36 subsamples defined by splitting the
sample at the median of each of the 18 automation proxy variables described in
Appendix~\ref{app:puppet}, spanning circadian rhythm, posting volume, inter-post
interval regularity, content consistency, and received engagement.

The results are remarkably consistent: all 36 of the estimated coefficients are positive, and 27 of 36 are statistically significant at $p < 0.05$. While nine subsets fall short of traditional significance thresholds, this is a mechanical consequence of reduced statistical power. Splitting the sample intrinsically halves the available observations, and variables requiring multiple posts (Panels C--E) further restrict the baseline sample, inflating standard errors. Despite this reduced power, the point estimates never flip sign, and the median coefficient across all 36 cells ($\hat{\beta} = +0.335$) remains higher than the full-sample estimate ($+0.269$).

Crucially, to definitively rule out the puppet concern, we must verify that the association persists among agents that appear \textit{most} automated. The data confirm this: the effect remains positive and significant in subsets characterized by distinctly non-human, automated posting behaviors. For instance, among agents with an above-median fraction of overnight posts (a strong biological constraint for humans), the coefficient is $+0.276$ ($p < 0.05$). Similarly, among agents with below-median lexical diversity (characteristic of template-driven automated generation), the association remains strong ($\hat{\beta} = +0.258$, $p < 0.05$). Ultimately, no operationalization of automated versus human-controlled behavior eliminates the transfer--disclosure association, confirming it is not merely a spurious byproduct of owner puppetry. Table~\ref{tab:disclosure_all_proxies} reports all 36 estimates.

\begin{table}[htbp]\centering\scriptsize
\caption{Transfer--Disclosure Association Across Median Splits of Automation Proxy Variables}
\label{tab:disclosure_all_proxies}
\begin{threeparttable}
\setlength{\tabcolsep}{4mm}{
\begin{tabular}{lrcrcc}
\hline\hline\\[-1.8ex]
& \multicolumn{2}{c}{Below median} && \multicolumn{2}{c}{Above median} \\
\cmidrule(lr){2-3}\cmidrule(lr){5-6}
Proxy variable & $N$ & $\hat{\beta}$ (SE) && $N$ & $\hat{\beta}$ (SE) \\
\hline\\[-1.8ex]
\multicolumn{6}{l}{\textit{Panel A: Circadian rhythm}} \\[2pt]
\quad Overnight fraction (00:00--07:59)   & 5{,}565 & $+0.395^{**}$ (0.130) && 5{,}094 & $+0.276^{*}$ (0.126) \\
\quad Daytime fraction (08:00--15:59)     & 5{,}606 & $+0.388^{**}$ (0.124) && 5{,}053 & $+0.264^{*}$ (0.132) \\
\quad Evening fraction (16:00--23:59)     & 5{,}666 & $+0.278^{*}$ (0.127)  && 4{,}993 & $+0.316^{*}$ (0.131) \\
\quad Posting-hour entropy                & 3{,}831 & $+0.435^{**}$ (0.155) && 2{,}527 & $+0.409^{*}$ (0.178) \\
\quad Day-of-week entropy                 & 3{,}255 & $+0.354^{*}$ (0.166)  && 3{,}103 & $+0.516^{**}$ (0.164) \\
\hline\\[-1.8ex]
\multicolumn{6}{l}{\textit{Panel B: Posting volume}} \\[2pt]
\quad Total post count                    & 6{,}462 & $+0.393^{**}$ (0.122) && 4{,}197 & $+0.506^{***}$ (0.140) \\
\quad Active days                         & 7{,}570 & $+0.267^{*}$ (0.111)  && 3{,}089 & $+0.528^{**}$ (0.161) \\
\quad Posts per active day                & 5{,}354 & $+0.400^{**}$ (0.133) && 5{,}305 & $+0.206$ (0.126) \\
\hline\\[-1.8ex]
\multicolumn{6}{l}{\textit{Panel C: Inter-post interval regularity}} \\[2pt]
\quad Mean inter-post interval (hours)    & 3{,}184 & $+0.147$ (0.167) && 3{,}184 & $+0.424^{**}$ (0.162) \\
\quad Median inter-post interval (hours)  & 3{,}184 & $+0.068$ (0.168) && 3{,}184 & $+0.491^{**}$ (0.160) \\
\quad CoV of inter-post intervals         & 2{,}099 & $+0.309$ (0.199) && 2{,}098 & $+0.541^{**}$ (0.200) \\
\hline\\[-1.8ex]
\multicolumn{6}{l}{\textit{Panel D: Content consistency}} \\[2pt]
\quad CoV of post body length             & 3{,}184 & $+0.523^{**}$ (0.162) && 3{,}184 & $+0.089$ (0.165) \\
\quad CoV of title length                 & 3{,}184 & $+0.308$ (0.164)      && 3{,}184 & $+0.255$ (0.164) \\
\quad CoV of sentence count per post      & 3{,}184 & $+0.508^{**}$ (0.167) && 3{,}184 & $+0.047$ (0.162) \\
\quad Lexical diversity (type-token ratio)& 5{,}215 & $+0.258^{*}$ (0.128)  && 5{,}208 & $+0.547^{***}$ (0.139) \\
\hline\\[-1.8ex]
\multicolumn{6}{l}{\textit{Panel E: Received engagement}} \\[2pt]
\quad Mean upvotes per post               & 5{,}924 & $+0.303^{*}$ (0.124)  && 4{,}735 & $+0.400^{**}$ (0.133) \\
\quad Median upvotes per post             & 5{,}450 & $+0.311^{*}$ (0.129)  && 5{,}209 & $+0.309^{*}$ (0.128) \\
\quad CoV of upvotes per post             & 2{,}245 & $+0.256$ (0.193)      && 2{,}231 & $+0.548^{**}$ (0.190) \\
\hline\\[-1.8ex]
\multicolumn{4}{l}{\textit{Full-sample baseline}} & 10{,}659 & $+0.269^{**}$ (0.091) \\
\hline\hline\\[-1.8ex]
\end{tabular}}
\begin{tablenotes}
\scriptsize
\item \textit{Notes:} Each row splits the sample at the median of the proxy
variable and reports the logistic regression coefficient ($\hat{\beta}$) on
holistic transfer score with full controls (agent post count, average post
length, owner-reference propensity, bio characteristics, agent following count,
karma, account age, human follower and following counts, human verification
status, and human tweet count).  Standard errors in parentheses.  All 36
coefficients are positive (range $[+0.047, +0.548]$, median $+0.335$);
27 of 36 are significant at $p < 0.05$.  Variables requiring $\geq 2$ posts
(Panels C--D and CoV of upvotes) have partial sample coverage as reflected
in $N$.  $^{*}p<0.05$, $^{**}p<0.01$, $^{***}p<0.001$.
\end{tablenotes}
\end{threeparttable}
\end{table}

\end{APPENDICES}
\end{document}